\documentclass{aa}  

\usepackage{graphicx}
\usepackage[svgnames]{xcolor}
\usepackage[pdftex,colorlinks,citecolor=DarkBlue]{hyperref}
\usepackage{txfonts}
\usepackage{mathrsfs}
\usepackage{diagbox}
\bibpunct{(}{)}{;}{a}{}{,}
\usepackage[switch]{lineno}
\usepackage{orcidlink}

\begin{document} 

\title{An improved view of cosmic-ray transport and the galactic outflow in NGC 253}


\titlerunning{Cosmic-ray transport in NGC~253}

\author{Shengtao Wang\orcidlink{0009-0003-9052-1976}\inst{1}
    \and
        George Heald\orcidlink{0000-0002-2155-6054}\inst{2,3} 
    \and
        Stefan W.~Duchesne\orcidlink{0000-0002-3846-0315}\inst{3}
    \and
        Xiaohui Sun\orcidlink{0000-0002-3464-5128}\inst{1} 
    \and
       Guangxing Li\orcidlink{0000-0003-3144-1952}\inst{4}
    \and
       Jiangtao Li\orcidlink{0000-0001-6239-3821}\inst{5}
    \and
       Chao-Wei Tsai\orcidlink{0000-0002-9390-9672}\inst{6,7,8}
    \and
       Andrew J. Battisti\orcidlink{0000-0003-4569-2285}\inst{9,10,11}
    \and
       Mark Seibert\orcidlink{0000-0002-1143-5515}\inst{12}
    \and 
       Kathryn Grasha\orcidlink{0000-0002-3247-5321}\inst{10,11,13}
    \and
       Jeff A. Rich\orcidlink{0000-0002-5807-5078}\inst{12}
    \and
       Rachael L. Beaton\orcidlink{0000-0002-1691-8217}\inst{12,14}
    \and
       Barry F. Madore\orcidlink{0000-0002-1576-1676}\inst{12,15}
    \and
       Jun Xu\orcidlink{0000-0003-1778-5580}\inst{6,16}
        }
\institute{School of Physics and Astronomy, Yunnan University, Kunming 650500, China\\
\email{wstfch@mail.ynu.edu.cn, xhsun@ynu.edu.cn} 
\and
SKA Observatory, SKA-Low Science Operations Centre, 26 Dick Perry Avenue, Kensington, WA 6151, Australia 
\and
CSIRO Space and Astronomy, PO Box 1130, Bentley, WA 6102, Australia 
\and
South-Western Institute for Astronomy Research, Yunnan University, Kunming 650091, China 
\and
Purple Mountain Observatory, Chinese Academy of Sciences, 10 Yuanhua Road, Nanjing 210023, China 
\and
National Astronomical Observatories, Chinese Academy of Sciences, 20A Datun Road, Beijing 100101, China 
\and
Institute for Frontiers in Astronomy and Astrophysics, Beijing Normal University, Beijing 102206, China 
\and
School of Astronomy and Space Science, University of Chinese Academy of Sciences, Beijing 100049, China 
\and 
International Centre for Radio Astronomy Research, University of Western Australia, 7 Fairway, Crawley, WA 6009, Australia 
\and
Research School of Astronomy and Astrophysics, Australian National University, Canberra, ACT 2611, Australia 
\and
ARC Centre of Excellence for All Sky Astrophysics in 3 Dimensions (ASTRO 3D), Australia 
\and
The Observatories, Carnegie Institution for Science, 813 Santa Barbara Street, Pasadena, CA 91101, USA 
\and
Visiting Fellow, Harvard-Smithsonian Center for Astrophysics, 60 Garden Street, Cambridge, MA 02138, USA 
\and
Department of Astrophysical Sciences, Princeton University, 4 Ivy Lane, Princeton, NJ 08544, USA 
\and
Department of Astronomy and Astrophysics, University of Chicago, Chicago, IL 60637, USA 
\and
National Key Laboratory for Radio Astronomy, Beijing 100101, China 
}



\abstract
{The nearly edge-on starburst galaxy NGC~253 has been observed to exhibit extended halo emission in multiple bands, making it an ideal laboratory for studying the transfer of matter from the disk to the halo.} 
{We aim to determine how the cosmic-ray electrons (CREs) flow from the disk to the halo and understand what drives their propagation.}
{By combining data from multiple observations, we generated improved total intensity images at 943~MHz with a resolution of $13\arcsec$ from the Australian SKA Pathfinder (ASKAP), and at 216~MHz with a resolution of $45\arcsec$ from the Murchison Widefield Array (MWA). The 1D advection and diffusion equations were solved, and the solutions were fitted to the observed synchrotron emission intensity and spectral-index profiles to constrain the propagation model and parameters.}
{The ASKAP total intensity map has an rms noise of 16~$\mu$Jy~beam$^{-1}$, reaching the classical confusion limit, and the MWA map has an rms noise of 1~mJy~beam$^{-1}$. The sensitivities are significantly improved in comparison to previous observations at similar frequencies. In the ASKAP image, we identify a clear loop-like structure in the northwestern radio spur, extending vertically up to $\sim9$\,kpc above the disk, while the southeastern spur reaches heights of $\sim8$\,kpc. The synchrotron emission intensity profiles perpendicular to the disk can be fitted with exponential components in the central regions and with Gaussian components in the outer regions. This result implies that CREs in these two regions propagate differently. By jointly fitting the vertical synchrotron emission intensity profiles at 943~MHz and 216~MHz, together with the corresponding synchrotron spectral-index profiles, our results provide the clearest evidence to date that CREs are transported from the disk by advection in the central region and by diffusion elsewhere in NGC~253. The advection speed in the central region increases exponentially with height and reaches the escape speed to form a superwind of CREs at about 5.5~kpc. This superwind is associated with regions in the disk with active star formation and X-ray emission, indicating a bulk motion of baryons caused by the advection. The combined thermal, magnetic, cosmic-ray, and ram pressures exceed the gravitational pressure below $z\lesssim5.5$~kpc, and this overpressure condition accelerates the superwind.}
{High-sensitivity low-frequency radio observations provide an important probe of the transport of CREs. With these observations, we have revealed a newly detailed view into the kinematic origin of the superwind from the center of NGC~253.}

\maketitle

\section{Introduction}
\label{sec:intro}

Galactic-scale winds are a common astrophysical phenomenon, observed in galaxies spanning from normal spirals to starbursts and active galactic nuclei (AGN) hosts~\citep{veilleux2005,heckman2017}. These powerful outflows are driven by intense stellar activity, including supernova and stellar winds, radiation pressure on dust grains, mechanical and radiative feedback from AGNs, and pressure exerted by cosmic rays (CRs) accelerated by supernova shocks or AGN jets~\citep{fabian2012,zweibel2017}. Investigating galactic winds is critical to understanding galaxy evolution, as these winds regulate the star formation rate (SFR) and influence the galaxy mass assembly~\citep{somerville2015}. They also play vital roles in chemically enriching the intergalactic medium (IGM) by distributing the metals produced in stars~\citep{tumlinson2017}, and in magnetizing the cosmic environment~\citep{beck2015}. 

Galactic winds have been observed and studied across multiple wavelengths ranging from millimeter to X-ray bands. Millimeter and sub-millimeter observations of CO emission lines have revealed massive molecular outflows driven by mechanical and radiation feedback from AGN and extreme starburst regions~\citep{cicone2014,fluetsch2019}. Infrared observations of dusty and star-forming galaxies have shown that radiation pressure on dust grains significantly accelerates the gas outward and thereby produces powerful radiation pressure-driven winds~\citep{hopkins2012,cicone2014}. Optical and ultraviolet observations trace warm ionized gas through emission lines such as H$\alpha$, [{\sc N\,ii}], [{\sc O\,iii}], and ultraviolet (UV) absorption lines such as \ion{Mg}{ii} and Ly$\alpha$, revealing filamentary structures extending out of galaxy disks~\citep{veilleux2005,rupke2005}. These winds are driven by thermal and radiation pressures generated by intense star formation activities, including supernova explosions and stellar radiation pressure on dust grains~\citep{murray2005}. X-ray observations probe hot and shock-heated plasma at temperatures of millions of Kelvin, resulting from energetic processes such as supernova shocks and AGN-driven jets~\citep{strickland2009}. These observations provide direct evidence of the high-energy processes and powerful mechanical energy sources underlying galactic wind phenomena.

In recent years, CRs have emerged as one of the key drivers of galactic winds, owing to their relatively soft equation of state, which causes their pressure to increase more slowly with density. As a result, CRs can establish large-scale pressure gradients in galactic disks and halos that act against gravity and contribute to the launching of pressure-supported outflows or CR–driven winds~\citep{breitschwerdt1991,breitschwerdt2002,recchia2016,zweibel2017}. Such CR-driven winds are expected to be most efficient in regions of intense star formation, where strong CR pressure gradients can develop~\citep{tabatabaei2022}.

CR electrons (CREs) are transported from the galactic disk to the halo through three mechanisms: diffusion, advection, and streaming. Diffusion describes the random walk of CREs caused by scattering off magnetic-field irregularities and turbulence, and it is commonly parameterized by an energy-dependent diffusion coefficient~\citep{jokipii1966,Schlickeiser2002,strong2007}. This process is expected to dominate in regions where large-scale bulk flows are weak and CRE transport is governed primarily by spatial gradients in the particle distribution~\citep{Schlickeiser2002,grenier2015}. Advection, by contrast, traces the bulk motion of magnetized plasma and transports CREs together with the background flow, and it can operate even when CR pressure is not the dominant driver of the outflow~\citep{breitschwerdt2002,heesen2016}.
Streaming describes how CREs propagate along magnetic-field lines at speeds regulated by interactions with self-generated Alfvén waves~\citep{kulsrud1969}. The theoretical framework of CREs propagating from the galactic disk to the halo has been established through the diffusion-advection equation, which can be applied to the local comoving coordinate system when the velocity of the background medium is non-relativistic~\citep{Schlickeiser2002}. 

CREs produce synchrotron emission when they spiral along the magnetic field, which results in strong radio continuum emission. Therefore, radio continuum observations can trace CR propagation within galactic halos~\citep{heesen_2018a} and provide a unique tool for studying galactic winds. Furthermore, polarized radio continuum observations reveal magnetic field structures and provide evidence of CRs as a driver of galactic-scale winds~\citep{beck2015,heesen2022}. A joint analysis of the spectral index and spatial distribution of the synchrotron emission can be used to determine how CREs are transported, and thus how the winds are driven. The toolkit Spectral Index Numerical Analysis of K(c)osmic-ray Electron Radio-emission, or \textrm{{\small SPINNAKER}}\footnote{\url{https://github.com/vheesen/Spinnaker}}, has been developed to simulate the 1D propagation of CREs and fit the observations~\citep{heesen2016,heesen_2018a}. 

Edge-on spiral galaxies are ideal laboratories for studying CR transport and galactic winds because the observable path of the galactic wind into the halo can be examined~\citep{stein2022}. The results from the simulations and observations indicate that no significant outflows are formed in diffusion-dominated halos~\citep{recchia2016}. \cite{krause2018} studied a sample of 12 edge-on spiral galaxies and found that the radio halos in most systems are better described by advection-dominated CRE transport rather than diffusion. If the advection speed of CREs is larger than the escape velocity of the galaxy, the outflowing gas can escape the gravitational potential into the circumgalactic medium (CGM) as a galactic wind, rather than circulating as in a galactic fountain. The existence of galactic winds has been reported for many of the galaxies in the CHANG-ES (Continuum Halos in Nearby Galaxies – an EVLA Survey) project~\citep{krause2018,miskolczi2019,stein_2019a,schmidt2019,mora2019}, as well as through observations of extraplanar ionized gas, such as $\rm H\alpha$ emission, in their targeted edge-on galaxies~\citep{vargas2019,lu2023}.

Among the nearby edge-on galaxies, NGC~253 is a unique target for studying galactic winds because it is one of the nearest and brightest starburst galaxies and harbors an outstanding outflow structure observed across nearly the entire electromagnetic spectrum. NGC~253 is a member of the Sculptor group at a distance of 3.5~Mpc~\citep{radburn2011}. It has a high inclination angle of $78\fdg3\pm 1\fdg0$~\citep{lucero2015}. Other parameters of NGC~253 are listed in Table~\ref{tab:parameters_table}. 

Multiwavelength observations of NGC~253 have revealed different aspects of its wind. The extraplanar CO-emitting clouds closely track the H$\alpha$ filaments of the outflow, and their kinematics imply a molecular mass outflow rate of $3-9\,M_{\odot}\,\rm yr^{-1}$ comparable to the ongoing SFR, suggesting the starburst is ejecting a significant fraction of its fuel in cold gas form~\citep{bolatto2013}. Deep {\sc H\,i} surveys confirmed that the neutral atomic gas is carried to a height of 9–10 kpc above the disk, consistent with a starburst-driven outflow~\citep{boomsma2005,lucero2015}. In the infrared, thermal dust emission was detected high above the disk, implying that radiation pressure from the intense starburst is driving dust and cold gas out of the galaxy~\citep{yamagishi2011}. Optical and UV observations revealed ionized nebular outflows and filamentary superbubbles along the minor axis~\citep{hoopes2005}. X-ray observations showed diffuse soft X-ray emission in an hourglass or X-shaped volume around the nucleus, extending at least 8 kpc out of the disk~\citep{strickland2002}.

Radio continuum images of NGC~253 at 333\,MHz, 1.5\,GHz, 5\,GHz, and 10\,GHz from VLA and Effelsberg observations~\citep{carilli1992,heesen_2009a}, together with earlier Murchison Widefield Array~\citep[MWA;][]{tingay2013} images from the GLEAM survey in the 76–227\,MHz band~\citep{kapinska2017}, clearly show extended radio halo emission. However, previous studies were limited by the resolution and sensitivity of the low-frequency observations and therefore could not derive the variation of the CRE bulk speed with height above the disk, and therefore limited their ability to assess the role of CR in driving the winds.
 
The Australian SKA Pathfinder (ASKAP), with its high sensitivity, high angular resolution, and wide-field capabilities~\citep{hotan2021} provides us with an opportunity to study the radio halo of NGC~253. Combining with low-frequency images from MWA Phase~I and the newly upgraded Phase~II allows us to build a solid picture of CRE propagation. This paper is organized as follows: data acquisition and reprocessing are described in Sect.~\ref{sec:ob_data}; results are presented in Sect.~\ref{sec:results}; discussions are made in Sect.~\ref{sec:discussions}; and conclusions are drawn in Sect.~\ref{sec:conclu}.

\begin{table}
	\centering
	\caption{Parameters of NGC 253.}
	\label{tab:parameters_table}
	\begin{tabular}{lccr} 
		\hline
		\hline
            RA (J2000)               &$\rm {00^h47^m33\fs12}$ \\
            Dec (J2000)              & $-25\degr17\arcmin7\farcs6$ \\
            $D$ (Mpc)\tablefootmark{a}               & $3.50$ \\
            $i$ (from face-on)\tablefootmark{b}      & $78.3\degr \pm 1.0\degr$\\
		Morphological type\tablefootmark{c}      & $\rm SAB(s)c$ \\
            $M_{\rm tot}$ ($ M_{\odot}$)\tablefootmark{d} & $10^{11}$ \\
            $D_{25}$\tablefootmark{d}                &25.8 \\
            ${V_\mathrm{rot}\,\rm (km\,s^{-1})}$\tablefootmark{e}  &$205$     \\
            Systemic velocity\,$\mathrm {(km\ s^{-1})}$\tablefootmark{f}  &$283\pm4$ \\
            PA\tablefootmark{g}                      & $52\degr$\\
            $\rm SFR$ ($ M_{\odot}\,\rm yr^{-1}$)\tablefootmark{h} & $4.9-6.5$ \\
            $\Sigma_{\rm SFR}$ ($M_{\odot}\, \rm yr^{-1}\,kpc^{-2}$)\tablefootmark{h} & $3.1\times10^{-2}$\\ 
            $\alpha_{\rm nt}$\tablefootmark{i}       &  $-0.71$ \\
            $B$ ($\rm \mu G$)\tablefootmark{j}         & $10.5$ \\
		\hline
	\end{tabular}
     \tablefoot{
    \textnormal{\tablefoottext{a}{The central coordinates and distance}{\citep{radburn2011}.} \tablefoottext{b}{\citet{lucero2015}.} \tablefoottext{c}{\citet{borlaff2023}.} \tablefoottext{d}{The total mass and the B-band diameter~\citep{heesen_2018b}.} \tablefoottext{e}{Maximum rotation speeds with the uncertainties of $\mathrm{\pm3\ km\ s^{-1}} $ are from \citet{pence1981}.} \tablefoottext{f}{The systemic velocity from \citet{lyu2023}.} \tablefoottext{g}{Position angle from \citet{heesen_2009a}.} \tablefoottext{h}{The star formation rate and star formation rate surface density~\citep{leroy2019,jarrett2019}.}\tablefoottext{i}{The non-thermal synchrotron spectral index~\citep{kapinska2017}.}} \tablefoottext{j}{The average magnetic field strength is derived from this work.}
      }
\end{table}

\section{Data acquisition and reprocessing}
\label{sec:ob_data}
\subsection{ASKAP 943~MHz observations}
\label{subsec:ASKAP_ob}

NGC~253 was covered in the field observed by the ASKAP Pilot Survey for Gravitational Wave Counterparts project (Project ID: AS111). The scheduling block IDs of the 8 observations used in our analysis are: SB9602, SB9649, SB9910, SB10463, SB15191, SB18912, SB18925, and SB27379. These observations were conducted between 2019 and 2020, and most of the observations lasted approximately 10.5~h each. The central frequency is 943~MHz with a total bandwidth of 288~MHz, split into 1-MHz channels. 

\begin{figure}	\includegraphics[width=\columnwidth]{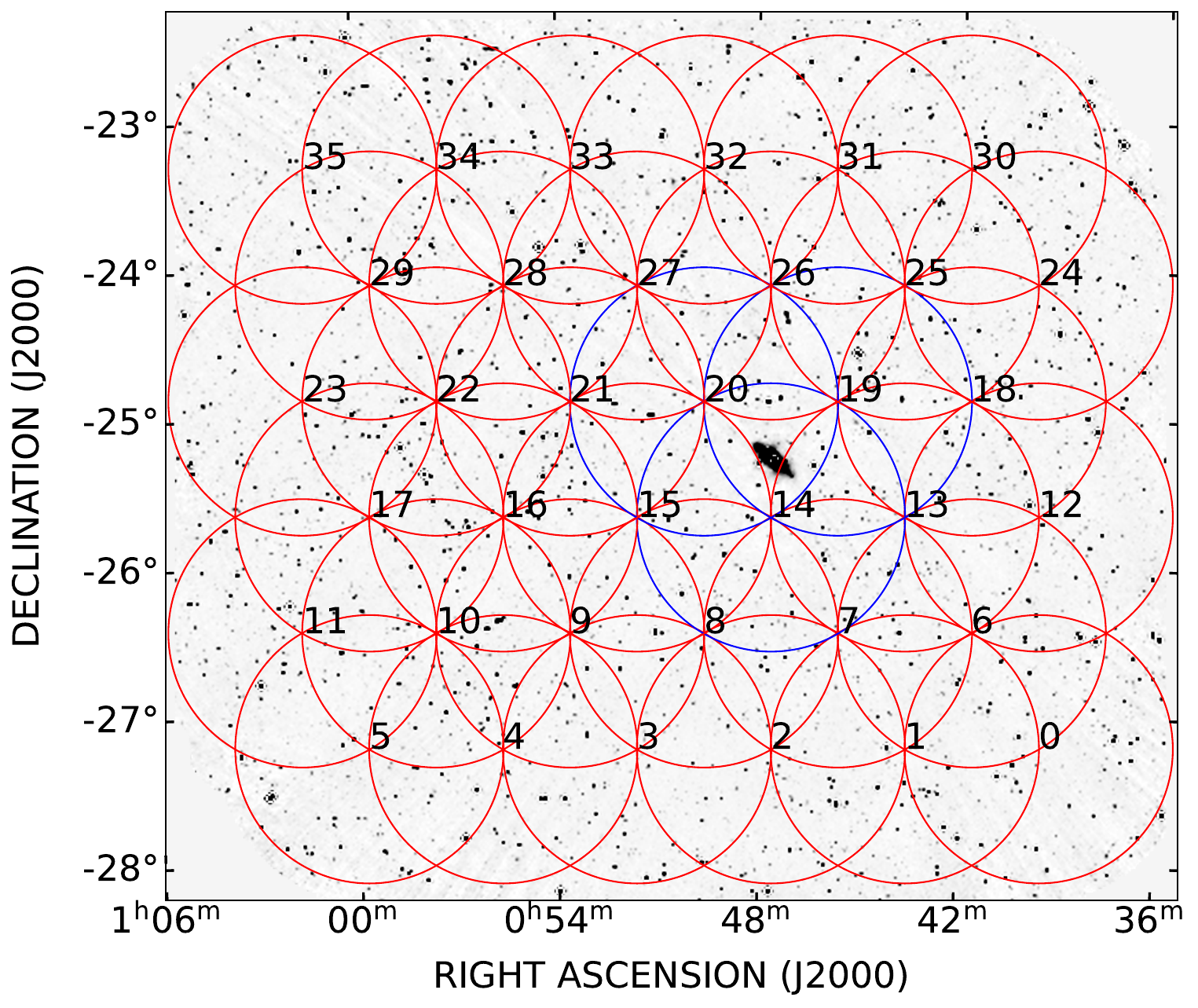}
    \caption{Layout of the 36 ASKAP beams with circles in the ``closepack36" footprint configuration, overlaid on the $I$ image from CASDA. The radius of the circles is $0\fdg9$, which is approximately the primary beam width. The blue circles indicate the beams that cover NGC~253.}
\label{fig:36beam_cover}
\end{figure}

Each ASKAP antenna is equipped with a phased array feed that forms 36 dual-polarization beams and delivers a field of view of about 30 square degrees. The footprints of the beams are in the ``closepack36'' configuration as shown in Fig.~\ref{fig:36beam_cover}. NGC~253 was fully covered by the three beams: 14, 19, and 20, outlined by blue circles in Fig.~\ref{fig:36beam_cover}. 

The calibrated visibilities for each individual beam and the total intensity ($I$) images for the whole field, processed with {\small ASKAPsoft}\footnote{ASKAPsoft is the suite of processing software developed by the ASKAP computing team to process ASKAP observations.}, are public and available from the CSIRO ASKAP Science Data Archive \textit \rm {(CASDA)}\footnote{\url{https://data.csiro.au/domain/casdaObservation}}. We found significant calibration and deconvolution errors for the area around NGC~253 in the $I$ images from CASDA, which is understandable since NGC~253 is a bright and extended source with complicated structures. We further processed the visibility data for beams 14, 19, and 20 from CASDA using the pipeline SASKAP\footnote{\url{https://gitlab.com/Sunmish/saskap/-/tree/petrichor}} to improve calibration and imaging~\citep{duchesne2024}. 

To improve the calibration, we performed two rounds of phase-only (p) and one round of amplitude-phase (ap) self-calibration, followed by a refined bandpass calibration. The solution interval was progressively reduced during each self-calibration round, starting from 120~s at the beginning to 20~s in the end. The solution interval of the final bandpass step was 15~minutes. These steps dramatically reduced the impact of calibration errors on the final images.

For imaging, the fast generic widefield imager \textit{\rm {\small \textsc{WSClean}}}\footnote{\url{https://wsclean.readthedocs.io/en/latest/changelogs/v3.2.html}}~\citep{offringa2014,offringa2017} was used to generate a multi-frequency and multiscale synthesis image. Briggs weighting~\citep{briggs1995} with ${\rm robust = 0.25}$ was employed to enhance sensitivity toward faint extended emission. Wgridder was employed to improve the accuracy of wide-field image reconstruction and to efficiently account for non-coplanar baseline effects~\citep{ye2022,arras2021}. To subtract compact point sources, we tapered the short baselines ({\tt minuv-l=3804} was used as an input parameter to \textsc{WSClean}) during the deconvolution process to obtain a model of small-scale structures (excluding features larger than about 1/minuv~$=2.63\times10^{-4}$~rad~$=54^{\prime\prime}$), subtracted the model from the visibility data, and re-imaged the visibility data to obtain images of the extended emission.

Primary beam correction is required to determine flux densities and combine images of different beams. Unfortunately, the primary beam measurement with the holographic method was not available during the observations for the project AS111. Following the method by~\citet{duchesne2024}, we built a robust model of the primary beam from the images themselves. We cross-matched the point sources with the Rapid ASKAP Continuum Survey low-frequency \textit \rm {(RACS-Low)}\footnote{\url{https://data.csiro.au/collection/csiro\%3A52217v3}}~\citep{mcconnell2020,hale2021}, and obtained the ratio of flux densities as a function of the relative position to the beam center. A two-dimensional elliptic Gaussian was fitted to the data to obtain the primary beam model, which was used to correct the images.

We imaged these data for all eight observations, each of which contained data from three beams, resulting in a total of 24 independent data sets. We then selected the eight highest-quality images with few spurious structures and linearly stacked them to form the final image. We found that when we included additional images beyond the 8 best-quality epochs, the noise level did not improve, but some artifacts were introduced.    
 
\begin{figure*}	\includegraphics[width=\linewidth]{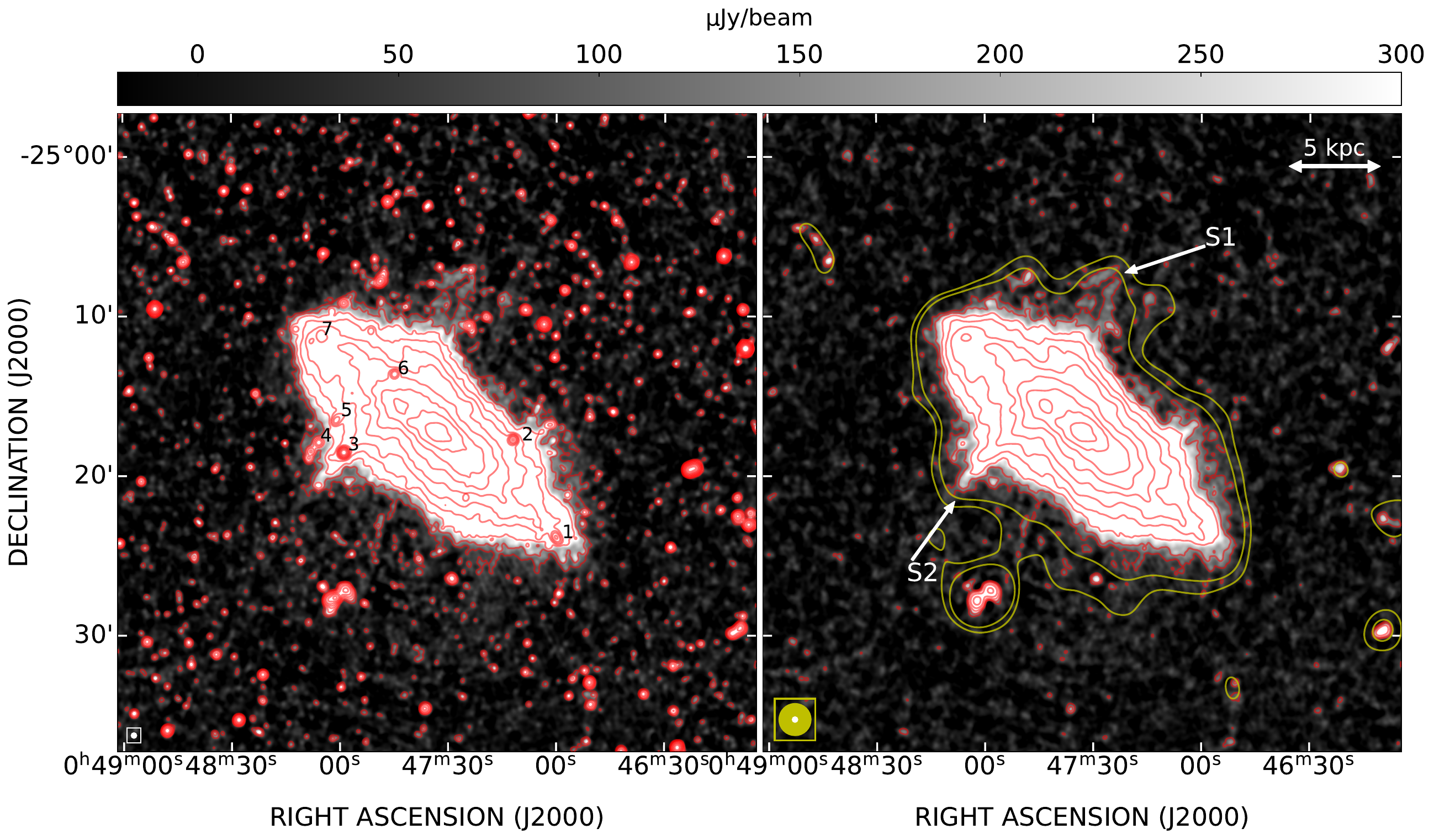}
\centering
    \caption{Total intensity images convolved to $20\arcsec$ angular resolution at 943~MHz from ASKAP: original (left) and with point sources subtracted (right), with contour levels of $3\sigma\times 2^n$ ($n=0,\,1,\,2,\,\ldots$), where $\sigma=27$~$\mu$Jy~beam$^{-1}$. The yellow contours on the right show the image further convolved to a resolution of $120\arcsec$ with $\sigma=264~\mu$Jy~beam$^{-1}$, at levels of 3 and $6\sigma$. 
    The white and green circles in the lower left corner indicate the beam size, and the scale bar in the upper right corner corresponds to 5 kpc. The two ``radio spurs'', S1 and S2, are indicated by white arrows. The black numbers indicate seven background point sources in NGC~253, which have been subtracted in the right panel.
    } 
\label{fig:ASKAP_total_intensity}
\end{figure*}

\subsection{MWA 216~MHz observations}

MWA operates in the 70–300 MHz frequency band~\citep{tingay2013}. Its Phase I configuration consisted of 128 tiles, each comprising a 4$\times$4 array of 16 dual-polarized dipole antennas, with maximum baselines of about 3~km. Phase II expanded this by adding 72 new core tiles in a hexagonal arrangement and 56 outer station tiles, extending the longest baselines to about 5.3 km~\citep{wayth2018}. 

NGC 253 was observed under the GaLactic and Extragalactic All-sky MWA \citep[GLEAM;][]{wayth2015,hurley2022} survey project (ID: G0008). We retrieved archival data covering the 200–231 MHz band (central frequency 216 MHz), consisting of 42 two-minute snapshots from Phase I and 51 two-minute snapshots from Phase II, resulting in a total on-source integration time of approximately three hours. 

Data reduction was performed using the pipeline piip\footnote{\url{https://gitlab.com/Sunmish/piip}}, and a more detailed description is presented in~\citet{duchesne2020,duchesne2025}. 
We provide a description of the most important data-processing steps for our case here. Each individual two-minute snapshot was processed through four sequential stages: pre-processing, initial calibration, self-calibration, and flux-density correction. Finally, all calibrated snapshots were corrected for the primary beam response at the location of the target~\citep{sokolowski2017}; these were then jointly deconvolved into a single image using \textsc{WSClean}, using Briggs weighting with $\rm robust = 0.5$ to improve sensitivity. 

\subsection{Auxiliary data}
In addition to the radio continuum data, we use data at other wavelengths, including H$\alpha$, infrared, and soft X-ray. The H$\alpha$ data are taken from the latest observations of the \textit \rm {TYPHOON}\footnote{\url{https://typhoon.datacentral.org.au/}}
 project, an integral-field spectroscopy survey of 44 nearby galaxies in the southern sky with angular sizes larger than $3\arcmin$ (\citealt{grasha2022}; Siebert et al., in preparation). The survey was conducted using the 2.5-m du Pont Telescope at Las Campanas Observatory in Chile. In this work, the H$\alpha$ data are used to estimate the thermal radio emission. Mid-infrared images at 3.6, 4.5, and 8.0~$\mu$m were obtained with the Infrared Array Camera (IRAC) on board Spitzer as part of the  Local Volume Legacy (LVL) survey~\citep{dale2009}, and are used to trace the stellar component and dust emission in NGC~253. The 24~$\mu$m image was obtained with the Multiband Imaging Photometer for Spitzer (MIPS) and is used for extinction correction. Soft X-ray data in the 0.1–2.4~keV band were taken from ROSAT/PSPC observations and are used to compare with the extended radio continuum emission~\citep{vogler1999}. The angular resolution and noise level of all the data are listed in Table~\ref{tab:data_table}

\begin{table}
        \begin{center}
	\caption{Parameters of the multiwavelength data of NGC~253.}
	\label{tab:data_table}
	\begin{tabular}{ccccc} 
		\hline
		\hline
            Band            & Instrument  & FWHM         & rms                                 & Ref \\\hline          
            139~cm          & MWA         & $45\arcsec$  & 1~$\rm mJy\,beam^{-1}$              &\tablefootmark{a} \\
            32~cm           &ASKAP        & $13\arcsec$  & 16~~$\rm \mu Jy\,beam^{-1}$         &\tablefootmark{a} \\
            $24\,\rm \mu m$ &Spitzer      & $6\arcsec$   & $\rm 4.9\times10^{-2}\,MJy\,sr^{-1}$&\tablefootmark{b}\\
            $8\,\rm \mu m$  &Spitzer      & $1\farcs98$& $\rm 4.4\times10^{-2}\,MJy\,sr^{-1}$&\tablefootmark{b}\\
            $4.5\,\rm \mu m$&Spitzer      & $1\farcs72$& $\rm 2.0\times10^{-2}\,MJy\,sr^{-1}$&\tablefootmark{b}\\
            $3.6\,\rm \mu m$&Spitzer      & $1\farcs66$& $\rm 2.5\times10^{-2}\,MJy\,sr^{-1}$&\tablefootmark{b}\\
            $\rm H\alpha$   &du Pont      & $1\farcs65$& $3\times10^{6}\,\rm Jy\,pix^{-1}$   &\tablefootmark{a}\\
            X-ray           &ROSAT        & $25\arcsec$  & 1.4~$\rm counts\,pix^{-1}$          &\tablefootmark{c}\\
 		\hline
	\end{tabular}
    \tablefoot{
    \textnormal{\tablefoottext{a}{From this work.}
    \tablefoottext{b}{The Spitzer IRAC observations at 3.6, 4.5, and $8\,\mu$m, together with the MIPS $24\,\mu$m imaging, were taken from the Local Volume Legacy (LVL) survey \citep{dale2009}.}
    \tablefoottext{c}{ROSAT/PSPC data of NGC~253~\citep{vogler1999}.}}
    }
        \end{center} 
\end{table} 

\begin{figure}	\includegraphics[width=\linewidth]{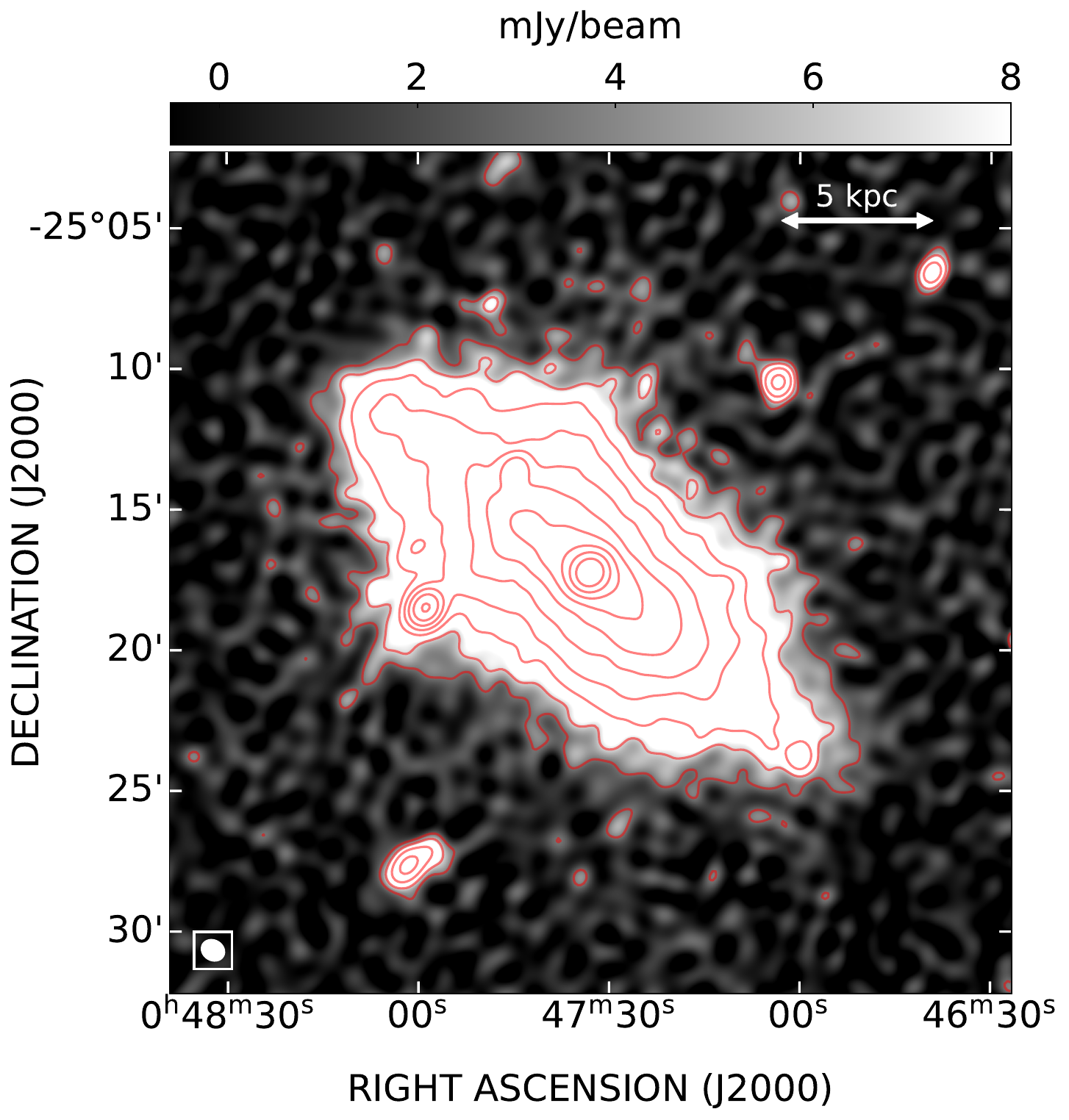}
\centering
    \caption{Total intensity map at 216~MHz from MWA, with contour levels of $3\sigma\times 2^n$ ($n=0,\,1,\,2,\,\ldots$). The rms is $\sigma = 1$~mJy~beam$^{-1}$ and the angular resolution is $45\arcsec$.} 
\label{fig:MWA_total_intensity}
\end{figure}

\section{Results}
\label{sec:results}
\subsection{Total intensity images}

Following the procedures described in Sect.~\ref{sec:ob_data}, we obtained the total intensity images from ASKAP and MWA observations, as displayed in Figs.~\ref{fig:ASKAP_total_intensity} and \ref{fig:MWA_total_intensity}. For ASKAP, we also derived the image with point sources subtracted. The ASKAP image at 943~MHz has an rms noise of 16~$\mu$Jy~beam$^{-1}$ and an angular resolution of $13\arcsec$. The rms reaches the expected classical confusion limit based on the estimates of~\citet{condon2012}, and indeed we detect a background-source density of $\sim4{,}100$~sources\,deg$^{-2}$ at $>3\sigma$. The MWA image at 216~MHz has an rms of 1~mJy~beam$^{-1}$ and a resolution of $45\arcsec$. The integrated flux densities measured within the $3\,\sigma$ regions are $16.6 \pm 1.2$~Jy at 216~MHz and $7.9 \pm 0.4$~Jy at 943~MHz. The uncertainty estimation is described in Sect.~\ref{sec:uncertain}.

These data provide the most sensitive total-intensity images of NGC~253 obtained to date at similar frequencies. For comparison, earlier VLA observations reached rms noise levels of $\sim350~\mu\mathrm{Jy\,beam^{-1}}$ at 1.5\,GHz with an angular resolution of $30^{\prime\prime}$ and $\sim8~\mathrm{mJy\,beam^{-1}}$ at 333\,MHz with a resolution of $70^{\prime\prime}$~\citep{carilli1992, heesen_2009a}. The Phase~I GLEAM MWA maps at 227\,MHz achieved an rms noise of $\sim13~\mathrm{mJy\,beam^{-1}}$ at a resolution of $102^{\prime\prime}$~\citep{kapinska2017}. We convolved the ASKAP total intensity image at 943~MHz to a resolution of $30\arcsec$, reaching an rms noise level of $\sim50~\mu\mathrm{Jy\,beam^{-1}}$. The MWA image at 216~MHz was convolved to resolutions of $70^{\prime\prime}$ and $102^{\prime\prime}$, resulting in rms noise levels of $\sim1.2~\mathrm{mJy\,beam^{-1}}$ and $\sim1.6~\mathrm{mJy\,beam^{-1}}$, respectively. The new images from ASKAP and MWA have considerably higher sensitivity, allowing us to detect emission that extends much farther from the disk into the halo than in previous images.

Compared with earlier radio observations, the most prominent common feature is that the total-intensity images of NGC~253 exhibit a characteristic ``dumbbell-shaped'' morphology, 
consistent with the 3.6, 6.2, 20, and 90\,cm maps presented by \citet{heesen_2009a} and \citet{carilli1992}. Emission protrudes from the southeastern, northeastern, and northwestern sides of the disk, forming structures that resemble a ``plate''. In Fig.~\ref{fig:ASKAP_total_intensity}, seven bright background sources coincident with NGC~253 are identified and marked in the left panel, and subtracted in the right panel.

\subsubsection{New features at 943~MHz}

After convolving the ASKAP image to a resolution of $20\arcsec$, the rms noise is $27~\mu\mathrm{Jy\,beam^{-1}}$, and we identify a clear loop-like structure in the northwestern ``radio spur'' S1, extending vertically up to $\sim9$\,kpc above the disk. On the southeastern side, the ``radio spur'' S2 reaches heights of $\sim 8$\,kpc above the disk. A more detailed discussion of the S1 and S2 features is presented in Sect.~\ref{subsec:superwind}. In the southeastern region of the galaxy, low-surface-brightness diffuse emission is visible extending far away from the disk. After convolving the map to $120\arcsec$, the rms noise decreases to $264~\mu\mathrm{Jy\,beam^{-1}}$ and the improved surface-brightness sensitivity reveals faint diffuse emission extending to $\sim 8$~kpc from the galactic midplane (right-hand panel of Fig.~\ref{fig:ASKAP_total_intensity}).

\subsection{Thermal emission estimate}
\begin{figure}	
    \includegraphics[width=\columnwidth]{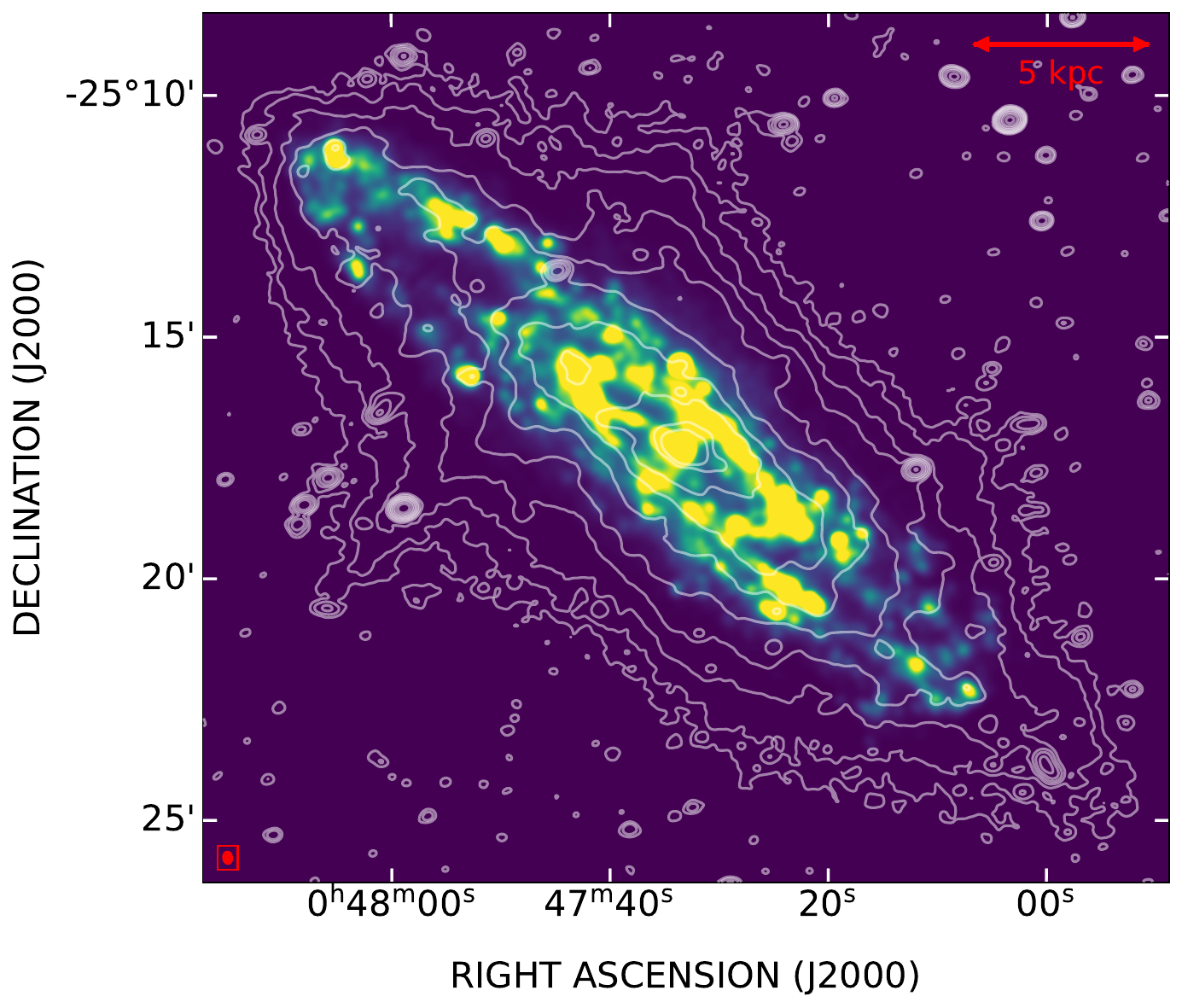}
    \caption{Image of H$\alpha$ emission from the TYPHOON project overlaid with the ASKAP total intensity contours. The angular resolution of $13\arcsec$ is indicated by the red circle in the lower-left corner.}
\label{fig:radio_Ha}
\end{figure}

The thermal free-free emission at the frequencies of ASKAP and MWA can be estimated based on $\rm H\alpha$ observations. We used the H\(\alpha\) data from the latest observations of the TYPHOON project. The H$\alpha$ image of NGC~253, convolved to an angular resolution of $13\arcsec$, is shown in Fig.~\ref{fig:radio_Ha}. Along the galactic major axis, the radio emission extends beyond the coverage of the H$\alpha$ image, owing to the limited field of view of the TYPHOON observations.

We used the mid-infrared $\rm 24\,\mu m$ data from the Spitzer observations~\citep{dale2009} to correct for dust extinction in the $\rm H\alpha$ emission following \cite{murphy2011}:
\begin{equation}
    {L(\rm H\alpha_{corr}}) = L({\rm H\alpha_{obs}}) + a \cdot \nu L_{\nu}(24\,\rm{\mu m}),
    \label{Halpha+24microns}
\end{equation}
where $L(\rm H\alpha_{corr})$ and $L({\rm H\alpha_{obs}})$ represent the corrected and uncorrected H$\alpha$ luminosity, $L_{\nu}(24\,\rm{\mu m})$ represents the 24~$\mu$m luminosity, and $a$ is the weighting factor for the $\rm 24\,\mu m$ contribution. Here, we adopted the value $a = 0.042$ derived by \cite{vargas2018} that is valid for edge-on or dusty galaxies. 

We calculated the thermal emission luminosity $L^T_\nu$ at the frequency $\nu$ according to \citet{murphy2011} and \citet{vargas2018}:
\begin{equation}
\frac{{L}^{{T}}_{\nu}}{\rm{erg} \cdot \rm{s}^{-1} \rm{Hz}^{-1}} =1.26\times 10^{-14} \left(\frac{{T_e}}{10^4 \rm{K}}\right)^{0.45} \left(\frac{\nu}{\rm{GHz}}\right)^{-0.1}\frac{L(\rm H\alpha_{corr})}{\rm erg\,s^{-1}}.
\label{radioequation}
\end{equation}

\begin{figure*}	\includegraphics[width=\linewidth]{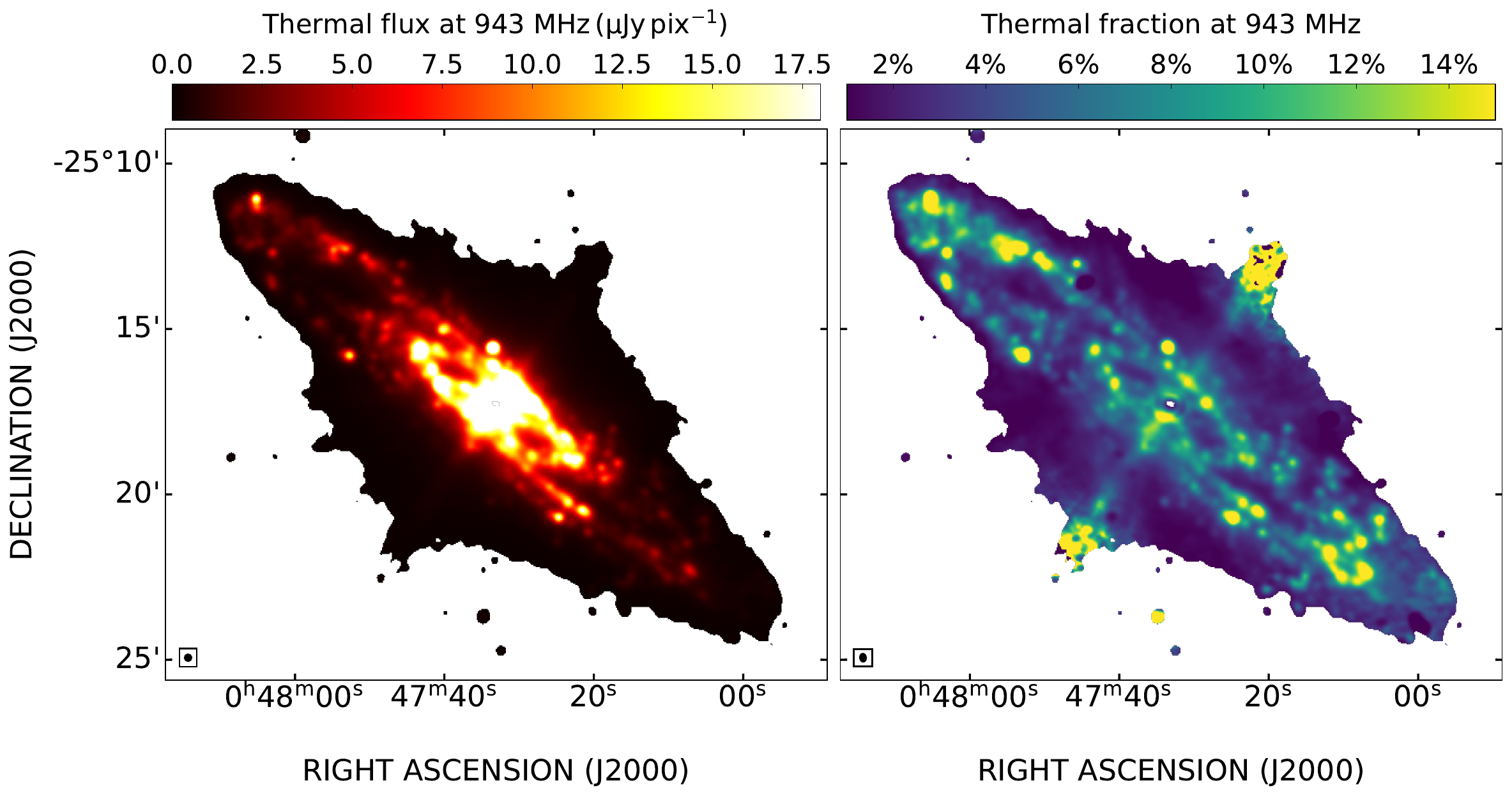}
\centering
    \caption{Thermal emission distribution at 943~MHz (left) and the corresponding thermal fraction (right).}
\label{fig:thermal_emission}
\end{figure*}

Assuming an electron temperature of 10,000~K, we derived the thermal emission at 943~MHz and the corresponding thermal fraction, as shown in Fig.~\ref{fig:thermal_emission}. The thermal emission is concentrated in the galactic disk, with enhanced contributions along the spiral arms. Based on the integrated flux densities at 943 and 216 MHz, the global thermal fractions are estimated to be 7.0\% and 4.3\%, respectively.

We calculated the thermal emission pixel by pixel, subtracted it from the total intensity, and derived images of synchrotron emission at 943~MHz and 216~MHz. The scale heights, magnetic field, and CRE transport were studied using the synchrotron emission, as shown later.

\subsection{Spectral index map}
\label{subsec:spex}
\begin{figure*}	\includegraphics[width=\linewidth]{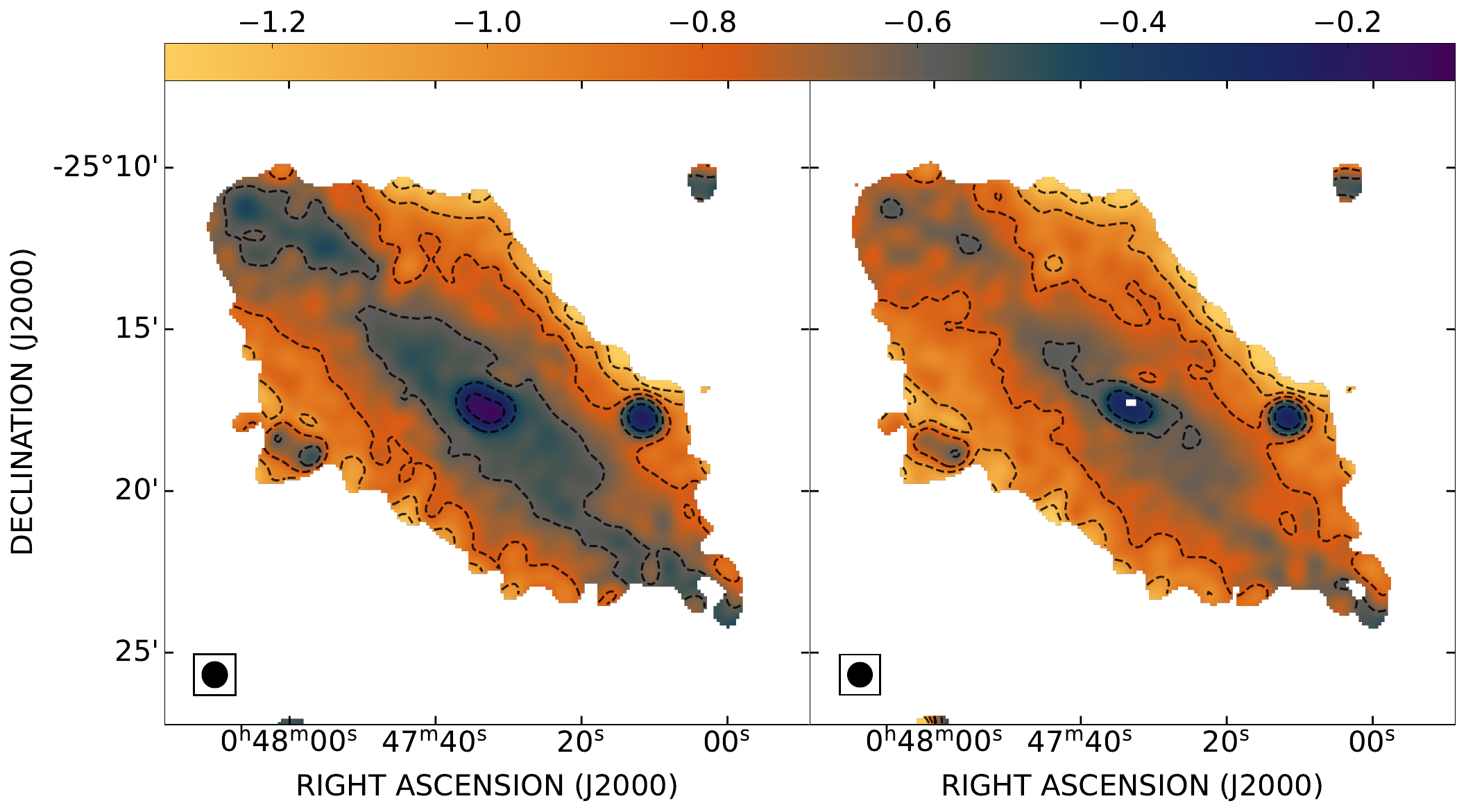}
\centering
    \caption{Images of total intensity spectral index (left) and synchrotron emission spectral index (right) between 216~MHz and 943~MHz with $45\arcsec$ resolution. Contour levels are: $-1.2$, $-1.0$, $-0.8$, $-0.6$, $-0.4$, and $-0.2$.}
\label{fig:spex_distrub}
\end{figure*}

The spectral index $\alpha$ ($S_\nu\propto\nu^\alpha$) in this work is defined as 
\begin{equation}
    \alpha = \frac{\mathrm{log}\,(S_{\nu_1}/S_{\nu_2})}{\mathrm{log}\,(\nu_1/\nu_2)},
    \label{spix}
\end{equation} 
where $S_\nu$ is the flux density at frequency $\nu$. We convolved the ASKAP image to $45\arcsec$, matching the angular resolution of the MWA image and achieving an rms noise level of $78~\mu\mathrm{Jy\,beam^{-1}}$. Using these images, we derived maps of the total intensity spectral index and synchrotron emission spectral index between 216~MHz and 943~MHz, as shown in Fig.~\ref{fig:spex_distrub}. Regions with intensities below 8 times the rms noise and fractional errors of the spectral index greater than 10\% were excluded. The uncertainty estimation is described in Sect.~\ref{sec:uncertain}. With increasing vertical distance from the galactic midplane, the spectral index gradually steepens from $\alpha \sim -0.4$ to $-1.2$, indicating progressive energy losses of CREs as they propagate away from the disk. Compared to the total intensity spectral index map, the synchrotron emission spectral index in the galactic disk is significantly steeper, as expected, since the contribution of thermal emission flattens the observed spectrum.

The spectral index within the central $2\arcmin$ of the galaxy is remarkably flat with $\alpha>-0.4$. This behavior is expected in a dense starburst nucleus and likely reflects a combination of effects. First, the central region hosts an intense nuclear starburst that produces a large amount of thermal free–free emission from compact \ion{H}{ii} regions and diffuse ionized gas. Under optically thin conditions, this thermal component has an intrinsically flat spectrum ($\alpha \approx -0.1$), which makes the total intensity spectral index flatter. Secondly, free–free absorption and synchrotron self-absorption in the nuclear starburst further flatten the spectrum, and at the lowest frequencies can even invert the non-thermal spectrum~\citep[see][]{kapinska2017}. The integrated flux density spectrum was systematically studied by~\citet{kapinska2017}, and in the following analysis, we adopt their spectral index of $\alpha = -0.71$ for the integrated spectrum, measured after subtracting the central starburst region.

ASKAP detects more extended emission above the disk plane (Fig.~\ref{fig:ASKAP_total_intensity}), whereas such emission is not detected in the MWA data because of their limited sensitivity. This prevents us from deriving reliable spectral indices at larger vertical heights and, consequently, from constraining CR transport models at these regions (see Sect.~\ref{subsec:spin_fit}). Future MWA observations with improved sensitivity will be essential to place stronger constraints on CR transport at large distances from the disk.

\subsection{Scale height of synchrotron emission}
\label{subsec:h_syn}
We derived the synchrotron scale heights using the synchrotron emission intensity maps at 943~MHz and 216~MHz, obtained by subtracting the thermal emission from the corresponding total-intensity maps.

\begin{figure}	\includegraphics[width=\columnwidth]{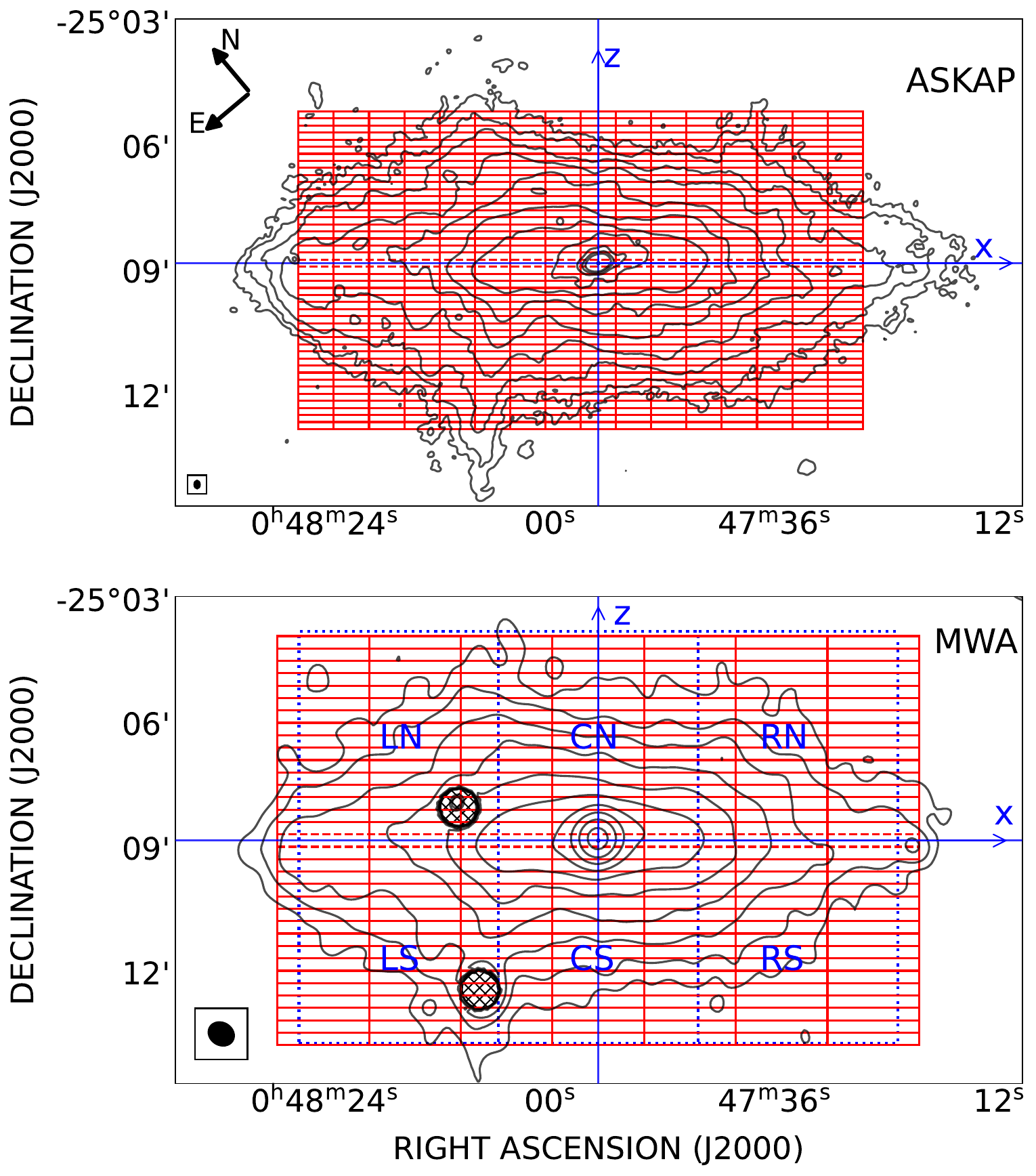}
    \caption{Rotated ASKAP (top panel) and MWA (bottom panel) synchrotron emission intensity images of NGC~253. The definition of $x$ and $z$ axes is outlined. The vertical strips split into many rectangles are also overlaid. Each red profile was fit to derive the intensity scale height. In the lower panel, we also mark the six blue dashed profiles used in the SPINNAKER fits. Two background point sources were masked by black circles with cross-hatching in the MWA image.}
\label{fig:cover}
\end{figure}

We rotated the ASKAP and MWA images of NGC~253 anticlockwise by its position angle of $52\degr$ to align its major axis horizontally and defined the $x$ and $z$ axes parallel and perpendicular to the major axis, respectively, as shown in Fig.~\ref{fig:cover}. For ASKAP, we made 16 vertical strips each with a width of $1\farcm1$ along the $x$ axis, which is five times the beamwidth of ASKAP image to allow for sufficient independent values. Each strip was divided into rectangles of $13\arcsec$ height, the same as the beamwidth of the ASKAP image. For MWA, we made seven profiles, each with a width of $2\farcm9$ (3.7 times the beamwidth). The height of each box was set to $23\arcsec$ (0.5 times the beamwidth). Intensities were averaged within each rectangle and vertical profiles of intensity versus $z$ were derived for all strips. 

We followed the procedure by \citet{heesen_2018b}, which was based on the method by \citet{dumke1995}, to account for the projection of the disk and the broadening of the intrinsic vertical profiles caused by the finite beamwidth $\Theta_0$ of the telescopes. A combined beamwidth $\Theta_c$ was introduced as $\Theta_c=\sqrt{\Theta_0^2+\Theta_d^2 \cos^2i}$, where $i$ is the inclination angle, $\Theta_d=R\cos(x/R\cdot\pi/2)$ is the equivalent beamwidth caused by the disk and $R$ is the radius of the disk~\citep{muller2017}. A Gaussian function with the effective beamwidth was convolved with an exponential or a Gaussian function to fit the observed synchrotron emission intensity profiles. The details are presented in Appendix~\ref{sec:intens_mode}.

We examined the CO(1–0) image by~\citet{kuno2007} and found that the maximum height above the disk is about 1~kpc, beyond which the emission is negligible. To further reduce the influence of the projection of the disk, only regions $|z|>1$~kpc were considered when deriving the scale heights.

\begin{figure*}[!h]	\includegraphics[width=\linewidth]{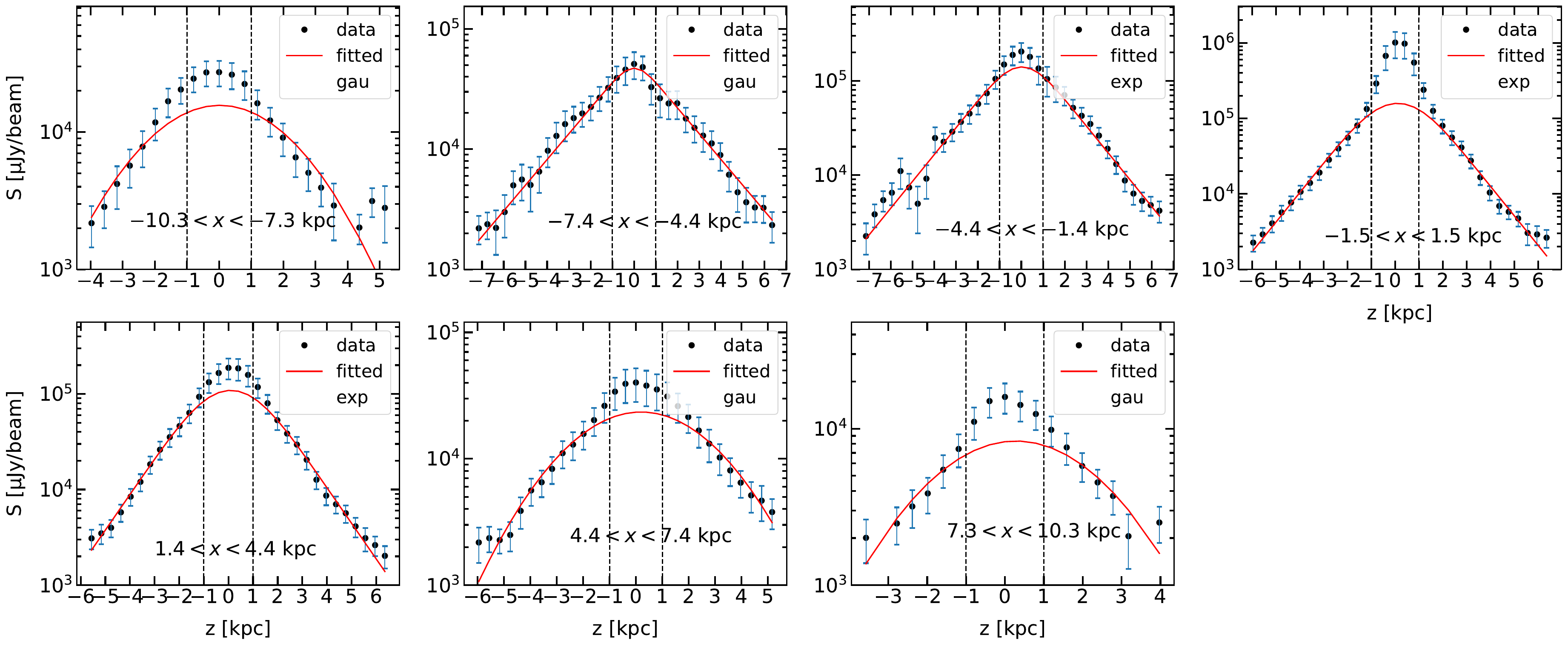}
    \caption{Vertical profiles derived from the thermal-subtracted MWA synchrotron emission intensity image at the original resolution of $45\arcsec$. The northern side is $z>0$ and the southern side is $z<0$. The red solid line represents the fitting with either exponential (exp) or Gaussian (gau) functions. Only data points outside the dashed region and above a $3\,\sigma$ threshold are included in the fits.}
\label{fig:intensity-fit-all}
\end{figure*}

Due to the high resolution of ASKAP, each profile exhibits pronounced north–south asymmetry. We therefore fitted the northern and southern sides separately; the results are shown in Figs.~\ref{fig:intensity-fit-north} and \ref{fig:intensity-fit-south}. The MWA northern and southern components were fitted jointly, and the results are presented in Fig.~\ref{fig:intensity-fit-all}.

\begin{figure}	\includegraphics[width=\columnwidth]{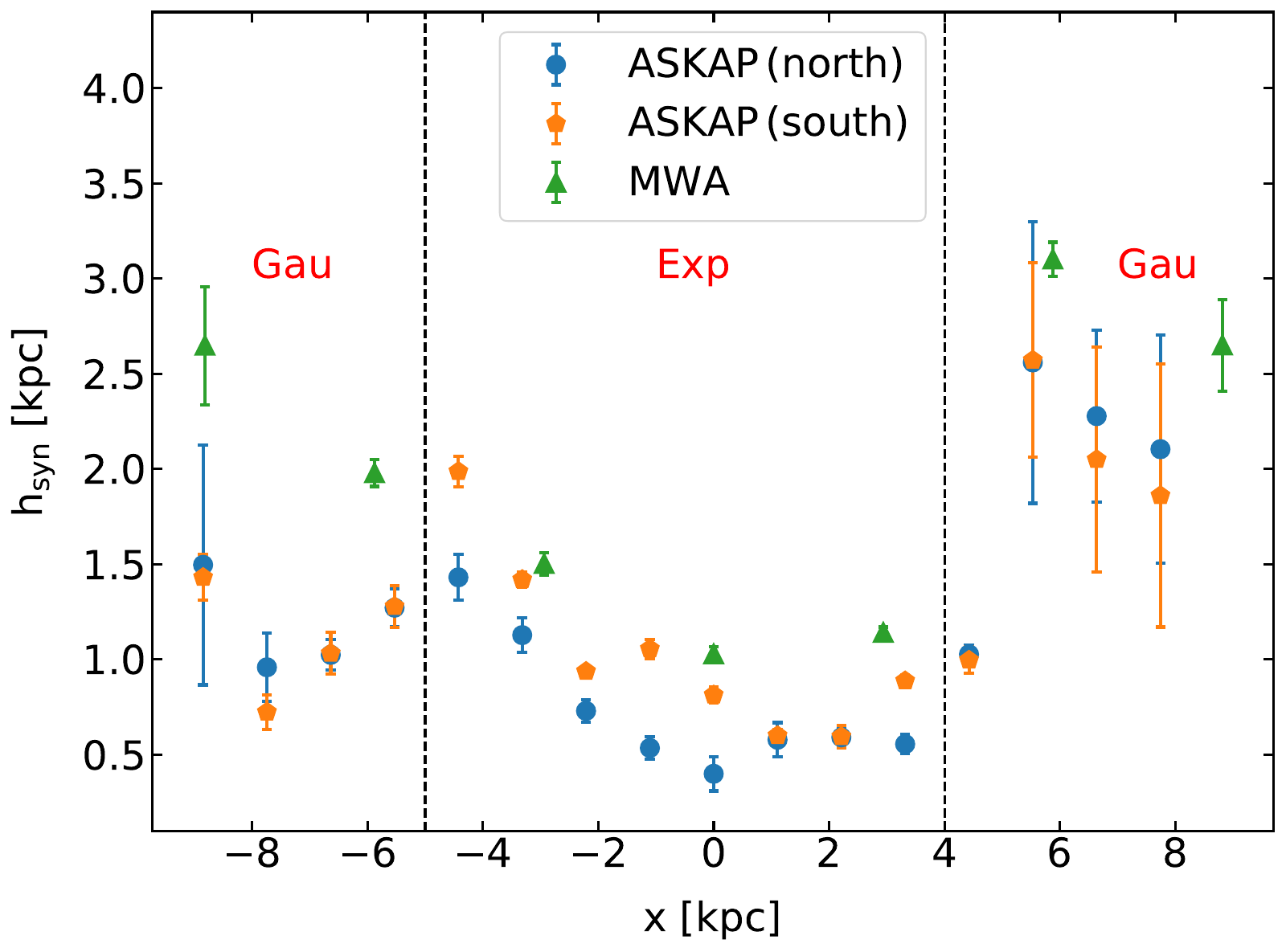}
    \caption{Synchrotron scale height $h_{\rm syn}$ versus $x$ for both north ($z > 0$) and south ($z < 0$) parts. The central dashed interval indicates the region fitted with an exponential function, while the outer parts are fitted with a Gaussian function.}  
\label{fig:scale_h}
\end{figure}

We found that in the central region of the galaxy, $|x|\lesssim 3$~kpc, the intensity profiles are better fitted by one exponential component, whereas in the outer regions they are better described by one Gaussian component. The fitting results are shown in Figs.~\ref{fig:intensity-fit-north}, \ref{fig:intensity-fit-south} and \ref{fig:intensity-fit-all}. We obtained the scale heights ($h_{\rm syn}$) of synchrotron emission, shown in Fig.~\ref{fig:scale_h}. 
 
The synchrotron scale heights gradually increase from the central region toward the outer parts of the galaxy. This trend is consistent with the results reported by \citet{heesen_2009a} based on 6.2 and 20~cm data. In the central region, the magnetic field strength and radiation energy density are higher, leading to stronger synchrotron and inverse-Compton (IC) losses and hence more rapid cooling of CREs. Toward the outer regions, where magnetic field strength and energy densities are lower, CREs can propagate to larger heights, resulting in more extended synchrotron halos. The synchrotron scale heights at 216~MHz are larger than those at 943~MHz. This behavior can be understood because the lower-energy CREs traced at 216~MHz experience weaker synchrotron and IC losses, allowing them to propagate to larger heights and produce a more extended radio halo. 

\subsection{Magnetic field strength}
\label{sec:equi_B}

We estimated the magnetic field strength assuming energy equipartition~\citep{beck2005}, which requires the synchrotron spectral index $\alpha$, the number density ratio between CR protons and electrons $K_0$, and the effective path length of emission $L$, in addition to the total intensity image. 

To estimate the average magnetic field strength for the entire galaxy, we adopted the spectral index $\alpha = -0.71$ and the ratio $K_0 = 100$. Following \citet{heesen_2009a}, we used the average of full width to the half power derived from fitting the vertical emission profiles, which is about 5.5~kpc as $L$. The resultant magnetic field is 11~$\mu$G, consistent with most of the nearby spiral galaxies~\citep{beck2000,beck2019,heesen2022}.

\begin{figure}	\includegraphics[width=\columnwidth]{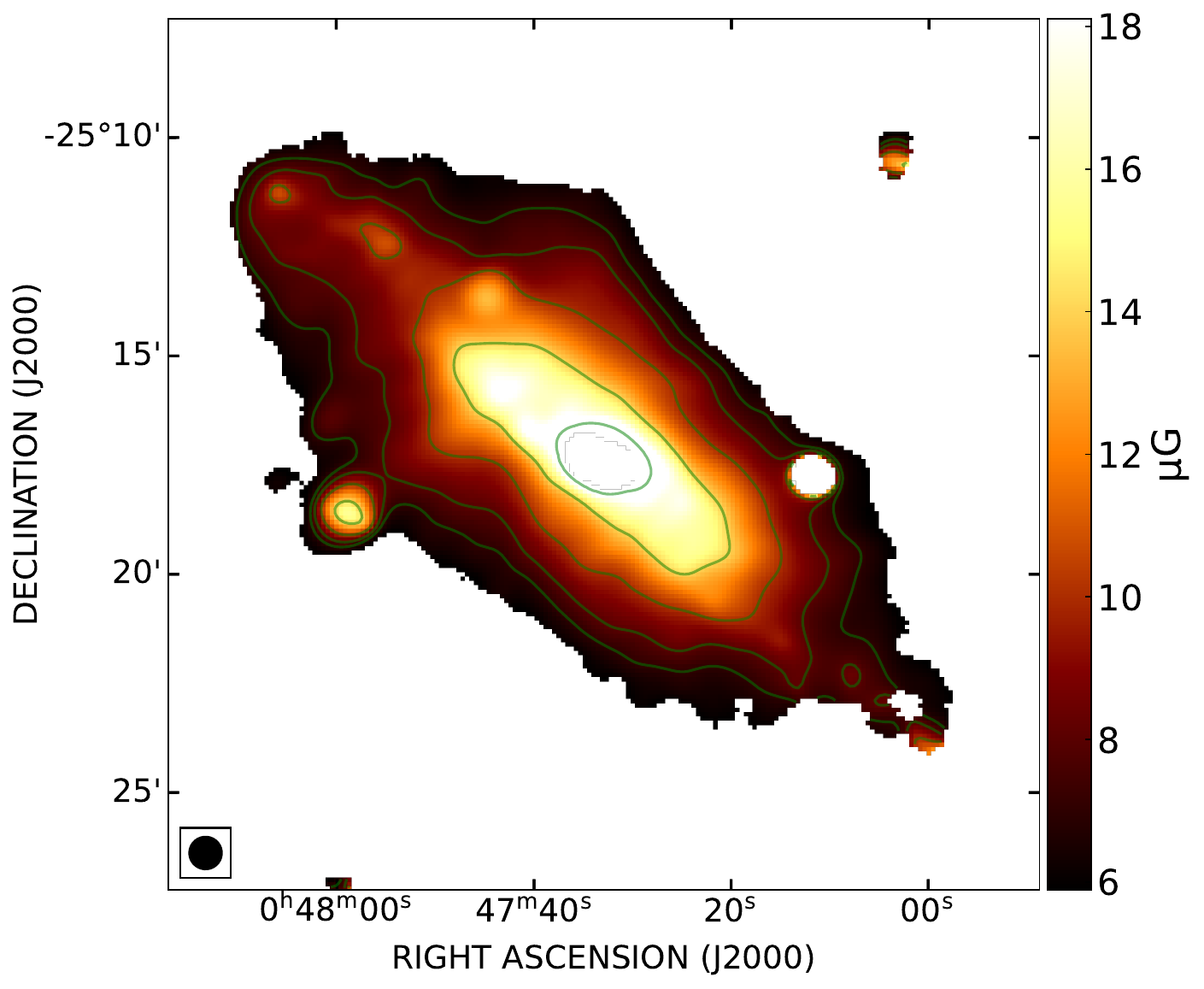}
    \caption{Map of the magnetic field strength with a resolution of $45\arcsec$. Contour levels are: 7, 8, 10, 15, and 20~$\mu$G.} 
\label{fig:B_distri}
\end{figure}

To derive the magnetic field strength at each position, we use the ASKAP synchrotron emission intensity and synchrotron emission spectral index maps at an angular resolution of $45\arcsec$. The same values of $K_0$ and $L$ as discussed above were adopted. To avoid overestimating the magnetic field strength, we constrained the spectral index to the range $-1.2 < \alpha < -0.6$, following the discussion by~\citet{heesen2022}. The resulting distribution of magnetic field strength is shown in Fig.~\ref{fig:B_distri}. It can be clearly seen that the magnetic field strength gradually decreases from 20~$\mu$G at the center of the galaxy to 6~$\mu$G at its outskirts, and this behavior is discussed in Sect.~\ref{subsec:adv_diff}.

\subsection{Uncertainties}
\label{sec:uncertain}
The uncertainty in the integrated flux density combines the contribution of relative calibration error and baseline error. The baselevel error $\sigma_{\rm N}=\sigma_{\rm rms}\sqrt{N_{\rm beam}}$, where \(N_{\rm beam}\) is the number of independent points within the region, calculated as \(N_{\rm beam} = A_{\text{region}} / A_{\text{beam}}\), where \(A_{\text{region}}\) is the area of the integration region and \(A_{\text{beam}}\) is the beam area. $\sigma_{\rm rms}$ is rms variation in the image:
\begin{equation}
    \sigma_{\rm rms}=\sqrt{\frac{\Sigma(p_i(x,y)-\mu)^2}{N}},
	\label{eq:sigma}
\end{equation}
where \(p_i(x, y)\) represents the value of the \(i\)-th pixel, $\mu$ is the mean value of all pixels within the region, and $N$ is the pixel number. The total error is $(\delta S_{\nu})^2=(\epsilon_{\rm cal}\,S_{\nu})^2+(\sigma_N)^2$, where $\epsilon_{\rm cal}$ denotes the relative flux-density calibration uncertainty. We adopt $\epsilon_{\rm cal}=0.05$ for ASKAP~\citep{duchesne2023}. For the MWA data, we adopt a calibration uncertainty of $\epsilon_{\rm cal}=0.10$ to account for additional systematic effects related to the wide field of view, including primary-beam and direction-dependent calibration uncertainties~\citep{duchesne2025}.

The error of the radio spectral index is calculated with the following equation:
\begin{equation}
\label{spix_error}
\Delta \alpha = \frac{1}{\ln\left(\nu_1/\nu_2\right)}
\sqrt{
\left(\frac{\delta S_{\nu_1}}{S_{\nu_1}}\right)^2 +
\left(\frac{\delta S_{\nu_2}}{S_{\nu_2}}\right)^2
},
\end{equation}
where $\delta S_{\nu_1}$ and $\delta S_{\nu_2}$ are the intensity uncertainties, and
$S_{\nu_1}$ and $S_{\nu_2}$ are the intensities measured at the observing frequencies
$\nu_1$ and $\nu_2$, respectively.

\section{Discussion}
\label{sec:discussions}

\subsection{Advection and diffusion: theory}
\label{subsec:adv_diff}
The transport of CREs from the disk into the halo can be described by two mechanisms, advection and diffusion, which can be expressed as~\citep{longair2011,heesen2016}:
\begin{equation}
  \frac{\partial N(E,z)} {\partial z} = \frac{1}{V}\left \lbrace
    \frac{\partial}{\partial E}\left [ b(E)
    N(E,z)\right ]\right \rbrace\qquad ({\rm Advection}),
\label{eq:adv}
\end{equation}

\begin{equation}
  \frac{\partial^2N(E,z)}{\partial z^2} = \frac{1}{D}\left \lbrace\frac{\partial}{\partial
    E}\left [ b(E) N(E,z)\right ]\right\rbrace\qquad ({\rm Diffusion}),
\label{eq:diff}
\end{equation}
where $V$ is the advection velocity, $D$ is the diffusion coefficient, and $N(E,z)$ denotes the number density of CREs as a function of energy $E$ and height $z$. The term $b(E)$ represents the energy loss rate of the CREs. We assume an energy-dependent diffusion coefficient of the form $D=D_0\,(E/{\rm GeV})^{\mu}$, while $D_0$ is the diffusion coefficient at $E=1$~GeV in the galactic midplane, and $\mu$ is the dependency factor on energy. The observing frequency $v$ can be related to the energy of CREs through the characteristic synchrotron frequency~\citep{beck2015}:
\begin{equation}
    E(\rm GeV)\approx\sqrt{\frac{\nu\,(\rm MHz)}{16\,\mathrm{MHz} \times B_{\perp}\,(\rm \mu G)}}.
\label{eq:v_to_E}
\end{equation}
where $B_{\perp}$ is the magnetic field component perpendicular to the line of sight.

CREs lose their energy mainly through synchrotron radiation and IC radiation. The combined rate of synchrotron and IC losses for CRE is described by \cite{longair2011}:
\begin{equation}
-\left (\frac{{\rm d}E}{{\rm d}t}\right )=b(E)=\frac{4}{3} \sigma_{\rm T} c \left (\frac{E}{m_{\rm e}c^2} \right )^2 (U_{\rm rad}+U_{\rm B}),
\label{eq:loss_rate}
\end{equation}
where $U_{\rm rad}$ is the radiation energy density, $U_{\rm B}=\mathrm{B^2/8\pi}$ is the magnetic energy density, $\sigma_{\mathrm{T}}=6.65\times 10^{-25}~\rm cm^2$ is the Thomson cross-section and $m_{\rm e}\,c^2=511$~keV. This energy is time-dependent, $E(t)=E_0(1+t/t_{\rm rad})^{-1}$. The CREs timescale $t_{\rm rad}$ can be determined through synchrotron and IC radiation losses \citep{heesen2016}:
\begin{equation}
t_{\rm rad} = 34.2 \left (\frac{\nu}{\rm 1~GHz}\right )^{-0.5}
\left (\frac{B}{\rm 10~\mu G}\right )^{-1.5} \left
  (1+\frac{U_{\rm rad}}{U_{\rm B}}\right )^{-1}~{\rm Myr}.
\label{eq:t_rad}
\end{equation}
where $U_{\rm rad}=U_{\rm CMB}+U_{\rm TIR}+U_{\rm star}$ is the total radiation energy density. The energy density of the cosmic microwave background (CMB) radiation is calculated as $U_{\rm CMB}=4\sigma/c\,T^{4}$, where $T=2.73$~K is the CMB temperature, $\sigma$ is the Stefan--Boltzmann constant, and $c$ is the speed of light. This yields $U_{\rm CMB}=4.2\times10^{-13}\,\mathrm{erg\,cm^{-3}}$. The starlight radiation energy density is related to the total infrared radiation energy density by $U_{\rm star}=1.73\,U_{\rm TIR}$, as found for the solar neighborhood \citep{draine2010}. The global infrared radiation energy density is estimated as $U_{\rm TIR}=L_{\rm FIR}/(2\pi r_{\rm int}^{2}c)$, where $r_{\rm int}=11.8$~kpc represents the galactocentric radius of the actively star-forming disk of NGC~253~\citep{heesen_2018b}. The total far-infrared luminosity of NGC~253 is
$L_{\rm FIR}=6.31\times10^{10}\,L_{\odot}$~\citep{dale2009}, which gives $U_{\rm TIR}=9.6\times10^{-13}\,\mathrm{erg\,cm^{-3}}$.
Consequently, the total radiation energy density is $U_{\rm rad}=3.0\times10^{-12}\,\mathrm{erg\,cm^{-3}}$. The magnetic field energy density of NGC~253 is
$U_{\rm B}=4.4\times10^{-12}\,\mathrm{erg\,cm^{-3}}$.
At 943~MHz, substituting these values into Eq.~\ref{eq:t_rad} gives a CRE radiative-loss timescale of $t_{\rm rad}=9.3\times10^{6}$~yr.

We used exponential functions to describe the variation with height $z$ of the advection velocity $V$ and the magnetic field strength $B$, as
\begin{equation}
  V=V_0\, \mathrm{exp}\left(\frac{|z|}{h_V}\right),
\label{eq:adv_v}
\end{equation}
and 
\begin{equation}
  B=B_0\,\mathrm{exp}\left(-\frac{|z|}{h_B}\right),
\label{eq:exp_b_1}
\end{equation}
where $V_0$ and $B_0$ are values at $z=0$, and $h_V$ and $h_B$ are scale heights. As shown later, these functional forms also provided a good fit to the observations. 

We assumed the transport to be symmetric across the disk and only solved the equations for $z\geq0$. We adopted a fixed inner boundary condition with $N(E,0)=N_0E^{-\gamma_{\mathrm{inj}}}$, where $\gamma_{\mathrm{inj}}$ is the energy spectral index of the injected CREs and $N_0$ is a normalization constant. 

\begin{figure} \includegraphics[width=\columnwidth]{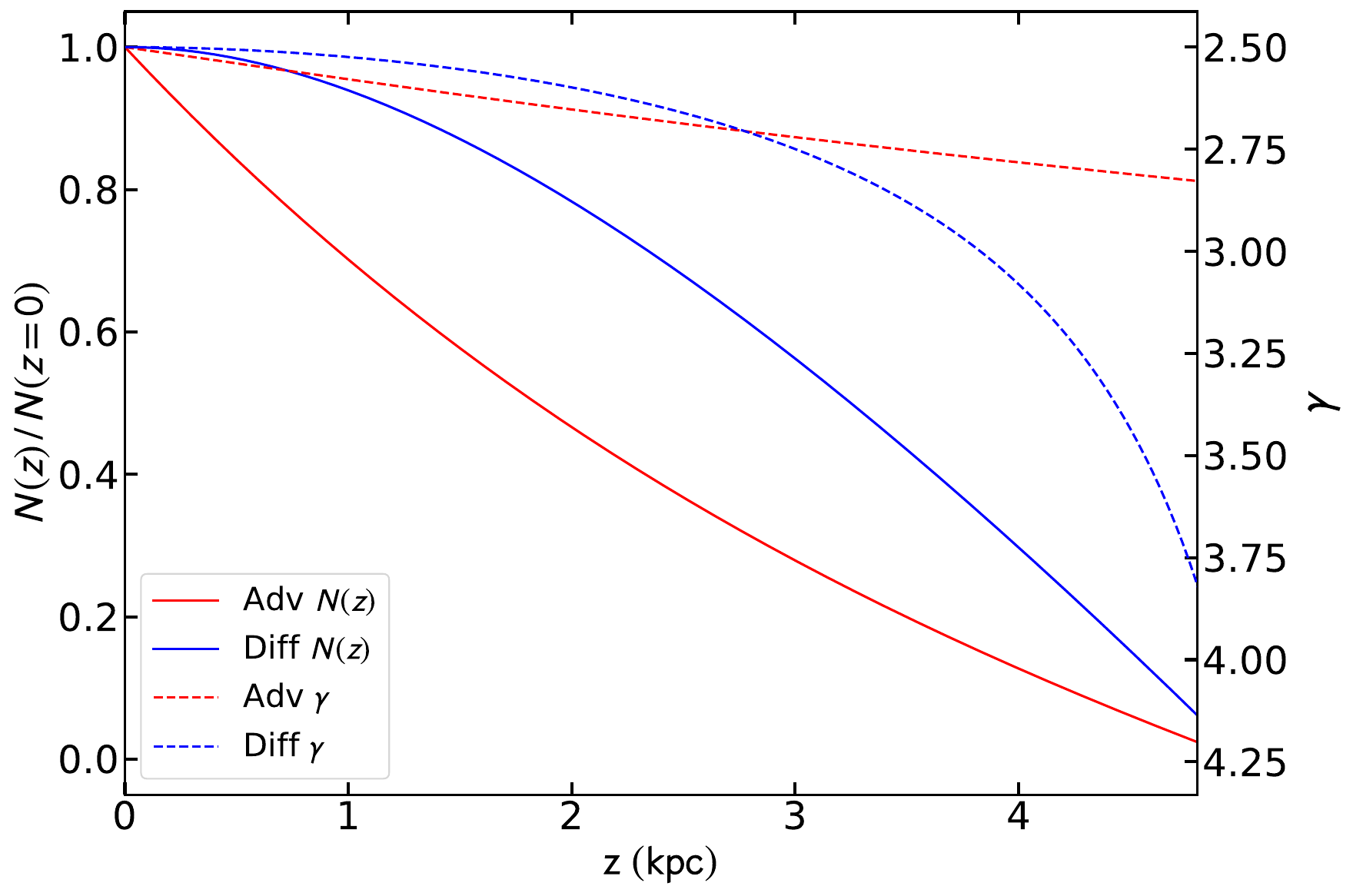}
    \caption{Normalized number density $N(z)/N(z=0)$ (solid lines) and energy spectral index $\gamma$ (dotted lines) of CREs versus $z$ derived by solving advection (red lines) and diffusion (blue lines) equations. The parameters are provided in Sect.~\ref{subsec:adv_diff}.} 
\label{fig:adv_diff}
\end{figure}

As we show later in Sect.~\ref{subsec:spin_fit}, the parameters $V_0$, $h_V$, $B_0$, $h_B$, $\mu$, and $\gamma_{\rm inj}$ can be derived from the best-fit models of the observation data. For demonstration, we solved the advection and diffusion equations using the following parameters: constant $V=V_0=$~150~km~s$^{-1}$, $B_0=13$~$\mu$G, $h_B=4$~kpc, $D_0=3\times10^{28}$~cm$^2$~s$^{-1}$, $\gamma_{\rm inj}=2.5$ and $\mu=0.5$ are consistent with values commonly adopted in models of CR propagation in the Milky Way~\citep{strong2007}. The solutions yield $N(E,z)=N(z)E^{-\gamma}$. The normalized number density $N(z)/N(z=0)$ and the energy spectral index $\gamma$ are shown in Fig.~\ref{fig:adv_diff}. There are clear distinctions of $N(E,z)$ and $\gamma$ between advection and diffusion. 

The synchrotron emission intensity can be derived by integrating the magnetic field multiplied by the CREs density over the entire frequency range as described by~\citet{heesen2016} and the spectral index of synchrotron emission $\alpha$ can be derived as $\alpha=(1-\gamma)/2$ . Fitting the solutions of advection and diffusion equations to the observed synchrotron emission intensity and spectral index $\alpha$ together enables us to differentiate the two processes and derive the parameters of transport models, which can be done with SPINNAKER. 

\subsection{Advection and diffusion: fitting NGC 253 observations}
\label{subsec:spin_fit}
We divided the rotated synchrotron emission intensity maps at 943~MHz and 216~MHz, together with the corresponding spectral index map of NGC~253, into 6 regions (blue dashed profiles in Fig.~\ref{fig:cover}), with 3 in the north ($z>0$) at the left, center and right: left north (LN, $-9$~kpc$<x<-3$~kpc), center north (CN, $|x|<3$~kpc), and right north (RN, $3$~kpc$<x<9$~kpc) and 3 in the south ($z<0)$ at the left, center and right: left south (LS, $-9$~kpc$<x<-3$~kpc), center south (CS, $|x|<3$~kpc), and right south (RS, $3$~kpc$<x<9$~kpc). We defined these regions for two reasons: (1) CREs probably propagate in different ways in the center and elsewhere because the intensity profile can be fitted with exponential functions for the former and Gaussian functions for the latter~(Figs.~\ref{fig:intensity-fit-north},  \ref{fig:intensity-fit-south}, and \ref{fig:intensity-fit-all}).  (2) The center regions (CN and CS) encompass the superbubble filled with X-ray emission as shown by~\cite[][their Fig.~19]{heesen_2009b}.

\begin{table*}[!htbp]
\caption{Best-fitting parameters for the transport models of CREs.}
\centering
\label{table:spin_fit_para}
\begin{tabular}{lcccccc}
\hline
                            & LN           & LS           & CN          & CS           & RN           & RS \\
                            &\multicolumn{2}{c}{$-$9~kpc$<x<-3$~kpc} & 
                            \multicolumn{2}{c}{$|x|<3$~kpc}&
                            \multicolumn{2}{c}{3~kpc$<x<9$~kpc}\\
                            &$z>0$ & $z<0$ & $z>0$ & $z<0$ &$z>0$ & $z<0$ \\
\hline
Model                                & diffusion   & diffusion     & advection & advection    & diffusion    & diffusion\\
$B_1\,(\rm \mu G)$\tablefootmark{*}  & $8.0$       & $8.0$         & $7.0$     & $7.0$        & $8.0$        & $8.0$   \\
$B_2\,(\rm \mu G)$\tablefootmark{*}  & $3.0$       & $3.0$         & $11.0$    & $11.0$       & $3.0$        & $3.0$    \\
$h_{B1}$ (kpc)                       & $3.0\pm0.5$ & $3.0\pm0.5$   & $2.0\pm0.3$& $1.5\pm0.5$ & $3.0\pm0.4$  & $2.5\pm0.3$    \\
$h_{B2}$ (kpc)                       & $10.0\pm1.0$& $10.0\pm1.0$  & $4.0\pm0.2$& $4.0\pm0.4$ & $10.0\pm1.0$ & $10.0\pm1.2$    \\
$D_0\,(10^{28}\,\rm cm^{2}\,s^{-1})$ & $3.0\pm0.3$ & $2.8\pm0.4$   & --         & --          & $3.0\pm0.3$  & $3.6\pm0.4$    \\
$\mu$                                & $0.40\pm0.05$& $0.45\pm0.04$& --         & --          & $0.45\pm0.03$ & $0.30\pm0.05$    \\
$\gamma_{\rm inj}$                   & $2.42\pm0.02$& $2.40\pm0.03$& $2.03\pm0.02$& $2.05\pm0.02$& $2.38\pm0.03$& $2.33\pm0.02$   \\
$V_0$ ($\rm km\,s^{-1}$)             & --          & --            & $135\pm10$  & $140\pm15$  & --          & --       \\
$h_V$ (kpc)                          & --          & --            & $4.5\pm0.4$ & $4.8\pm0.5$ & --          & --       \\ 
$\chi^2$                             & 0.98        & 0.76          & 0.93        & 0.86        & 1.18        & 0.76     \\
\hline
\end{tabular}
\tablefoot{\tablefoottext{*}{$B_1$ and $B_2$ were fixed when fitting the profiles with SPINNAKER.}}
\end{table*}

\begin{figure*}[!htbp]	
\centering \includegraphics[width=0.9\textwidth]{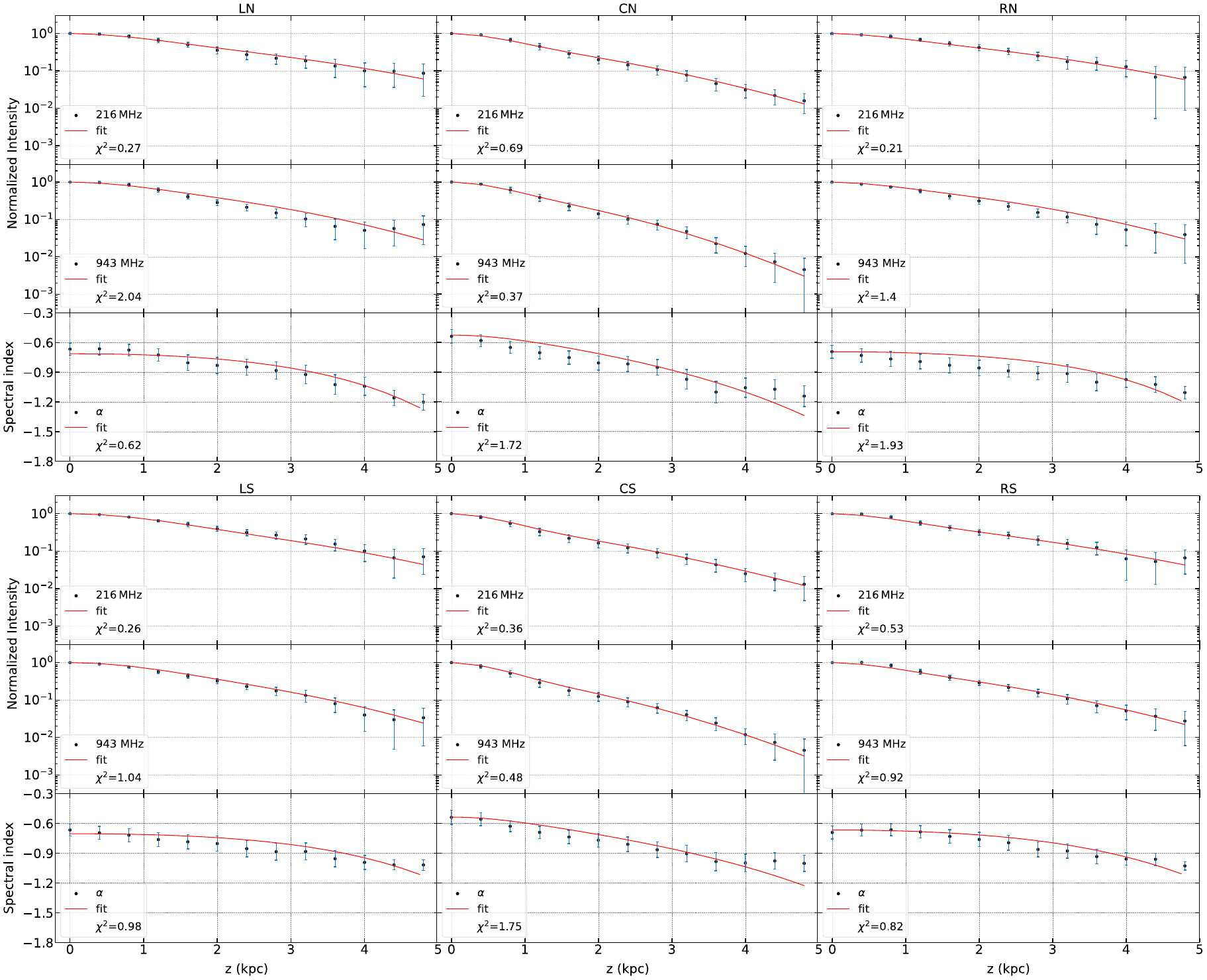}
    \caption{The SPINNAKER fitting results, with the upper and lower panels corresponding to the northern and southern, respectively. The central panels display the advection models, while the left and right panels show the diffusion models. In each panel, the rows from top to bottom present the synchrotron emission intensity profiles at 216~MHz (MWA) and 943~MHz (ASKAP), followed by the corresponding synchrotron spectral-index profile.} 
\label{fig:profiles_fit}
\end{figure*}

We derived the $z$ profiles of the synchrotron emission intensities at 943~MHz and 216~MHz with a convolved $45\arcsec$ resolution and the corresponding spectral index for all the 6 regions. The intensity profiles were normalized by the values at $z=0$. We then used SPINNAKER to fit these profiles to obtain the best solutions for the advection and diffusion equations. We found that two exponential components for the magnetic field were needed with parameters $B_1$, $h_{B1}$, $B_2$, and $h_{B2}$, as in Eq.~(\ref{eq:exp_b_1}). We fitted the magnetic field derived from energy equipartition in Fig.~\ref{fig:B_distri} and obtained these parameters as initial values. For fitting with SPINNAKER, we fixed $B_1$ and $B_2$ and thus the total magnetic field at $z=0$, but relaxed $h_{B1}$ and $h_{B2}$. 

The main degeneracy that needs to be resolved arises because either a high advection speed or a large diffusion coefficient can increase the CRE density in the halo. This increase can compensate for a weaker magnetic field, resulting in a radio continuum intensity that still matches the observations. Conversely, a stronger magnetic field can offset a lower CRE density in the halo, producing the same observed radio intensity. This degeneracy can be broken by using the radio spectral index. A higher advection speed or diffusion coefficient suppresses the radiative ageing of CREs, leading to a flatter radio spectral index profile with increasing height. Consequently, fitting both the radio intensity and the spectral index profiles allows us to distinguish between different transport scenarios. Synchrotron emission intensity profiles are generally better described by Gaussian functions for diffusion-dominated transport and by exponential functions for advection-dominated transport~\citep{heesen2016}.

For a given set of $h_{B1}$, $h_{B2}$, $D_0$, $\mu$, $\gamma_{\rm inj}$, $V_0$ and $h_V$, we solved the advection or diffusion equation to obtain the synchrotron emission intensity and spectral-index profiles and compared them with the observations. To account for projection effects from the inclined disk, the simulated synchrotron emission intensity profiles are convolved with the effective beam using the Gaussian function described in Eq.~\ref{eq:gauss_kernel}. By trial and error to minimize the reduced $\chi^2$ values resulting from fits to the synchrotron emission intensity and spectral index profiles, we derived the transport models and the parameters that can best reproduce the observations. The parameter uncertainties were estimated by varying each parameter around its best-fitting value. The quoted uncertainties represent the ranges over which the model profiles remained consistent with the observations within their corresponding $1\sigma$ measurement uncertainties. The profiles and fits are shown in Fig.~\ref{fig:profiles_fit}, and the parameters for the transport models of CREs are listed in Table~\ref{table:spin_fit_para}.

From the fitting results, it can be seen that CREs propagate differently in the center as compared to elsewhere. From the analysis by \cite{heesen_2009b}, it was shown that for $-$6~kpc$<x<$0~kpc, the transport is through advection and for 1.5~kpc$<x<6$~kpc, the transport is more likely through diffusion. Our higher-sensitivity and higher-resolution data confirm that CRE transport is advection-dominated in the central region and diffusion-dominated in the outer region, consistent with earlier results.

\begin{figure} \includegraphics[width=\columnwidth]{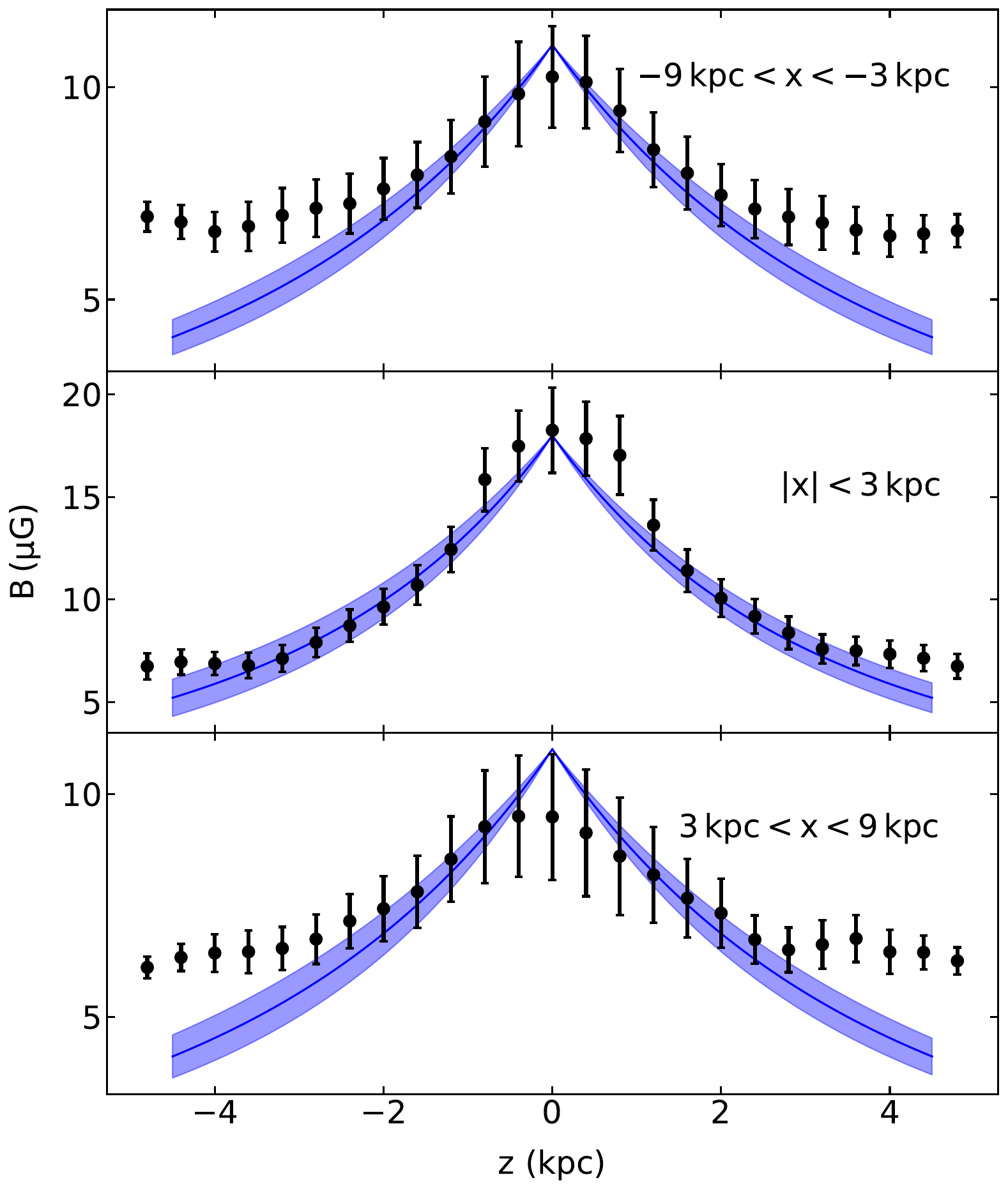}
    \caption{Equipartition magnetic field strength (dots) in comparison to that from SPINNAKER simulation (blue shaded area for $1\sigma$ error) in the left, center, and right regions.} 
\label{fig:B_model}
\end{figure}

For consistency check, we plot the $z$ profiles of equipartition magnetic field (Fig.~\ref{fig:B_distri}) versus that derived from fitting with SPINNAKER in Fig.~\ref{fig:B_model} for $-9$~kpc$<x<-3$~kpc (left), $|x|<3$~kpc (center), and 3~kpc$<x<$9~kpc (right). For the center regions where CREs are transported by advection, the fitting from SPINNAKER agrees well with the equipartition values. In the diffusion-dominated outer regions of NGC~253, the magnetic field strengths derived from the CR transport modeling are lower than those estimated under the equipartition assumption. Because of reduced synchrotron and IC losses at large heights, CREs can efficiently populate the halo even in the presence of relatively weak magnetic fields. Note that the difference is less than a factor of about 2, meaning that energy equipartition still provides a reasonable first-order estimate of the magnetic field strength.

\subsection{Superwind from the center}
\label{subsec:superwind}
The existence of a superwind, with speed reaching hundreds to thousands of km~s$^{-1}$, in the center of NGC~253 has been confirmed by multiple observations~\citep{strickland2000,bolatto2013,lopez2023,thompson2024,cronin2025}. The radio observations can provide an independent examination of the superwind scenario and its associated emission profile in our analysis.

\begin{figure}	\includegraphics[width=\columnwidth]{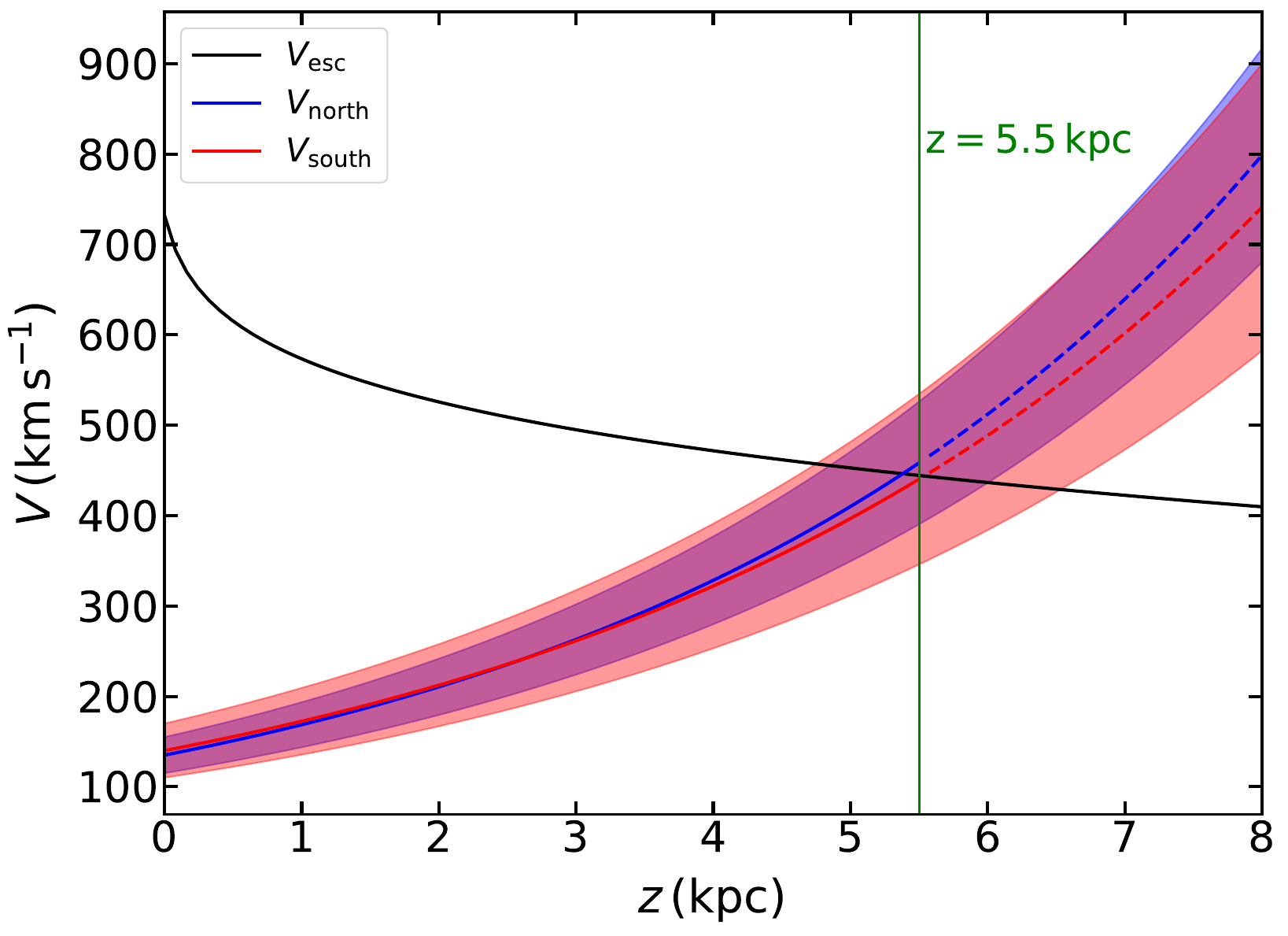}
    \caption{Advection velocity of the CREs for the northern halo (blue line) and southern halo (red line) of NGC~253, with the shaded areas representing $1\sigma$ uncertainties. The black line shows the escape velocity, while the green line marks the height at which the CREs reach the escape speed ($\rm z=5.5\,kpc$). Dashed segments of the velocity curves denote the extrapolated portions of the profiles beyond the range directly constrained by the radio data.} 
\label{fig:gal_wind}
\end{figure}

From the fitting with SPINNAKER, we find that the CREs are transported by advection in the center region, and the advection speed increases exponentially, which can potentially form winds. In Fig.~\ref{fig:gal_wind}, we plot $z$ profiles of the advection velocity of CREs for the southern halo ($z<0$) and the northern halo ($z>0$) using the parameters from Table~\ref{table:spin_fit_para}. Following \cite{veilleux2005} and \cite{stein2022}, we also obtained the escape velocity $V_{\rm esc}$ as
\begin{equation}
  V_{\rm esc}=\sqrt{2}V_{\rm rot}\sqrt{1+\mathrm{ln}\left(\frac{R_{\rm max}}{r}\right)},
\label{eq:esc_speed}
\end{equation}
where $V_{\rm rot}=200$~km~s$^{-1}$ is the rotational velocity of the galaxy~\citep{pence1981}, $r$ is the distance from the galactic center and $R_{\rm max}$ is the radius of the truncated isothermal dark matter halo, assumed to be 30~kpc~\citep{greggio2014}. The profile of $V_{\rm esc}$ is also shown in Fig.~\ref{fig:gal_wind}.

In both the northern and southern halos of the galaxy, the propagation speed of CREs increases very rapidly. At the height of approximately 5.5~kpc, the speed of CREs reaches the escape velocity of 462~km~s$^{-1}$. At larger heights, the lack of detectable radio emission means that we only have extrapolated velocities, which indicate that the CREs continue to accelerate outward and may escape into intergalactic space.

NGC 253 is a prototypical nuclear starburst hosting a multiphase, nuclear-outflow–driven superwind. Fig.~\ref{fig:Three_color_map_optical} presents a three-color composite of [{\sc O\,iii}] (blue), [{\sc N\,ii}] (green), and $\rm H\alpha$ (red) from the latest TYPHOON observations, clearly revealing the nuclear-starburst-driven conical outflow of ionized gas traced by [{\sc O\,iii}] and [{\sc N\,ii}] (outlined by the red trapezoid). Chandra soft X-ray imaging reveals a limb-brightened, kiloparsec-scale cone that is broadly co-spatial with the optical ionized gas, tracing the hot, shock-heated phase of the wind~\citep{strickland2000,lopez2023}. Atacama Large Millimeter Array (ALMA) detects a massive molecular outflow emerging from the central few hundred parsecs, implying substantial mass loading~\citep{bolatto2013}. Taken together, these data show that the nuclear outflow entrains and expels ionized, neutral, and molecular gas while coupling to the hot plasma, magnetic fields, and CRs, thereby feeding and sustaining the galaxy-scale superwind and contributing to chemical enrichment of the inner halo.

\begin{figure} \includegraphics[width=\columnwidth]{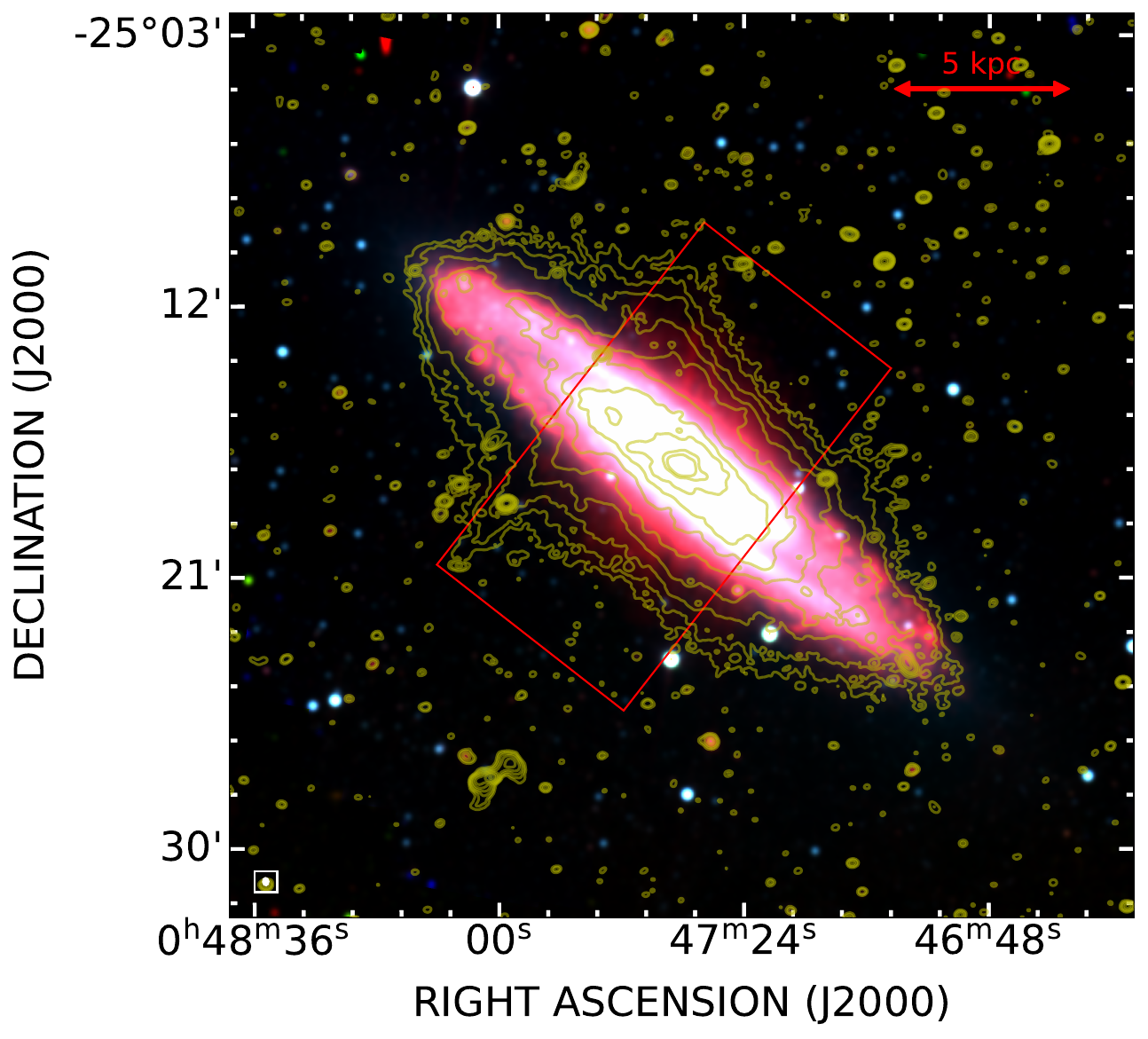}
    \caption{Three-color IRAC image ($3.6\,\mu$m in blue, $4.5\,\mu$m in green, and $8\,\mu$m in red) overlaid with ASKAP total intensity contours. The red box marks the central superwind region, and all images are convolved to $13\arcsec$ resolution, shown by the white circle.} 
\label{fig:w1_w3_color_map}
\end{figure}

\begin{figure} \includegraphics[width=\columnwidth]{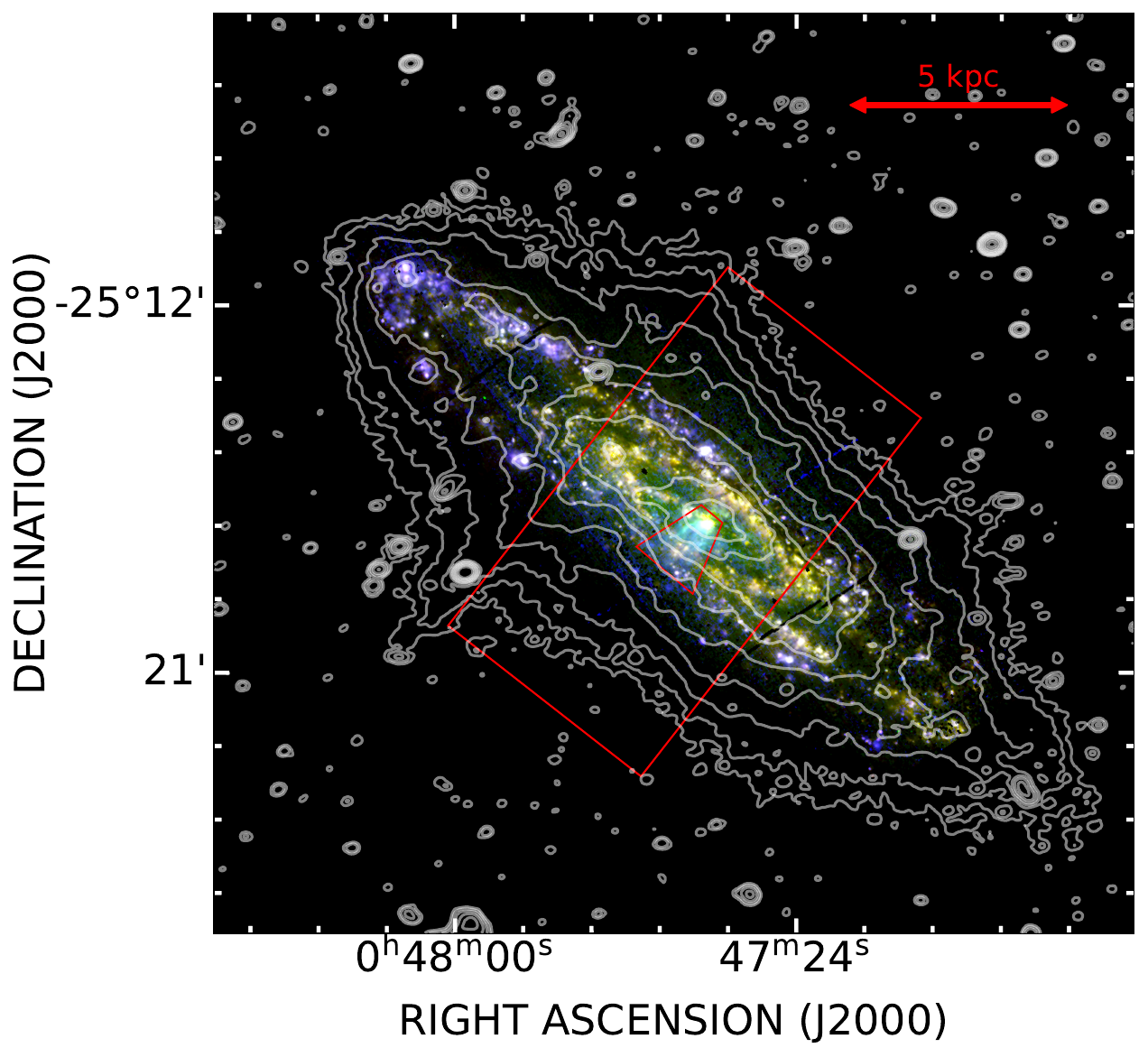}
    \caption{Three-color composite TYPHOON image combining [{\sc O\,iii}] (blue), [{\sc N\,ii}] (green), and $\rm H\alpha$ (red), overlaid with the ASKAP 943~MHz total-intensity contours. The nuclear outflow is outlined by the red trapezoid, and the red box (6~kpc$\times11$~kpc) outlines the center region with superwind.} 
\label{fig:Three_color_map_optical}
\end{figure}

Beyond the nucleus, sustaining a kiloparsec-scale superwind requires distributed star formation across the inner disk. In Fig.~\ref{fig:w1_w3_color_map}, we present the three-color image of NGC~253 convolved to a resolution of $13\arcsec$, with the $3.6\,\mu$m emission shown in blue, the $4.5\,\mu$m emission in green, and the $8\,\mu$m emission in red, overlaid with the ASKAP total intensity contours. The center region of $\rm 6~kpc\times11~kpc$ harboring the superwind of CREs is outlined with a red rectangle. The pronounced $8\,\rm \mu m$ brightness in this region indicates strong Polycyclic aromatic hydrocarbon (PAH) excitation by a hard UV radiation field, consistent with recent massive star formation. The associated concentration of young OB stars likely provides substantial stellar feedback that can contribute to the large-scale galactic wind.

To trace faint, large-scale halo emission, we combine the 943~MHz ASKAP and ROSAT 0.1–2.4~keV data. The ASKAP map is convolved to a resolution of $45\arcsec$ to enhance the diffuse emission, and the ROSAT image retrieved from the NASA Extragalactic Database (NED)\footnote{\url{https://ned.ipac.caltech.edu/}} is convolved to the same resolution, as shown in Fig.~\ref{fig:NGC253_xray_radio}. Both bands show a clear spatial correspondence in the S1 and S2 regions. In addition, {\sc H\,i} emission extends to vertical heights similar to X-ray emission, with slight spatial offsets~\citep{boomsma2005}; $\rm H\alpha$ emission also reaches nearly the same extent as radio emission~\citep{hoopes1996,hoopes2005,heesen_2009b}. The correspondence of multi-wavelength emission confirms the existence of the superwind with CREs and other gas components propagating together in a bulk motion. 

\subsubsection{Origin and nature of the loops}

\begin{figure} \includegraphics[width=\columnwidth]{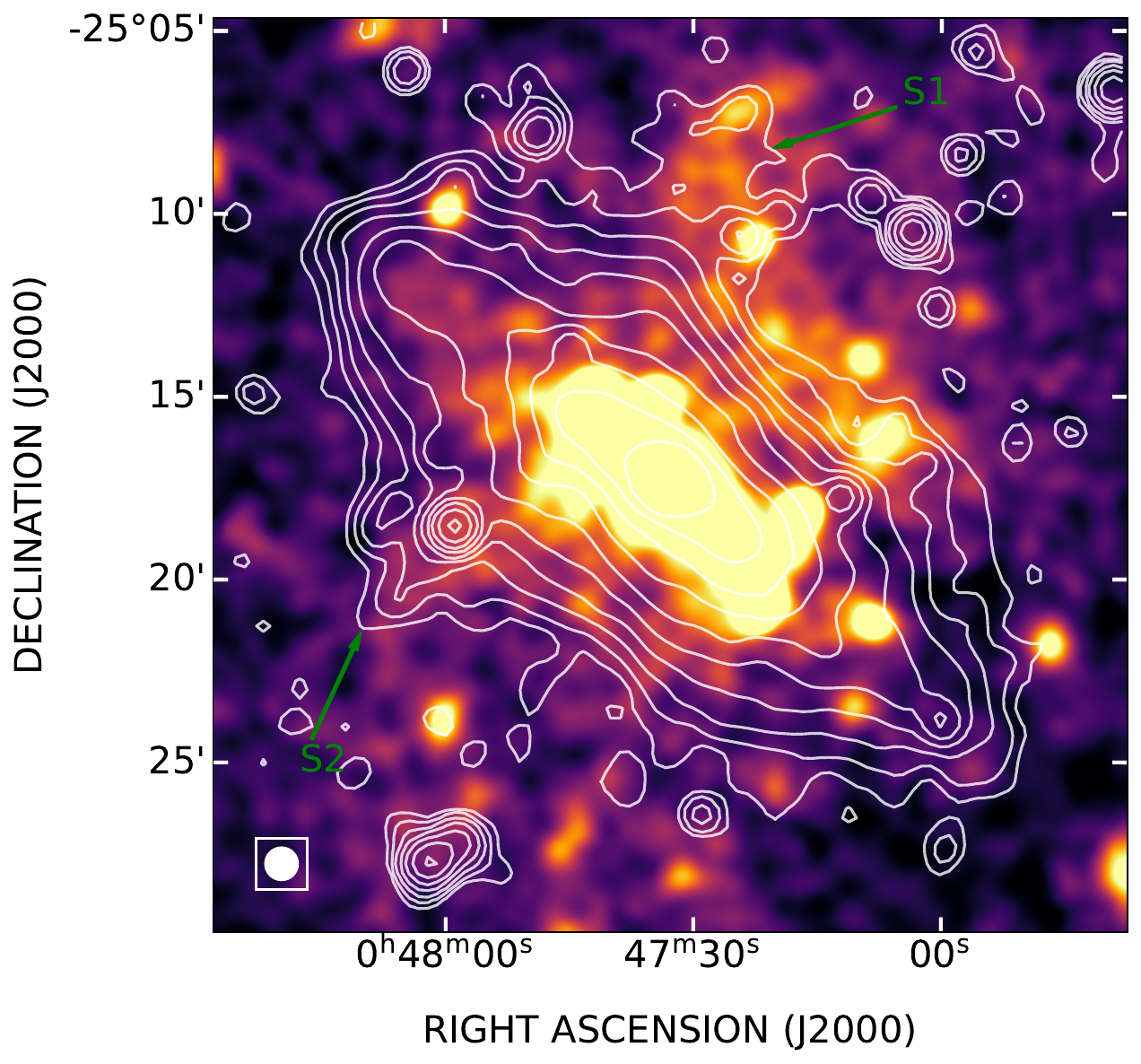}
    \caption{ASKAP total intensity contours overlaid on the ROSAT 0.1-2.4~keV X-ray image. Two extended ``radio spurs'' are labeled as S1 and S2. The resolution is $45\arcsec$.} 
\label{fig:NGC253_xray_radio}
\end{figure}

The loop structure in region S1 may be part of the wall of a large-scale superbubble or wind-blown cavity produced by an outflow from the disk. The multiwavelength studies of NGC~253 ({\sc H\,i}, $\rm H\alpha$, soft X-rays, radio continuum and polarization) have established a coherent picture of the starburst-driven superbubbles and superwind in this galaxy~\citep{strickland2000,strickland2004,boomsma2005,bauer2008,heesen_2009b,romero2018}. Region S1 lies along the rim of the known superbubble and shows an excellent spatial correspondence with the soft X-ray emission in Fig.~\ref{fig:NGC253_xray_radio}, suggesting that the loop traces a limb-brightened segment of the bubble wall where the outflow has opened a funnel or chimney-like cavity toward the halo, rather than a small, closed, spherical bubble. In a nearly edge-on view, the emission from {\sc H\,i}, $\rm H\alpha$ and soft X-rays is integrated along the line of sight, so radiation from the front and back sides of the shell, the hot interior, and unrelated foreground/background material is mixed in projection. As a result, any density cavity associated with the bubble may not appear as a clear ``hole'' in the existing data. Higher-resolution and higher-sensitivity observations in these bands will be required to test this scenario more stringently.

Another possible explanation for the loop is that it traces a large magnetic arch produced by Parker-type magnetic buoyancy instability in the stratified disk–halo system. In this picture, horizontal magnetic fields and CR pressure that support the gas against gravity become buoyantly unstable, causing field lines to arch into the halo and lifting magnetized plasma and CRs along them, thereby forming large-scale magnetic loops with synchrotron emission enhanced along the loop compared to the surrounding halo~\citep{Parker1966,matsumoto1988,Basu1997,Tharakkal2023}. In the Parker-type magnetic buoyancy scenario, the growth time of a large-scale magnetic loop can be approximated as $t_{\rm grow}\simeq h_e/V_A$, where $h_e$ is a characteristic density scale height of the magnetized ionized gas and $V_A=B/\sqrt{4\pi\rho}$ is the Alfvén speed. Assuming that the hot ionized gas dominates in the S1 loop region of NGC~253, we adopt a total particle number density of $n\simeq6.8\times10^{-3}\,\rm cm^{-3}$~\citep{strickland2002,romero2018}. For a mean molecular weight of $\mu=0.6$, the corresponding mass density is $\rho=\mu m_p n$. Adopting a halo magnetic-field strength of $B\simeq6\,\rm\mu G$, we obtain an Alfv\'en speed of $V_A\simeq205\,\rm km\,s^{-1}$. Adopting $h_e=1.8$~kpc, the typical Milky Way value reported by~\citet{gaensler2008}, we obtain a Parker growth time of $t_{\rm grow}\simeq8.6\,\rm Myr$, comparable to the CRE radiative-loss timescale derived in Sect.~\ref{subsec:adv_diff}. Using the H\,\textsc{i} kinematics of NGC~253, the anomalous extraplanar gas is found to lag the thin disk by $\Delta V_\phi \sim 80-100~{\rm km\,s^{-1}}$ at radii of a few kpc \citep{boomsma2005,lucero2015}. At the galactocentric radius of S1, $R \sim 5$~kpc, this implies a characteristic shear timescale
\begin{equation}
    t_{\rm shear} \sim \frac{1}{\Delta\Omega_z} = \frac{R}{|\Delta V_\phi(z)|}\sim (5\text{-}6)\times 10^{7}~{\rm yr},
\end{equation}
where $\Delta\Omega_z = |V_\phi(z_1)-V_\phi(z_2)|/R$ is the difference in angular velocity between heights $z_1$ and $z_2$. Consequently, $t_{\rm shear}$ is an order of magnitude longer than the Parker growth time. Differential rotation is therefore expected mainly to twist and stretch the magnetic arch on longer timescales, rather than to prevent the formation of the S1 loop.

In the Parker-type picture, the S1 loop is supported primarily by non-thermal components, namely the magnetic field and CRs. Approximating the loop as a cylindrical structure with a projected radius of $R=2$~kpc and adopting a line-of-sight depth of $l=1$~kpc, we obtain a volume $V_{\rm loop}=3.7\times10^{65}\,\rm cm^3$. For a magnetic-field strength of $B\simeq6\,\rm \mu G$, the magnetic energy density is $u_B=B^2/(8\pi)\simeq1.4\times10^{-12}\,\rm erg\,cm^{-3}$. Assuming approximate energy equipartition between CRs and magnetic fields, the total non-thermal energy stored in the loop is $E_{\rm nt}=(u_B+u_{\rm CR})V_{\rm loop}\simeq1.1\times10^{54}$~erg. If this non-thermal energy was ultimately supplied by the supernova- and stellar-wind-driven galactic outflow, the available mechanical luminosity can be estimated as $L_{\rm w}=7\times10^{41}\epsilon\,(\mathrm{SFR}/M_\odot\,\rm yr^{-1})\,erg\,s^{-1}$~\citep{veilleux2005}, where $\epsilon = 0.75$ is the thermalization efficiency~\citep{romero2018}. Adopting ${\rm SFR} \simeq 5~M_\odot\,{\rm yr^{-1}}$ for NGC~253 gives $L_{\rm w} \approx 2.6\times10^{42}\ {\rm erg\,s^{-1}}$. Over the characteristic Parker growth timescale of $t_{\rm grow} \simeq 8.6$~Myr, the total injected mechanical energy is $E_{\rm w}= L_{\rm w}t_{\rm grow}\simeq7\times 10^{56}$~erg. Thus, a coupling fraction of only $E_{\rm nt}/E_{\rm w}\sim0.2\%$ would be required, making it energetically plausible that the magnetic and CR energy currently stored in the S1 loop was ultimately supplied by the galactic wind.

Like S1, S2 may trace another limb-brightened segment of the same large-scale superbubble or wind-blown cavity, corresponding to a different side of the bubble wall seen in projection. In the nearly edge-on geometry of NGC~253, emission from different parts of the bubble shell (front, back, and lateral walls) is projected along the line of sight and can appear as spatially separated radio loops or spurs. Variations in local gas density, magnetic field strength, and CRE distribution may naturally lead to differences in brightness and morphology between S1 and S2. Deeper, higher-resolution multiwavelength and polarization observations will be required to determine whether S1 and S2 indeed trace different segments of the same expanding structure.

\subsection{Pressure}
\label{subsec:pressure}

There are various pressure components that maintain the superwind. In the following discussion, we considered thermal pressure, magnetic pressure, CR pressure, ram pressure, and opposing gravitational pressure. 

The thermal pressure is expressed as $P_{\mathrm{therm}} = 2\,n_e(z)\,k_B\,T(z)$, where $n_e(z)$ is the thermal electron density, $k_B$ is the Boltzmann constant, and $T(z)$ is the temperature of the gas at height of $z$ perpendicular to the disk of NGC~253. In the halo of star-forming galaxies, the hot ($\sim10^6-10^{7}$\,K) X-ray–emitting plasma is known to dominate the thermal pressure of the multiphase medium \citep[e.g.][]{strickland2000,strickland2004,li2013}. We therefore estimate the thermal pressure using the density and temperature profiles derived from the X-ray data. We adopted $n_e(0) \approx 0.025\,\mathrm{cm}^{-3}$ as a first-order estimate of the disk-plane density of the hot ionized gas, based on the southeastern nuclear-outflow region closest to the disk~\citep{bauer2007}, and assumed an exponential vertical distribution, $n_e(z) = n_e(0)\,\exp(-|z|/h_e)$, with $h_e=1.8$~kpc. Based on X-ray observations, the gas temperature has been estimated in multiple regions, and the X-ray data trace the halo out to approximately $5$–$8$\,kpc above the disk~\citep{bauer2008}. We extracted the temperature values along the central vertical profile of NGC~253 and performed a double exponential fit to obtain $T(z)$. 

Assuming local energy equipartition, the CR and magnetic energy densities at each height satisfy $u_{\rm CR}(z)=u_B(z)=B^2(z)/(8\pi)$. Treating CRs as a relativistic fluid with an adiabatic index of
$\gamma_{\rm CR}=4/3$, the corresponding pressures are
$P_{\rm CR}(z)=(\gamma_{\rm CR}-1)u_{\rm CR}(z)
=B^2(z)/(24\pi)$ and
$P_B(z)=B^2(z)/(8\pi)$~\citep{rodriguez2024}, where $B(z)$ is derived from the equipartition magnetic-field strength distribution shown in Fig.~\ref{fig:B_distri}. Although the superwind regions are not strictly in steady state, CREs are expected to remain strongly coupled to the magnetic field through resonant scattering with Alfv\'en waves, even in the presence of large-scale outflows. In the halo of NGC~253, the characteristic advection timescale $t_{\rm adv}\sim(0.7-3.3)\times10^7$~yr for $z\sim3-5$~kpc, $V\sim150-400\,\rm km\,s^{-1}$, is comparable to the synchrotron and IC cooling timescale $9.3\times10^6$~yr. The similarity of these timescales implies that CRs and magnetic fields evolve in a coupled manner, making the assumption of approximate energy equipartition a reasonable first-order approximation. Importantly, in the central superwind region of NGC~253, the magnetic-field strengths derived from the CREs transport modeling with SPINNAKER, which does not assume local energy equipartition, are in good agreement with those obtained independently from the equipartition method. This consistency supports the equipartition approximation on kiloparsec scales in the thick disk, halo and superwind regions of NGC~253. 

The ram pressure associated with the bulk outflow is estimated as $P_{\mathrm{ram}} = \mu_e\,n_e(z)\,m_p\,V(z)^2$, where $\mu_e\simeq1.18$ is the mean molecular weight per electron adopted for the ionized plasma~\citep{sharma2012}, and $V(z)$ is the advection speed inferred from the SPINNAKER modeling. This estimate assumes that the CREs remain sufficiently coupled to the magnetized thermal plasma for their inferred advection speed to trace the bulk outflow velocity.

The gravitational pressure can be modeled as~\citep{ostriker2010}:
\begin{align}
P_{\mathrm{grav}}(z) \approx\ 
&\ 2\pi G\,\Sigma_g\,\rho_{g,0}\,h_g\, 
    \mathrm{sech}^2\left( \frac{z}{2h_g} \right) \nonumber \\
&\ 
+ 2\pi G\,\Sigma_\star\,\rho_{g,0}
    \int_{z}^{\infty} 
        \mathrm{sech}^2\left( \frac{z}{2h_g} \right)
        \tanh\left( \frac{z}{2h_\star} \right) dz,
\label{eq:grav_pre}        
\end{align}
where $G$ is the gravitational constant, $\Sigma_g=4\,M_\odot\,\rm pc^{-2}$~\citep{puche1991} and $\Sigma_\star=350\,M_\odot\,\rm pc^{-2}$~\citep{iodice2014,pence1981} are the gas and stellar surface density. $\rho_{g,0}=\pi G \Sigma_g^2/2\sigma_g^2$ is the gas density in the disk. The vertical scale heights of the gas and stellar disks are given by $h_g=\sigma_g^2/2\pi G\Sigma_g$ and $h_\star=\sigma_\star^2/2\pi G\Sigma_\star$, where $\sigma_g=20\,\rm km\,s^{-1}$~\citep{sakamoto2011} and $\sigma_\star=70\,\rm km\,s^{-1}$~\citep{prada1998} are the velocity dispersions of the gas and stars, respectively.

\begin{figure*}[!htbp]
    \centering
    \includegraphics[width=0.47\textwidth]{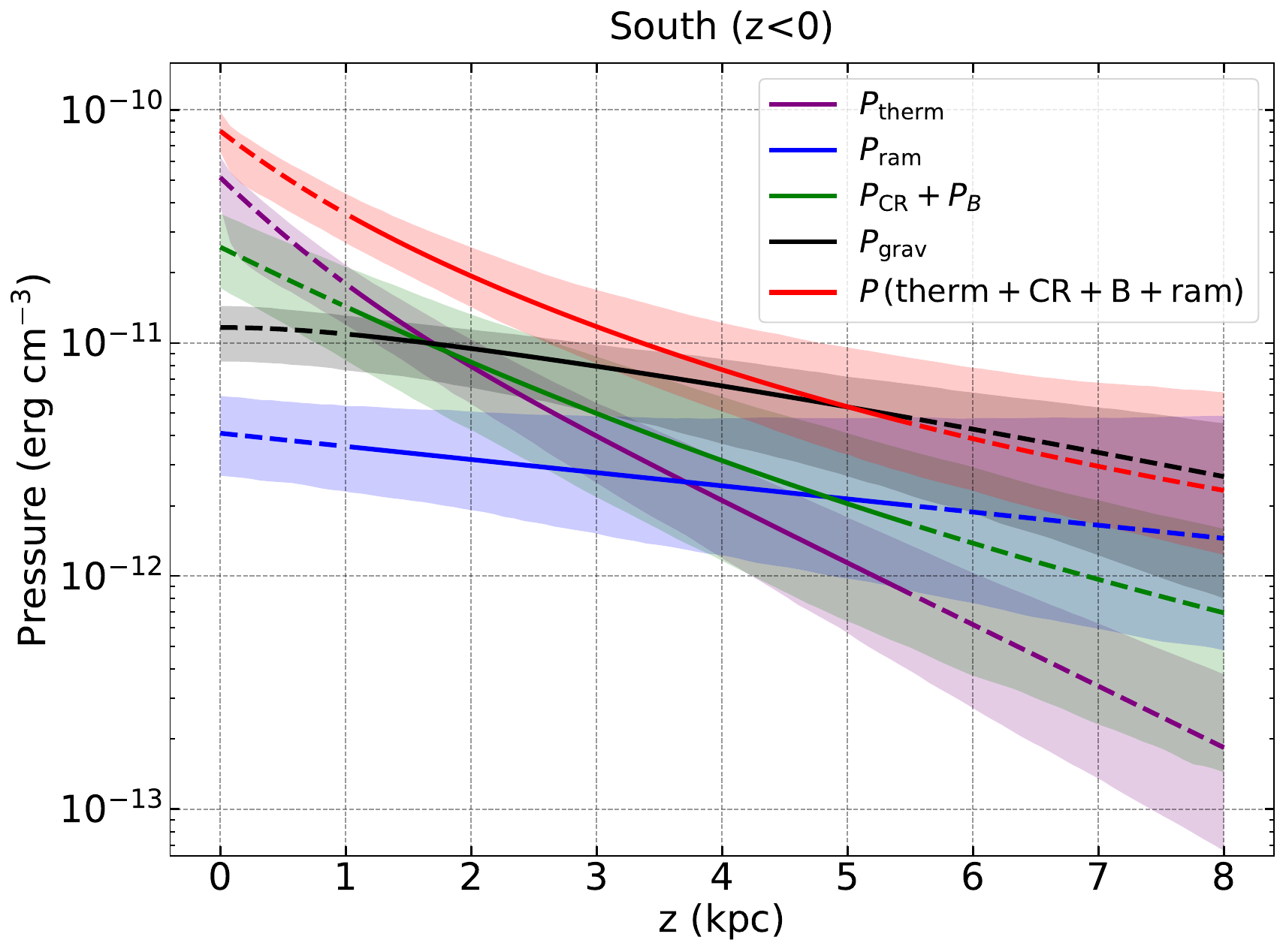}
    \includegraphics[width=0.47\textwidth]{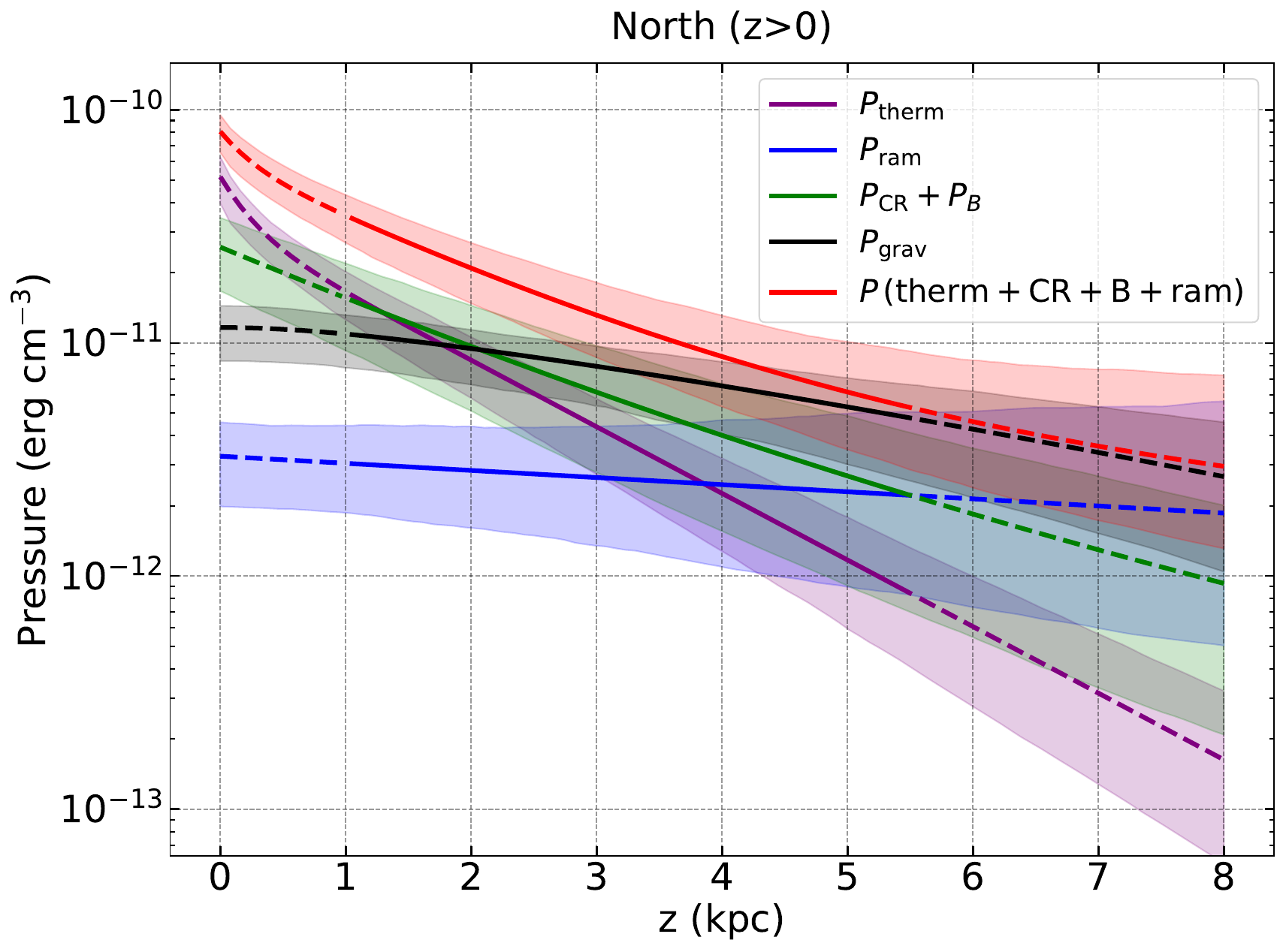}
    \caption{Pressure versus $z$ for the southern halo ($z<0$) and the northern halo ($z$>0). The red curve shows the total pressure—thermal (purple), ram (blue), magnetic, and cosmic-ray (green), which is greater than or comparable to the gravitational pressure (black) at all heights above the galactic midplane. The shaded region indicates the $1\sigma$ uncertainties for each component, and dashed segments denote the extrapolated portions of the profiles. At $z<1$~kpc, the emission is affected by projections of disk contamination, whereas at $z>5.5$~kpc the profiles extend beyond the range directly constrained by radio data.}
    \label{fig:pressure}
\end{figure*}

We show all pressures as functions of $z$ for the southern halo ($z<0$) and the northern halo ($z>0$) in Fig.~\ref{fig:pressure}. The thermal pressure $P_{\rm therm}$ is higher than the combined CR and magnetic pressure $P_{\rm CR}+P_{\rm B}$ near the disk and becomes smaller far away from the disk. The outward pressure, $P = P_{\rm therm} + P_{\rm CR} + P_{\rm B} + P_{\rm ram}$, exceeds the inward gravitational pressure $P_{\rm grav}$ at all heights below $\sim5$~kpc. This is consistent with the SPINNAKER results and implies that the gas is persistently accelerated within this region, which can be seen in Fig.~\ref{fig:gal_wind} showing increasing speed of the superwind. At $|z|<1$~kpc, the measurements are affected by projected emission from the inclined disk, while at heights exceeding 5~kpc the pressure can only be estimated by extrapolation owing to the limitations of the radio data. Above $\sim5$~kpc, if the extrapolation in Fig.~\ref{fig:pressure} is valid, the flow is expected to approach a terminal velocity. At greater heights, if the gravitational pressure exceeds the total outward pressure, part of the material may stall and eventually fall back to the disk in a galactic fountain flow; otherwise, it can escape into the circumgalactic medium (CGM) or even into the intergalactic medium (IGM). The ASKAP total intensity images reveal faint, extended emission at heights exceeding 5~kpc, particularly in the southern halo, supporting the presence of large-scale outflows.

In the midplane of NGC~253, the advection of CREs is most plausibly associated with a starburst-driven thermal wind. Our pressure analysis shows that, close to the disk, the thermal pressure dominates or is at least comparable to the non-thermal components, indicating that the outflow is primarily launched by hot gas heated by supernova feedback. CRs and magnetic fields contribute significantly to the total pressure budget and may help sustain or further accelerate the flow at larger heights, but they are unlikely to dominate the driving of the outflow in the launch region. The CREs advection speeds derived from the transport modeling therefore provide an important tracer of the bulk velocity of the thermally driven superwind in the central region.

Future broadband, low-frequency, and higher sensitivity radio continuum observations will be crucial for probing the outer halo and possibly the CGM, and for extending the constraints on the advection and diffusion models to larger heights. Our ASKAP data do not reveal significant polarized emission, most likely due to strong Faraday depolarization at 943~MHz; nevertheless, polarization measurements remain essential, as they will map the ordered magnetic-field geometry and directly test whether the galactic wind advects a coherent vertical field into the halo, as well as provide further support for our interpretation of the newly detected loop structure. 

\section{Conclusions}
\label{sec:conclu}
We re-processed the observations of NGC~253 by ASKAP and MWA by improving the calibration and imaging, and obtained total intensity images at 943~MHz and 216~MHz. We achieved an rms noise of 16~$\mu$Jy~beam$^{-1}$ and a resolution of $13\arcsec$ for the ASKAP image and an rms noise of 1~mJy~beam$^{-1}$ and a resolution of $45\arcsec$ for the MWA image. The rms noise of the ASKAP image has reached the confusion limit. These images have a much better sensitivity than the previous images, allowing us to detect weak emission further away from the disk, and thus study the propagation of CREs. 

Based on the synchrotron emission intensity maps obtained with ASKAP at 943~MHz and MWA at 216~MHz, we examined the synchrotron emission intensity profiles perpendicular to the galactic disk. We found that these intensity profiles can be fitted with exponential components in the central region ($x\lesssim3$~kpc) and Gaussian components elsewhere (3~kpc$<|x|<$9~kpc), implying that CREs propagate differently. We also obtained the scale heights for synchrotron emission.  

Using SPINNAKER, the advection and diffusion equations were solved to obtain vertical profiles of the synchrotron emission intensity at 943~MHz and 216~MHz and the spectral index. By comparing these profiles with observations, the propagation models of CREs can be determined. We found that CREs were transported by advection in the central regions ($|x|<$3~kpc) and by diffusion elsewhere. The advection speed, the diffusion coefficient, the spectral index of injected CREs, and the scale heights of the magnetic field were also determined.  

The advection speed increases exponentially with $z$ in the central region and reaches the escape velocity of about 462~km~s$^{-1}$ at a height of 5.5~kpc. Beyond this height, CREs move freely into intergalactic space. There is a good correspondence of multiwavelength emission extending from disk to halo, such as radio, X-ray, {\sc H\,i}, and H$\alpha$, confirming the existence of a superwind causing the bulk motion of all the materials.  

We estimated the thermal, CR, magnetic, ram, and gravitational pressure as a function of $z$. We found that the combination of outward pressure from the disk is larger than the inward gravitational pressure at $z \lesssim 5$\,kpc, which accelerated the speed of the wind.

\begin{acknowledgements}
We thank the referee for the constructive comments, which have significantly improved the manuscript. This research has been supported by the National SKA Program of China (2022SKA0120101, 2022SKA0120103). This work is also supported by the 16th Graduate Research and Innovation Project of Yunnan Province, China (Project No. KC-242410116). J.T.L. acknowledges the financial support from the National Natural Science Foundation of China (NSFC) through the grants 12321003 and 12273111, the science research grants from the China Manned Space Program with grant no. CMS-CSST-2025-A04 and CMS-CSST-2025-A10, and Jiangsu Innovation and Entrepreneurship Talent Team Program through the grant JSSCTD202436. This scientific work uses data obtained from Inyarrimanha Ilgari Bundara / the Murchison Radio-astronomy Observatory. We acknowledge the Wajarri Yamaji People as the Traditional Owners and native title holders of the Observatory site. The Australian SKA Pathfinder is part of the Australia Telescope National Facility, which is managed by CSIRO. Operation of ASKAP is funded by the Australian Government with support from the National Collaborative Research Infrastructure Strategy. Establishment of ASKAP, the Murchison Radio-astronomy Observatory and the Pawsey Supercomputing Centre are initiatives of the Australian Government, with support from the Government of Western Australia and the Science and Industry Endowment Fund. 

Support for the operation of the MWA is provided by the Australian Government (NCRIS), under a contract to Curtin University administered by Astronomy Australia Limited. The MWA Phase II upgrade project was supported by Australian Research Council LIEF grant LE160100031 and the Dunlap Institute for Astronomy and Astrophysics at the University of Toronto. We acknowledge the Pawsey Supercomputing Centre which is supported by the Western Australian and Australian Governments.

\end{acknowledgements}
\bibliographystyle{aa}
\bibliography{ngc253}

@ARTICLE{arras2021,
       author = {{Arras}, Philipp and {Reinecke}, Martin and {Westermann}, R{\"u}diger and {En{\ss}lin}, Torsten A.},
        title = "{Efficient wide-field radio interferometry response}",
      journal = {\aap},
         year = 2021,
        month = feb,
       volume = {646},
          eid = {A58},
        pages = {A58},
          doi = {10.1051/0004-6361/202039723},
archivePrefix = {arXiv},
       eprint = {2010.10122},
 primaryClass = {astro-ph.IM},
       adsurl = {https://ui.adsabs.harvard.edu/abs/2021A&A...646A..58A}
}

@ARTICLE{breitschwerdt1991,
       author = {{Breitschwerdt}, D. and {McKenzie}, J.~F. and {Voelk}, H.~J.},
        title = "{Galactic winds. I. Cosmic ray and wave-driven winds from the galaxy.}",
      journal = {\aap},
         year = 1991,
        month = may,
       volume = {245},
        pages = {79},
       adsurl = {https://ui.adsabs.harvard.edu/abs/1991A&A...245...79B}
}

@ARTICLE{breitschwerdt2002,
       author = {{Breitschwerdt}, D. and {Dogiel}, V.~A. and {V{\"o}lk}, H.~J.},
        title = "{The gradient of diffuse gamma -ray emission in the Galaxy}",
      journal = {\aap},
         year = 2002,
        month = apr,
       volume = {385},
        pages = {216-238},
          doi = {10.1051/0004-6361:20020152},
archivePrefix = {arXiv},
       eprint = {astro-ph/0201345},
 primaryClass = {astro-ph},
       adsurl = {https://ui.adsabs.harvard.edu/abs/2002A&A...385..216B}
}

@INPROCEEDINGS{briggs1995,
       author = {{Briggs}, D.~S.},
        title = "{High Fidelity Interferometric Imaging: Robust Weighting and NNLS Deconvolution}",
    booktitle = {American Astronomical Society Meeting Abstracts},
         year = 1995,
       series = {American Astronomical Society Meeting Abstracts},
       volume = {187},
        month = dec,
          eid = {112.02},
        pages = {112.02},
       adsurl = {https://ui.adsabs.harvard.edu/abs/1995AAS...18711202B}
}

@ARTICLE{basu1997,
       author = {{Basu}, Shantanu and {Mouschovias}, Telemachos Ch. and {Paleologou}, Efthimios V.},
        title = "{Dynamical Effects of the Parker Instability in the Interstellar Medium}",
      journal = {\apjl},
         year = 1997,
        month = may,
       volume = {480},
       number = {1},
        pages = {L55-L58},
          doi = {10.1086/310620},
       adsurl = {https://ui.adsabs.harvard.edu/abs/1997ApJ...480L..55B}
}

@ARTICLE{bauer2007,
       author = {{Bauer}, M. and {Pietsch}, W. and {Trinchieri}, G. and {Breitschwerdt}, D. and {Ehle}, M. and {Read}, A.},
        title = "{High-resolution X-ray spectroscopy and imaging of the nuclear outflow of the starburst galaxy NGC 253}",
      journal = {\aap},
         year = 2007,
        month = jun,
       volume = {467},
       number = {3},
        pages = {979-989},
          doi = {10.1051/0004-6361:20066340},
archivePrefix = {arXiv},
       eprint = {astro-ph/0610302},
 primaryClass = {astro-ph},
       adsurl = {https://ui.adsabs.harvard.edu/abs/2007A&A...467..979B}
}

@ARTICLE{bauer2008,
       author = {{Bauer}, M. and {Pietsch}, W. and {Trinchieri}, G. and {Breitschwerdt}, D. and {Ehle}, M. and {Freyberg}, M.~J. and {Read}, A.~M.},
        title = "{XMM-Newton observations of the diffuse X-ray emission in the starburst galaxy NGC 253}",
      journal = {\aap},
         year = 2008,
        month = oct,
       volume = {489},
       number = {3},
        pages = {1029-1046},
          doi = {10.1051/0004-6361:20078935},
archivePrefix = {arXiv},
       eprint = {0711.3182},
 primaryClass = {astro-ph},
       adsurl = {https://ui.adsabs.harvard.edu/abs/2008A&A...489.1029B}
}

@ARTICLE{bolatto2013,
       author = {{Bolatto}, Alberto D. and {Warren}, Steven R. and {Leroy}, Adam K. and {Walter}, Fabian and {Veilleux}, Sylvain and {Ostriker}, Eve C. and {Ott}, J{\"u}rgen and {Zwaan}, Martin and {Fisher}, David B. and {Weiss}, Axel and {Rosolowsky}, Erik and {Hodge}, Jacqueline},
        title = "{Suppression of star formation in the galaxy NGC 253 by a starburst-driven molecular wind}",
      journal = {\nat},
         year = 2013,
        month = jul,
       volume = {499},
       number = {7459},
        pages = {450-453},
          doi = {10.1038/nature12351},
archivePrefix = {arXiv},
       eprint = {1307.6259},
 primaryClass = {astro-ph.CO},
       adsurl = {https://ui.adsabs.harvard.edu/abs/2013Natur.499..450B}
}

@ARTICLE{borlaff2023,
       author = {{Borlaff}, Alejandro S. and {Lopez-Rodriguez}, Enrique and {Beck}, Rainer and {Clark}, Susan E. and {Ntormousi}, Evangelia and {Tassis}, Konstantinos and {Martin-Alvarez}, Sergio and {Tahani}, Mehrnoosh and {Dale}, Daniel A. and {del Moral-Castro}, Ignacio and {Roman-Duval}, Julia and {Marcum}, Pamela M. and {Beckman}, John E. and {Subramanian}, Kandaswamy and {Eftekharzadeh}, Sarah and {Proudfit}, Leslie},
        title = "{Extragalactic Magnetism with SOFIA (SALSA Legacy Program). V. First Results on the Magnetic Field Orientation of Galaxies}",
      journal = {\apj},
         year = 2023,
        month = jul,
       volume = {952},
       number = {1},
          eid = {4},
        pages = {4},
          doi = {10.3847/1538-4357/acd934},
archivePrefix = {arXiv},
       eprint = {2303.13586},
 primaryClass = {astro-ph.GA},
       adsurl = {https://ui.adsabs.harvard.edu/abs/2023ApJ...952....4B}
}

@ARTICLE{beck2005,
       author = {{Beck}, R. and {Krause}, M.},
        title = "{Revised equipartition and minimum energy formula for magnetic field strength estimates from radio synchrotron observations}",
      journal = {Astronomische Nachrichten},
         year = 2005,
        month = jul,
       volume = {326},
       number = {6},
        pages = {414-427},
          doi = {10.1002/asna.200510366},
archivePrefix = {arXiv},
       eprint = {astro-ph/0507367},
 primaryClass = {astro-ph},
       adsurl = {https://ui.adsabs.harvard.edu/abs/2005AN....326..414B}
}

@INPROCEEDINGS{beck2000,
       author = {{Beck}, Rainer},
        title = "{Magnetic fields in normal galaxies}",
    booktitle = {Astronomy, physics and chemistry of H$^{+}$$_{3}$},
         year = 2000,
       volume = {358},
        month = feb,
        pages = {777-796},
          doi = {10.1098/rsta.2000.0558},
       adsurl = {https://ui.adsabs.harvard.edu/abs/2000RSPTA.358..777B}
}

@INPROCEEDINGS{beck2015,
       author = {{Beck}, R. and {Bomans}, D. and {Colafrancesco}, S. and {Dettmar}, R.~J. and {Ferri{\`e}re}, K. and {Fletcher}, A. and {Heald}, G. and {Heesen}, V. and {Horellou}, C. and {Krause}, M. and {Lou}, Y.~Q. and {Mao}, S.~A. and {Paladino}, R. and {Schinnerer}, E. and {Sokoloff}, D. and {Stil}, J. and {Tabatabaei}, F.},
        title = "{Structure, dynamical impact and origin of magnetic fields in nearby galaxies in the SKA era}",
    booktitle = {Advancing Astrophysics with the Square Kilometre Array (AASKA14)},
         year = 2015,
        month = apr,
          eid = {94},
        pages = {94},
          doi = {10.22323/1.215.0094},
archivePrefix = {arXiv},
       eprint = {1501.00385},
 primaryClass = {astro-ph.GA},
       adsurl = {https://ui.adsabs.harvard.edu/abs/2015aska.confE..94B}
}

@ARTICLE{beck2019,
       author = {{Beck}, Rainer and {Chamandy}, Luke and {Elson}, Ed and {Blackman}, Eric G.},
        title = "{Synthesizing Observations and Theory to Understand Galactic Magnetic Fields: Progress and Challenges}",
      journal = {Galaxies},
         year = 2019,
        month = dec,
       volume = {8},
       number = {1},
          eid = {4},
        pages = {4},
          doi = {10.3390/galaxies8010004},
archivePrefix = {arXiv},
       eprint = {1912.08962},
 primaryClass = {astro-ph.GA},
       adsurl = {https://ui.adsabs.harvard.edu/abs/2019Galax...8....4B}
}

@ARTICLE{boomsma2005,
       author = {{Boomsma}, R. and {Oosterloo}, T.~A. and {Fraternali}, F. and {van der Hulst}, J.~M. and {Sancisi}, R.},
        title = "{Extra-planar H I in the starburst galaxy NGC 253}",
      journal = {\aap},
         year = 2005,
        month = feb,
       volume = {431},
        pages = {65-72},
          doi = {10.1051/0004-6361:20041715},
archivePrefix = {arXiv},
       eprint = {astro-ph/0410055},
 primaryClass = {astro-ph},
       adsurl = {https://ui.adsabs.harvard.edu/abs/2005A&A...431...65B}
}

@ARTICLE{cicone2014,
       author = {{Cicone}, C. and {Maiolino}, R. and {Sturm}, E. and {Graci{\'a}-Carpio}, J. and {Feruglio}, C. and {Neri}, R. and {Aalto}, S. and {Davies}, R. and {Fiore}, F. and {Fischer}, J. and {Garc{\'\i}a-Burillo}, S. and {Gonz{\'a}lez-Alfonso}, E. and {Hailey-Dunsheath}, S. and {Piconcelli}, E. and {Veilleux}, S.},
        title = "{Massive molecular outflows and evidence for AGN feedback from CO observations}",
      journal = {\aap},
         year = 2014,
        month = feb,
       volume = {562},
          eid = {A21},
        pages = {A21},
          doi = {10.1051/0004-6361/201322464},
archivePrefix = {arXiv},
       eprint = {1311.2595},
 primaryClass = {astro-ph.CO},
       adsurl = {https://ui.adsabs.harvard.edu/abs/2014A&A...562A..21C}
}

@ARTICLE{carilli1992,
       author = {{Carilli}, C.~L. and {Holdaway}, M.~A. and {Ho}, P.~T.~P. and {de Pree}, C.~G.},
        title = "{Discovery of a Synchrotron-emitting Halo around NGC 253}",
      journal = {\apjl},
         year = 1992,
        month = nov,
       volume = {399},
        pages = {L59},
          doi = {10.1086/186606},
       adsurl = {https://ui.adsabs.harvard.edu/abs/1992ApJ...399L..59C}
}

@ARTICLE{condon2012,
       author = {{Condon}, J.~J. and {Cotton}, W.~D. and {Fomalont}, E.~B. and {Kellermann}, K.~I. and {Miller}, N. and {Perley}, R.~A. and {Scott}, D. and {Vernstrom}, T. and {Wall}, J.~V.},
        title = "{Resolving the Radio Source Background: Deeper Understanding through Confusion}",
      journal = {\apj},
         year = 2012,
        month = oct,
       volume = {758},
       number = {1},
          eid = {23},
        pages = {23},
          doi = {10.1088/0004-637X/758/1/23},
archivePrefix = {arXiv},
       eprint = {1207.2439},
 primaryClass = {astro-ph.CO},
       adsurl = {https://ui.adsabs.harvard.edu/abs/2012ApJ...758...23C}
}

@ARTICLE{cronin2025,
       author = {{Cronin}, Serena A. and {Bolatto}, Alberto D. and {Congiu}, Enrico and {Donaghue}, Keaton and {Kreckel}, Kathryn and {Leroy}, Adam K. and {Levy}, Rebecca C. and {Veilleux}, Sylvain and {Walter}, Fabian and {Nolasco}, Lenin},
        title = "{Physical Conditions of the Ionized Superwind in NGC 253 with VLT/MUSE}",
      journal = {\apj},
         year = 2025,
        month = jul,
       volume = {987},
       number = {1},
          eid = {92},
        pages = {92},
          doi = {10.3847/1538-4357/add738},
archivePrefix = {arXiv},
       eprint = {2505.04707},
 primaryClass = {astro-ph.GA},
       adsurl = {https://ui.adsabs.harvard.edu/abs/2025ApJ...987...92C}
}

@ARTICLE{dale2009,
       author = {{Dale}, D.~A. and {Cohen}, S.~A. and {Johnson}, L.~C. and {Schuster}, M.~D. and {Calzetti}, D. and {Engelbracht}, C.~W. and {Gil de Paz}, A. and {Kennicutt}, R.~C. and {Lee}, J.~C. and {Begum}, A. and {Block}, M. and {Dalcanton}, J.~J. and {Funes}, J.~G. and {Gordon}, K.~D. and {Johnson}, B.~D. and {Marble}, A.~R. and {Sakai}, S. and {Skillman}, E.~D. and {van Zee}, L. and {Walter}, F. and {Weisz}, D.~R. and {Williams}, B. and {Wu}, S.-Y. and {Wu}, Y.},
        title = "{The Spitzer Local Volume Legacy: Survey Description and Infrared Photometry}",
      journal = {\apj},
         year = 2009,
        month = sep,
       volume = {703},
       number = {1},
        pages = {517-556},
          doi = {10.1088/0004-637X/703/1/517},
archivePrefix = {arXiv},
       eprint = {0907.4722},
 primaryClass = {astro-ph.CO},
       adsurl = {https://ui.adsabs.harvard.edu/abs/2009ApJ...703..517D}
}

@ARTICLE{dumke1995,
       author = {{Dumke}, M. and {Krause}, M. and {Wielebinski}, R. and {Klein}, U.},
        title = "{Polarized radio emission at 2.8cm from a selected sample of edge-on galaxies.}",
      journal = {\aap},
         year = 1995,
        month = oct,
       volume = {302},
        pages = {691},
       adsurl = {https://ui.adsabs.harvard.edu/abs/1995A&A...302..691D}
}

@ARTICLE{duchesne2020,
       author = {{Duchesne}, S.~W. and {Johnston-Hollitt}, M. and {Zhu}, Z. and {Wayth}, R.~B. and {Line}, J.~L.~B.},
        title = "{Murchison Widefield Array detection of steep-spectrum, diffuse, non-thermal radio emission within Abell 1127}",
      journal = {\pasa},
         year = 2020,
        month = sep,
       volume = {37},
          eid = {e037},
        pages = {e037},
          doi = {10.1017/pasa.2020.29},
archivePrefix = {arXiv},
       eprint = {2007.15199},
 primaryClass = {astro-ph.GA},
       adsurl = {https://ui.adsabs.harvard.edu/abs/2020PASA...37...37D}
}

@ARTICLE{duchesne2023,
       author = {{Duchesne}, S.~W. and {Thomson}, A.~J.~M. and {Pritchard}, J. and {Lenc}, E. and {Moss}, V.~A. and {McConnell}, D. and {Wieringa}, M.~H. and {Whiting}, M.~T. and {Wang}, Z. and {Wang}, Y. and {Rose}, K. and {Raja}, W. and {Murphy}, Tara and {Leung}, J.~K. and {Huynh}, M.~T. and {Hotan}, A.~W. and {Hodgson}, T. and {Heald}, G.~H.},
        title = "{The Rapid ASKAP Continuum Survey IV: continuum imaging at 1367.5 MHz and the first data release of RACS-mid}",
      journal = {\pasa},
         year = 2023,
        month = aug,
       volume = {40},
          eid = {e034},
        pages = {e034},
          doi = {10.1017/pasa.2023.31},
archivePrefix = {arXiv},
       eprint = {2306.07194},
 primaryClass = {astro-ph.IM},
       adsurl = {https://ui.adsabs.harvard.edu/abs/2023PASA...40...34D}
}

@ARTICLE{duchesne2024,
       author = {{Duchesne}, S.~W. and {Botteon}, A. and {Koribalski}, B.~S. and {Loi}, F. and {Rajpurohit}, K. and {Riseley}, C.~J. and {Rudnick}, L. and {Vernstrom}, T. and {Andernach}, H. and {Hopkins}, A.~M. and {Kapinska}, A.~D. and {Norris}, R.~P. and {Zafar}, T.},
        title = "{Evolutionary Map of the Universe (EMU): A pilot search for diffuse, non-thermal radio emission in galaxy clusters with the Australian SKA Pathfinder}",
      journal = {\pasa},
         year = 2024,
        month = jan,
       volume = {41},
          eid = {e026},
        pages = {e026},
          doi = {10.1017/pasa.2024.10},
archivePrefix = {arXiv},
       eprint = {2402.06192},
 primaryClass = {astro-ph.CO},
       adsurl = {https://ui.adsabs.harvard.edu/abs/2024PASA...41...26D}
}

@ARTICLE{duchesne2025,
       author = {{Duchesne}, Stefan William and {Cook}, Jaiden H. and {Hurley-Walker}, Natasha and {Thomson}, Alec J.~M. and {Paterson}, Sean and {Riseley}, Christopher J. and {McSweeney}, Sammy J. and {Mantovanini}, Silvia and {Heald}, George and {Franzen}, Thomas M.~O. and {Ross}, Kathryn and {Seymour}, Nicholas and {Wayth}, Randall B. and {Galvin}, Timothy James},
        title = "{GLEAM-300: The GaLactic and Extragalactic All-sky Murchison Widefield Array (GLEAM) survey at 300 MHz}",
      journal = {\pasa},
         year = 2025,
        month = nov,
       volume = {42},
          eid = {e158},
        pages = {e158},
          doi = {10.1017/pasa.2025.10115},
archivePrefix = {arXiv},
       eprint = {2510.18511},
 primaryClass = {astro-ph.GA},
       adsurl = {https://ui.adsabs.harvard.edu/abs/2025PASA...42..158D}
}

@book{draine2010,
  title={Physics of the interstellar and intergalactic medium},
  author={Draine, Bruce T},
  volume={19},
  year={2010},
  publisher={Princeton University Press}
}

@ARTICLE{fabian2012,
       author = {{Fabian}, A.~C.},
        title = "{Observational Evidence of Active Galactic Nuclei Feedback}",
      journal = {\araa},
         year = 2012,
        month = sep,
       volume = {50},
        pages = {455-489},
          doi = {10.1146/annurev-astro-081811-125521},
archivePrefix = {arXiv},
       eprint = {1204.4114},
 primaryClass = {astro-ph.CO},
       adsurl = {https://ui.adsabs.harvard.edu/abs/2012ARA&A..50..455F}
}

@ARTICLE{fluetsch2019,
       author = {{Fluetsch}, A. and {Maiolino}, R. and {Carniani}, S. and {Marconi}, A. and {Cicone}, C. and {Bourne}, M.~A. and {Costa}, T. and {Fabian}, A.~C. and {Ishibashi}, W. and {Venturi}, G.},
        title = "{Cold molecular outflows in the local Universe and their feedback effect on galaxies}",
      journal = {\mnras},
         year = 2019,
        month = mar,
       volume = {483},
       number = {4},
        pages = {4586-4614},
          doi = {10.1093/mnras/sty3449},
archivePrefix = {arXiv},
       eprint = {1805.05352},
 primaryClass = {astro-ph.GA},
       adsurl = {https://ui.adsabs.harvard.edu/abs/2019MNRAS.483.4586F}
}

@ARTICLE{gaensler2008,
       author = {{Gaensler}, B.~M. and {Madsen}, G.~J. and {Chatterjee}, S. and {Mao}, S.~A.},
        title = "{The Vertical Structure of Warm Ionised Gas in the Milky Way}",
      journal = {\pasa},
         year = 2008,
        month = nov,
       volume = {25},
       number = {4},
        pages = {184-200},
          doi = {10.1071/AS08004},
archivePrefix = {arXiv},
       eprint = {0808.2550},
 primaryClass = {astro-ph},
       adsurl = {https://ui.adsabs.harvard.edu/abs/2008PASA...25..184G}
}

@ARTICLE{grasha2022,
       author = {{Grasha}, K. and {Chen}, Q.~H. and {Battisti}, A.~J. and {Acharyya}, A. and {Ridolfo}, S. and {Poehler}, E. and {Mably}, S. and {Verma}, A.~A. and {Hayward}, K.~L. and {Kharbanda}, A. and {Poetrodjojo}, H. and {Seibert}, M. and {Rich}, J.~A. and {Madore}, B.~F. and {Kewley}, L.~J.},
        title = "{Metallicity, Ionization Parameter, and Pressure Variations of H II Regions in the TYPHOON Spiral Galaxies: NGC 1566, NGC 2835, NGC 3521, NGC 5068, NGC 5236, and NGC 7793}",
      journal = {\apj},
         year = 2022,
        month = apr,
       volume = {929},
       number = {2},
          eid = {118},
        pages = {118},
          doi = {10.3847/1538-4357/ac5ab2},
archivePrefix = {arXiv},
       eprint = {2203.02522},
 primaryClass = {astro-ph.GA},
       adsurl = {https://ui.adsabs.harvard.edu/abs/2022ApJ...929..118G}
}

@dataset{greggio2014,
       author = {{Greggio}, L. and {Rejkuba}, M. and {Gonzales}, O.~A. and {Arnaboldi}, M. and {Iodice}, E. and {Irwin}, M. and {Neeser}, M.~J. and {Emerson}, J.},
        title = "{VizieR Online Data Catalog: ZJ VISTA photometry in NGC253 stellar halo (Greggio+, 2014)}",
 howpublished = {VizieR On-line Data Catalog: J/A+A/562/A73. Originally published in: 2014A\&A...562A..73G},
         year = 2014,
        month = feb,
          eid = {J/A+A/562/A73},
          doi = {10.26093/cds/vizier.35620073},
       adsurl = {https://ui.adsabs.harvard.edu/abs/2014yCat..35620073G}
}

@ARTICLE{grenier2015,
       author = {{Grenier}, Isabelle A. and {Black}, John H. and {Strong}, Andrew W.},
        title = "{The Nine Lives of Cosmic Rays in Galaxies}",
      journal = {\araa},
         year = 2015,
        month = aug,
       volume = {53},
        pages = {199-246},
          doi = {10.1146/annurev-astro-082214-122457},
       adsurl = {https://ui.adsabs.harvard.edu/abs/2015ARA&A..53..199G}
}

@INCOLLECTION{heckman2017,
       author = {{Heckman}, Timothy M. and {Thompson}, Todd A.},
        title = "{Galactic Winds and the Role Played by Massive Stars}",
    booktitle = {Handbook of Supernovae},
         year = 2017,
       editor = {{Alsabti}, Athem W. and {Murdin}, Paul},
        pages = {2431},
          doi = {10.1007/978-3-319-21846-5_23},
       adsurl = {https://ui.adsabs.harvard.edu/abs/2017hsn..book.2431H}
}

@ARTICLE{hoopes1996,
       author = {{Hoopes}, Charles G. and {Walterbos}, Rene A.~M. and {Greenwalt}, Bruce E.},
        title = "{Diffuse Ionized Gas in Three Sculptor Group Galaxies}",
      journal = {\aj},
         year = 1996,
        month = oct,
       volume = {112},
        pages = {1429},
          doi = {10.1086/118111},
archivePrefix = {arXiv},
       eprint = {astro-ph/9607048},
 primaryClass = {astro-ph},
       adsurl = {https://ui.adsabs.harvard.edu/abs/1996AJ....112.1429H}
}

@ARTICLE{hale2021,
       author = {{Hale}, Catherine L. and {McConnell}, D. and {Thomson}, A.~J.~M. and {Lenc}, E. and {Heald}, G.~H. and {Hotan}, A.~W. and {Leung}, J.~K. and {Moss}, V.~A. and {Murphy}, T. and {Pritchard}, J. and {Sadler}, E.~M. and {Stewart}, A.~J. and {Whiting}, M.~T.},
        title = "{The Rapid ASKAP Continuum Survey Paper II: First Stokes I Source Catalogue Data Release}",
      journal = {\pasa},
         year = 2021,
        month = nov,
       volume = {38},
          eid = {e058},
        pages = {e058},
          doi = {10.1017/pasa.2021.47},
archivePrefix = {arXiv},
       eprint = {2109.00956},
 primaryClass = {astro-ph.GA},
       adsurl = {https://ui.adsabs.harvard.edu/abs/2021PASA...38...58H}
}

@ARTICLE{hurley2022,
       author = {{Hurley-Walker}, N. and {Galvin}, T.~J. and {Duchesne}, S.~W. and {Zhang}, X. and {Morgan}, J. and {Hancock}, P.~J. and {An}, T. and {Franzen}, T.~M.~O. and {Heald}, G. and {Ross}, K. and {Vernstrom}, T. and {Anderson}, G.~E. and {Gaensler}, B.~M. and {Johnston-Hollitt}, M. and {Kaplan}, D.~L. and {Riseley}, C.~J. and {Tingay}, S.~J. and {Walker}, M.},
        title = "{GaLactic and Extragalactic All-sky Murchison Widefield Array survey eXtended (GLEAM-X) I: Survey description and initial data release}",
      journal = {\pasa},
         year = 2022,
        month = aug,
       volume = {39},
          eid = {e035},
        pages = {e035},
          doi = {10.1017/pasa.2022.17},
archivePrefix = {arXiv},
       eprint = {2204.12762},
 primaryClass = {astro-ph.GA},
       adsurl = {https://ui.adsabs.harvard.edu/abs/2022PASA...39...35H}
}

@ARTICLE{heesen_2009a,
       author = {{Heesen}, V. and {Beck}, R. and {Krause}, M. and {Dettmar}, R.-J.},
        title = "{Cosmic rays and the magnetic field in the nearby starburst galaxy NGC 253. I. The distribution and transport of cosmic rays}",
      journal = {\aap},
         year = 2009,
        month = feb,
       volume = {494},
       number = {2},
        pages = {563-577},
          doi = {10.1051/0004-6361:200810543},
archivePrefix = {arXiv},
       eprint = {0812.0346},
 primaryClass = {astro-ph},
       adsurl = {https://ui.adsabs.harvard.edu/abs/2009A&A...494..563H}
}

@ARTICLE{heesen_2009b,
       author = {{Heesen}, V. and {Krause}, M. and {Beck}, R. and {Dettmar}, R.-J.},
        title = "{Cosmic rays and the magnetic field in the nearby starburst galaxy NGC 253. II. The magnetic field structure}",
      journal = {\aap},
         year = 2009,
        month = nov,
       volume = {506},
       number = {3},
        pages = {1123-1135},
          doi = {10.1051/0004-6361/200911698},
archivePrefix = {arXiv},
       eprint = {0908.2985},
 primaryClass = {astro-ph.CO},
       adsurl = {https://ui.adsabs.harvard.edu/abs/2009A&A...506.1123H}
}

@ARTICLE{heesen2016,
       author = {{Heesen}, Volker and {Dettmar}, Ralf-J{\"u}rgen and {Krause}, Marita and {Beck}, Rainer and {Stein}, Yelena},
        title = "{Advective and diffusive cosmic ray transport in galactic haloes}",
      journal = {\mnras},
         year = 2016,
        month = may,
       volume = {458},
       number = {1},
        pages = {332-353},
          doi = {10.1093/mnras/stw360},
archivePrefix = {arXiv},
       eprint = {1602.04085},
 primaryClass = {astro-ph.GA},
       adsurl = {https://ui.adsabs.harvard.edu/abs/2016MNRAS.458..332H}
}

@ARTICLE{heesen_2018a,
       author = {{Heesen}, V. and {Croston}, J.~H. and {Morganti}, R. and {Hardcastle}, M.~J. and {Stewart}, A.~J. and {Best}, P.~N. and {Broderick}, J.~W. and {Br{\"u}ggen}, M. and {Brunetti}, G. and {Chy{\.Z}y}, K.~T. and {Harwood}, J.~J. and {Haverkorn}, M. and {Hess}, K.~M. and {Intema}, H.~T. and {Jamrozy}, M. and {Kunert-Bajraszewska}, M. and {McKean}, J.~P. and {Orr{\'u}}, E. and {R{\"o}ttgering}, H.~J.~A. and {Shimwell}, T.~W. and {Shulevski}, A. and {White}, G.~J. and {Wilcots}, E.~M. and {Williams}, W.~L.},
        title = "{LOFAR reveals the giant: a low-frequency radio continuum study of the outflow in the nearby FR I radio galaxy 3C 31}",
      journal = {\mnras},
         year = 2018,
        month = mar,
       volume = {474},
       number = {4},
        pages = {5049-5067},
          doi = {10.1093/mnras/stx2869},
archivePrefix = {arXiv},
       eprint = {1710.09746},
 primaryClass = {astro-ph.GA},
       adsurl = {https://ui.adsabs.harvard.edu/abs/2018MNRAS.474.5049H}
}

@ARTICLE{heesen_2018b,
       author = {{Heesen}, V. and {Krause}, M. and {Beck}, R. and {Adebahr}, B. and {Bomans}, D.~J. and {Carretti}, E. and {Dumke}, M. and {Heald}, G. and {Irwin}, J. and {Koribalski}, B.~S. and {Mulcahy}, D.~D. and {Westmeier}, T. and {Dettmar}, R.-J.},
        title = "{Radio haloes in nearby galaxies modelled with 1D cosmic ray transport using SPINNAKER}",
      journal = {\mnras},
         year = 2018,
        month = may,
       volume = {476},
       number = {1},
        pages = {158-183},
          doi = {10.1093/mnras/sty105},
archivePrefix = {arXiv},
       eprint = {1801.05211},
 primaryClass = {astro-ph.GA},
       adsurl = {https://ui.adsabs.harvard.edu/abs/2018MNRAS.476..158H}
}

@ARTICLE{heesen2022,
       author = {{Heesen}, V. and {Klocke}, T.-L. and {Br{\"u}ggen}, M. and {Tabatabaei}, F.~S. and {Basu}, A. and {Beck}, R. and {Drabent}, A. and {Nikiel-Wroczy{\'n}ski}, B. and {Paladino}, R. and {Schulz}, S. and {Stein}, M.},
        title = "{Nearby galaxies in the LOFAR Two-metre Sky Survey. II. The magnetic field-gas relation}",
      journal = {\aap},
         year = 2023,
        month = jan,
       volume = {669},
          eid = {A8},
        pages = {A8},
          doi = {10.1051/0004-6361/202243328},
archivePrefix = {arXiv},
       eprint = {2208.11068},
 primaryClass = {astro-ph.GA},
       adsurl = {https://ui.adsabs.harvard.edu/abs/2023A&A...669A...8H}
}

@ARTICLE{hotan2021,
       author = {{Hotan}, A.~W. and {Bunton}, J.~D. and {Chippendale}, A.~P. and {Whiting}, M. and {Tuthill}, J. and {Moss}, V.~A. and {McConnell}, D. and {Amy}, S.~W. and {Huynh}, M.~T. and {Allison}, J.~R. and {Anderson}, C.~S. and {Bannister}, K.~W. and {Bastholm}, E. and {Beresford}, R. and {Bock}, D.~C.-J. and {Bolton}, R. and {Chapman}, J.~M. and {Chow}, K. and {Collier}, J.~D. and {Cooray}, F.~R. and {Cornwell}, T.~J. and {Diamond}, P.~J. and {Edwards}, P.~G. and {Feain}, I.~J. and {Franzen}, T.~M.~O. and {George}, D. and {Gupta}, N. and {Hampson}, G.~A. and {Harvey-Smith}, L. and {Hayman}, D.~B. and {Heywood}, I. and {Jacka}, C. and {Jackson}, C.~A. and {Jackson}, S. and {Jeganathan}, K. and {Johnston}, S. and {Kesteven}, M. and {Kleiner}, D. and {Koribalski}, B.~S. and {Lee-Waddell}, K. and {Lenc}, E. and {Lensson}, E.~S. and {Mackay}, S. and {Mahony}, E.~K. and {McClure-Griffiths}, N.~M. and {McConigley}, R. and {Mirtschin}, P. and {Ng}, A.~K. and {Norris}, R.~P. and {Pearce}, S.~E. and {Phillips}, C. and {Pilawa}, M.~A. and {Raja}, W. and {Reynolds}, J.~E. and {Roberts}, P. and {Roxby}, D.~N. and {Sadler}, E.~M. and {Shields}, M. and {Schinckel}, A.~E.~T. and {Serra}, P. and {Shaw}, R.~D. and {Sweetnam}, T. and {Troup}, E.~R. and {Tzioumis}, A. and {Voronkov}, M.~A. and {Westmeier}, T.},
        title = "{Australian square kilometre array pathfinder: I. system description}",
      journal = {\pasa},
         year = 2021,
        month = mar,
       volume = {38},
          eid = {e009},
        pages = {e009},
          doi = {10.1017/pasa.2021.1},
archivePrefix = {arXiv},
       eprint = {2102.01870},
 primaryClass = {astro-ph.IM},
       adsurl = {https://ui.adsabs.harvard.edu/abs/2021PASA...38....9H}
}

@ARTICLE{hopkins2012,
       author = {{Hopkins}, Philip F. and {Quataert}, Eliot and {Murray}, Norman},
        title = "{Stellar feedback in galaxies and the origin of galaxy-scale winds}",
      journal = {\mnras},
         year = 2012,
        month = apr,
       volume = {421},
       number = {4},
        pages = {3522-3537},
          doi = {10.1111/j.1365-2966.2012.20593.x},
archivePrefix = {arXiv},
       eprint = {1110.4638},
 primaryClass = {astro-ph.CO},
       adsurl = {https://ui.adsabs.harvard.edu/abs/2012MNRAS.421.3522H}
}

@ARTICLE{hoopes2005,
       author = {{Hoopes}, Charles G. and {Heckman}, Timothy M. and {Strickland}, David K. and {Seibert}, Mark and {Madore}, Barry F. and {Rich}, R. Michael and {Bianchi}, Luciana and {Gil de Paz}, Armando and {Burgarella}, Denis and {Thilker}, David A. and {Friedman}, Peter G. and {Barlow}, Tom A. and {Byun}, Yong-Ik and {Donas}, Jose and {Forster}, Karl and {Jelinsky}, Patrick N. and {Lee}, Young-Wook and {Malina}, Roger F. and {Martin}, D. Christopher and {Milliard}, Bruno and {Morrissey}, Patrick F. and {Neff}, Susan G. and {Schiminovich}, David and {Siegmund}, Oswald H.~W. and {Small}, Todd and {Szalay}, Alex S. and {Welsh}, Barry Y. and {Wyder}, Ted K.},
        title = "{GALEX Observations of the Ultraviolet Halos of NGC 253 and M82}",
      journal = {\apjl},
         year = 2005,
        month = jan,
       volume = {619},
       number = {1},
        pages = {L99-L102},
          doi = {10.1086/423032},
archivePrefix = {arXiv},
       eprint = {astro-ph/0411309},
 primaryClass = {astro-ph},
       adsurl = {https://ui.adsabs.harvard.edu/abs/2005ApJ...619L..99H}
}

@ARTICLE{iodice2014,
       author = {{Iodice}, E. and {Arnaboldi}, M. and {Rejkuba}, M. and {Neeser}, M.~J. and {Greggio}, L. and {Gonzalez}, O.~A. and {Irwin}, M. and {Emerson}, J.~P.},
        title = "{The near-infrared structure of the barred galaxy NGC 253 from VISTA{\ensuremath{\star}}}",
      journal = {\aap},
         year = 2014,
        month = jul,
       volume = {567},
          eid = {A86},
        pages = {A86},
          doi = {10.1051/0004-6361/201423480},
archivePrefix = {arXiv},
       eprint = {1405.7301},
 primaryClass = {astro-ph.GA},
       adsurl = {https://ui.adsabs.harvard.edu/abs/2014A&A...567A..86I}
}

@ARTICLE{jokipii1966,
       author = {{Jokipii}, J.~R.},
        title = "{Cosmic-Ray Propagation. I. Charged Particles in a Random Magnetic Field}",
      journal = {\apj},
         year = 1966,
        month = nov,
       volume = {146},
        pages = {480},
          doi = {10.1086/148912},
       adsurl = {https://ui.adsabs.harvard.edu/abs/1966ApJ...146..480J}
}

@ARTICLE{jarrett2019,
       author = {{Jarrett}, T.~H. and {Cluver}, M.~E. and {Brown}, M.~J.~I. and {Dale}, D.~A. and {Tsai}, C.~W. and {Masci}, F.},
        title = "{The WISE Extended Source Catalog (WXSC). I. The 100 Largest Galaxies}",
      journal = {\apjs},
         year = 2019,
        month = dec,
       volume = {245},
       number = {2},
          eid = {25},
        pages = {25},
          doi = {10.3847/1538-4365/ab521a},
archivePrefix = {arXiv},
       eprint = {1910.11793},
 primaryClass = {astro-ph.GA},
       adsurl = {https://ui.adsabs.harvard.edu/abs/2019ApJS..245...25J}
}

@ARTICLE{kulsrud1969,
       author = {{Kulsrud}, Russell and {Pearce}, William P.},
        title = "{The Effect of Wave-Particle Interactions on the Propagation of Cosmic Rays}",
      journal = {\apj},
         year = 1969,
        month = may,
       volume = {156},
        pages = {445},
          doi = {10.1086/149981},
       adsurl = {https://ui.adsabs.harvard.edu/abs/1969ApJ...156..445K}
}

@ARTICLE{kuno2007,
       author = {{Kuno}, Nario and {Sato}, Naoko and {Nakanishi}, Hiroyuki and {Hirota}, Akihiko and {Tosaki}, Tomoka and {Shioya}, Yasuhiro and {Sorai}, Kazuo and {Nakai}, Naomasa and {Nishiyama}, Kota and {Vila-Vilar{\'o}}, Baltsar},
        title = "{Nobeyama CO Atlas of Nearby Spiral Galaxies: Distribution of Molecular Gas in Barred and Nonbarred Spiral Galaxies}",
      journal = {\pasj},
         year = 2007,
        month = feb,
       volume = {59},
        pages = {117-166},
          doi = {10.1093/pasj/59.1.117},
archivePrefix = {arXiv},
       eprint = {0705.2678},
 primaryClass = {astro-ph},
       adsurl = {https://ui.adsabs.harvard.edu/abs/2007PASJ...59..117K}
}

@ARTICLE{krause2018,
       author = {{Krause}, Marita and {Irwin}, Judith and {Wiegert}, Theresa and {Miskolczi}, Arpad and {Damas-Segovia}, Ancor and {Beck}, Rainer and {Li}, Jiang-Tao and {Heald}, George and {M{\"u}ller}, Peter and {Stein}, Yelena and {Rand}, Richard J. and {Heesen}, Volker and {Walterbos}, Rene A.~M. and {Dettmar}, Ralf-J{\"u}rgen and {Vargas}, Carlos J. and {English}, Jayanne and {Murphy}, Eric J.},
        title = "{CHANG-ES. IX. Radio scale heights and scale lengths of a consistent sample of 13 spiral galaxies seen edge-on and their correlations}",
      journal = {\aap},
         year = 2018,
        month = mar,
       volume = {611},
          eid = {A72},
        pages = {A72},
          doi = {10.1051/0004-6361/201731991},
archivePrefix = {arXiv},
       eprint = {1712.03780},
 primaryClass = {astro-ph.GA},
       adsurl = {https://ui.adsabs.harvard.edu/abs/2018A&A...611A..72K}
}

@ARTICLE{kapinska2017,
       author = {{Kapi{\'n}ska}, A.~D. and {Staveley-Smith}, L. and {Crocker}, R. and {Meurer}, G.~R. and {Bhandari}, S. and {Hurley-Walker}, N. and {Offringa}, A.~R. and {Hanish}, D.~J. and {Seymour}, N. and {Ekers}, R.~D. and {Bell}, M.~E. and {Callingham}, J.~R. and {Dwarakanath}, K.~S. and {For}, B.-Q. and {Gaensler}, B.~M. and {Hancock}, P.~J. and {Hindson}, L. and {Johnston-Hollitt}, M. and {Lenc}, E. and {McKinley}, B. and {Morgan}, J. and {Procopio}, P. and {Wayth}, R.~B. and {Wu}, C. and {Zheng}, Q. and {Barry}, N. and {Beardsley}, A.~P. and {Bowman}, J.~D. and {Briggs}, F. and {Carroll}, P. and {Dillon}, J.~S. and {Ewall-Wice}, A. and {Feng}, L. and {Greenhill}, L.~J. and {Hazelton}, B.~J. and {Hewitt}, J.~N. and {Jacobs}, D.~J. and {Kim}, H.-S. and {Kittiwisit}, P. and {Line}, J. and {Loeb}, A. and {Mitchell}, D.~A. and {Morales}, M.~F. and {Neben}, A.~R. and {Paul}, S. and {Pindor}, B. and {Pober}, J.~C. and {Riding}, J. and {Sethi}, S.~K. and {Udaya Shankar}, N. and {Subrahmanyan}, R. and {Sullivan}, I.~S. and {Tegmark}, M. and {Thyagarajan}, N. and {Tingay}, S.~J. and {Trott}, C.~M. and {Webster}, R.~L. and {Wyithe}, S.~B. and {Cappallo}, R.~J. and {Deshpande}, A.~A. and {Kaplan}, D.~L. and {Lonsdale}, C.~J. and {McWhirter}, S.~R. and {Morgan}, E. and {Oberoi}, D. and {Ord}, S.~M. and {Prabu}, T. and {Srivani}, K.~S. and {Williams}, A. and {Williams}, C.~L.},
        title = "{Spectral Energy Distribution and Radio Halo of NGC 253 at Low Radio Frequencies}",
      journal = {\apj},
         year = 2017,
        month = mar,
       volume = {838},
       number = {1},
          eid = {68},
        pages = {68},
          doi = {10.3847/1538-4357/aa5f5d},
archivePrefix = {arXiv},
       eprint = {1702.02434},
 primaryClass = {astro-ph.GA},
       adsurl = {https://ui.adsabs.harvard.edu/abs/2017ApJ...838...68K}
}

@ARTICLE{li2013,
       author = {{Li}, Jiang-Tao and {Wang}, Q. Daniel},
        title = "{Chandra survey of nearby highly inclined disc galaxies - I. X-ray measurements of galactic coronae}",
      journal = {\mnras},
         year = 2013,
        month = jan,
       volume = {428},
       number = {3},
        pages = {2085-2108},
          doi = {10.1093/mnras/sts183},
archivePrefix = {arXiv},
       eprint = {1210.2997},
 primaryClass = {astro-ph.CO},
       adsurl = {https://ui.adsabs.harvard.edu/abs/2013MNRAS.428.2085L}
}

@ARTICLE{lucero2015,
       author = {{Lucero}, D.~M. and {Carignan}, C. and {Elson}, E.~C. and {Randriamampandry}, T.~H. and {Jarrett}, T.~H. and {Oosterloo}, T.~A. and {Heald}, G.~H.},
        title = "{H I observations of the nearest starburst galaxy NGC 253 with the SKA precursor KAT-7}",
      journal = {\mnras},
         year = 2015,
        month = jul,
       volume = {450},
       number = {4},
        pages = {3935-3951},
          doi = {10.1093/mnras/stv856},
       adsurl = {https://ui.adsabs.harvard.edu/abs/2015MNRAS.450.3935L}
}

@ARTICLE{leroy2019,
       author = {{Leroy}, Adam K. and {Sandstrom}, Karin M. and {Lang}, Dustin and {Lewis}, Alexia and {Salim}, Samir and {Behrens}, Erica A. and {Chastenet}, J{\'e}r{\'e}my and {Chiang}, I-Da and {Gallagher}, Molly J. and {Kessler}, Sarah and {Utomo}, Dyas},
        title = "{A z = 0 Multiwavelength Galaxy Synthesis. I. A WISE and GALEX Atlas of Local Galaxies}",
      journal = {\apjs},
         year = 2019,
        month = oct,
       volume = {244},
       number = {2},
          eid = {24},
        pages = {24},
          doi = {10.3847/1538-4365/ab3925},
archivePrefix = {arXiv},
       eprint = {1910.13470},
 primaryClass = {astro-ph.GA},
       adsurl = {https://ui.adsabs.harvard.edu/abs/2019ApJS..244...24L}
}

@ARTICLE{lyu2023,
       author = {{Lyu}, Xuanyi and {Westmeier}, T. and {Meurer}, Gerhardt R. and {Hanish}, D.~J.},
        title = "{On the origin of the anomalous gas, non-declining rotation curve, and disc asymmetries in NGC 253}",
      journal = {\mnras},
         year = 2023,
        month = sep,
       volume = {524},
       number = {1},
        pages = {1169-1190},
          doi = {10.1093/mnras/stad1772},
archivePrefix = {arXiv},
       eprint = {2306.06869},
 primaryClass = {astro-ph.GA},
       adsurl = {https://ui.adsabs.harvard.edu/abs/2023MNRAS.524.1169L}
}

@ARTICLE{lopez2023,
       author = {{Lopez}, Sebastian and {Lopez}, Laura A. and {Nguyen}, Dustin D. and {Thompson}, Todd A. and {Mathur}, Smita and {Bolatto}, Alberto D. and {Vulic}, Neven and {Sardone}, Amy},
        title = "{X-Ray Properties of NGC 253's Starburst-driven Outflow}",
      journal = {\apj},
         year = 2023,
        month = jan,
       volume = {942},
       number = {2},
          eid = {108},
        pages = {108},
          doi = {10.3847/1538-4357/aca65e},
archivePrefix = {arXiv},
       eprint = {2209.09260},
 primaryClass = {astro-ph.HE},
       adsurl = {https://ui.adsabs.harvard.edu/abs/2023ApJ...942..108L}
}

@BOOK{longair2011,
       author = {{Longair}, Malcolm S.},
        title = "{High Energy Astrophysics}",
         year = 2011,
       adsurl = {https://ui.adsabs.harvard.edu/abs/2011hea..book.....L}
}

@ARTICLE{lu2023,
       author = {{Lu}, Li-Yuan and {Li}, Jiang-Tao and {Vargas}, Carlos J. and {Beck}, Rainer and {Bregman}, Joel N. and {Dettmar}, Ralf-J{\"u}rgen and {English}, Jayanne and {Fang}, Taotao and {Heald}, George H. and {Li}, Hui and {Qu}, Zhijie and {Rand}, Richard J. and {Stein}, Michael and {Wang}, Q. Daniel and {Wang}, Jing and {Wiegert}, Theresa and {Zheng}, Yun},
        title = "{eDIG-CHANGES I: extended H{\ensuremath{\alpha}} emission from the extraplanar diffuse ionized gas (eDIG) around CHANG-ES galaxies}",
      journal = {\mnras},
         year = 2023,
        month = mar,
       volume = {519},
       number = {4},
        pages = {6098-6110},
          doi = {10.1093/mnras/stad006},
archivePrefix = {arXiv},
       eprint = {2212.14824},
 primaryClass = {astro-ph.GA},
       adsurl = {https://ui.adsabs.harvard.edu/abs/2023MNRAS.519.6098L}
}

@ARTICLE{matsumoto1988,
       author = {{Matsumoto}, Ryoji and {Horrnchi}, Toshiro and {Shibata}, Kazunari and {Hanawa}, Tomoyuki},
        title = "{Parker Instability in Nonuniform Gravitational Fields. II. Nonlinear Time Evolution}",
      journal = {\pasj},
         year = 1988,
        month = may,
       volume = {40},
       number = {2},
        pages = {171-195},
          doi = {10.1093/pasj/40.2.171},
       adsurl = {https://ui.adsabs.harvard.edu/abs/1988PASJ...40..171M}
}

@ARTICLE{mora2019,
       author = {{Mora-Partiarroyo}, Silvia Carolina and {Krause}, Marita and {Basu}, Aritra and {Beck}, Rainer and {Wiegert}, Theresa and {Irwin}, Judith and {Henriksen}, Richard and {Stein}, Yelena and {Vargas}, Carlos J. and {Heesen}, Volker and {Walterbos}, Ren{\'e} A.~M. and {Rand}, Richard J. and {Heald}, George and {Li}, Jiangtao and {Kamieneski}, Patrick and {English}, Jayanne},
        title = "{CHANG-ES. XV. Large-scale magnetic field reversals in the radio halo of NGC 4631}",
      journal = {\aap},
         year = 2019,
        month = dec,
       volume = {632},
          eid = {A11},
        pages = {A11},
          doi = {10.1051/0004-6361/201935961},
archivePrefix = {arXiv},
       eprint = {1910.07590},
 primaryClass = {astro-ph.GA},
       adsurl = {https://ui.adsabs.harvard.edu/abs/2019A&A...632A..11M}
}

@ARTICLE{miskolczi2019,
       author = {{Miskolczi}, A. and {Heesen}, V. and {Horellou}, C. and {Bomans}, D.-J. and {Beck}, R. and {Heald}, G. and {Dettmar}, R.-J. and {Blex}, S. and {Nikiel-Wroczy{\'n}ski}, B. and {Chy{\.z}y}, K.~T. and {Stein}, Y. and {Irwin}, J.~A. and {Shimwell}, T.~W. and {Wang}, Q.~D.},
        title = "{CHANG-ES XII. A LOFAR and VLA view of the edge-on star-forming galaxy NGC 3556}",
      journal = {\aap},
         year = 2019,
        month = feb,
       volume = {622},
          eid = {A9},
        pages = {A9},
          doi = {10.1051/0004-6361/201833931},
archivePrefix = {arXiv},
       eprint = {1811.04015},
 primaryClass = {astro-ph.GA},
       adsurl = {https://ui.adsabs.harvard.edu/abs/2019A&A...622A...9M}
}

@ARTICLE{murphy2011,
       author = {{Murphy}, E.~J. and {Condon}, J.~J. and {Schinnerer}, E. and {Kennicutt}, R.~C. and {Calzetti}, D. and {Armus}, L. and {Helou}, G. and {Turner}, J.~L. and {Aniano}, G. and {Beir{\~a}o}, P. and {Bolatto}, A.~D. and {Brandl}, B.~R. and {Croxall}, K.~V. and {Dale}, D.~A. and {Donovan Meyer}, J.~L. and {Draine}, B.~T. and {Engelbracht}, C. and {Hunt}, L.~K. and {Hao}, C.-N. and {Koda}, J. and {Roussel}, H. and {Skibba}, R. and {Smith}, J.-D.~T.},
        title = "{Calibrating Extinction-free Star Formation Rate Diagnostics with 33 GHz Free-free Emission in NGC 6946}",
      journal = {\apj},
         year = 2011,
        month = aug,
       volume = {737},
       number = {2},
          eid = {67},
        pages = {67},
          doi = {10.1088/0004-637X/737/2/67},
archivePrefix = {arXiv},
       eprint = {1105.4877},
 primaryClass = {astro-ph.CO},
       adsurl = {https://ui.adsabs.harvard.edu/abs/2011ApJ...737...67M}
}

@ARTICLE{murray2005,
       author = {{Murray}, Norman and {Quataert}, Eliot and {Thompson}, Todd A.},
        title = "{On the Maximum Luminosity of Galaxies and Their Central Black Holes: Feedback from Momentum-driven Winds}",
      journal = {\apj},
         year = 2005,
        month = jan,
       volume = {618},
       number = {2},
        pages = {569-585},
          doi = {10.1086/426067},
archivePrefix = {arXiv},
       eprint = {astro-ph/0406070},
 primaryClass = {astro-ph},
       adsurl = {https://ui.adsabs.harvard.edu/abs/2005ApJ...618..569M}
}

@ARTICLE{muller2017,
       author = {{M{\"u}ller}, Peter and {Krause}, Marita and {Beck}, Rainer and {Schmidt}, Philip},
        title = "{The NOD3 software package: A graphical user interface-supported reduction package for single-dish radio continuum and polarisation observations}",
      journal = {\aap},
         year = 2017,
        month = oct,
       volume = {606},
          eid = {A41},
        pages = {A41},
          doi = {10.1051/0004-6361/201731257},
archivePrefix = {arXiv},
       eprint = {1707.05573},
 primaryClass = {astro-ph.IM},
       adsurl = {https://ui.adsabs.harvard.edu/abs/2017A&A...606A..41M}
}

@ARTICLE{mcconnell2020,
       author = {{McConnell}, D. and {Hale}, C.~L. and {Lenc}, E. and {Banfield}, J.~K. and {Heald}, George and {Hotan}, A.~W. and {Leung}, James K. and {Moss}, Vanessa A. and {Murphy}, Tara and {O'Brien}, Andrew and {Pritchard}, Joshua and {Raja}, Wasim and {Sadler}, Elaine M. and {Stewart}, Adam and {Thomson}, Alec J.~M. and {Whiting}, M. and {Allison}, James R. and {Amy}, S.~W. and {Anderson}, C. and {Ball}, Lewis and {Bannister}, Keith W. and {Bell}, Martin and {Bock}, Douglas C.-J. and {Bolton}, Russ and {Bunton}, J.~D. and {Chippendale}, A.~P. and {Collier}, J.~D. and {Cooray}, F.~R. and {Cornwell}, T.~J. and {Diamond}, P.~J. and {Edwards}, P.~G. and {Gupta}, N. and {Hayman}, Douglas B. and {Heywood}, Ian and {Jackson}, C.~A. and {Koribalski}, B{\"a}rbel S. and {Lee-Waddell}, Karen and {McClure-Griffiths}, N.~M. and {Ng}, Alan and {Norris}, Ray P. and {Phillips}, Chris and {Reynolds}, John E. and {Roxby}, Daniel N. and {Schinckel}, Antony E.~T. and {Shields}, Matt and {Tremblay}, Chenoa and {Tzioumis}, A. and {Voronkov}, M.~A. and {Westmeier}, Tobias},
        title = "{The Rapid ASKAP Continuum Survey I: Design and first results}",
      journal = {\pasa},
         year = 2020,
        month = nov,
       volume = {37},
          eid = {e048},
        pages = {e048},
          doi = {10.1017/pasa.2020.41},
archivePrefix = {arXiv},
       eprint = {2012.00747},
 primaryClass = {astro-ph.IM},
       adsurl = {https://ui.adsabs.harvard.edu/abs/2020PASA...37...48M}
}

@ARTICLE{ostriker2010,
       author = {{Ostriker}, Eve C. and {McKee}, Christopher F. and {Leroy}, Adam K.},
        title = "{Regulation of Star Formation Rates in Multiphase Galactic Disks: A Thermal/Dynamical Equilibrium Model}",
      journal = {\apj},
         year = 2010,
        month = oct,
       volume = {721},
       number = {2},
        pages = {975-994},
          doi = {10.1088/0004-637X/721/2/975},
archivePrefix = {arXiv},
       eprint = {1008.0410},
 primaryClass = {astro-ph.CO},
       adsurl = {https://ui.adsabs.harvard.edu/abs/2010ApJ...721..975O}
}

@ARTICLE{offringa2014,
       author = {{Offringa}, A.~R. and {McKinley}, B. and {Hurley-Walker}, N. and {Briggs}, F.~H. and {Wayth}, R.~B. and {Kaplan}, D.~L. and {Bell}, M.~E. and {Feng}, L. and {Neben}, A.~R. and {Hughes}, J.~D. and {Rhee}, J. and {Murphy}, T. and {Bhat}, N.~D.~R. and {Bernardi}, G. and {Bowman}, J.~D. and {Cappallo}, R.~J. and {Corey}, B.~E. and {Deshpande}, A.~A. and {Emrich}, D. and {Ewall-Wice}, A. and {Gaensler}, B.~M. and {Goeke}, R. and {Greenhill}, L.~J. and {Hazelton}, B.~J. and {Hindson}, L. and {Johnston-Hollitt}, M. and {Jacobs}, D.~C. and {Kasper}, J.~C. and {Kratzenberg}, E. and {Lenc}, E. and {Lonsdale}, C.~J. and {Lynch}, M.~J. and {McWhirter}, S.~R. and {Mitchell}, D.~A. and {Morales}, M.~F. and {Morgan}, E. and {Kudryavtseva}, N. and {Oberoi}, D. and {Ord}, S.~M. and {Pindor}, B. and {Procopio}, P. and {Prabu}, T. and {Riding}, J. and {Roshi}, D.~A. and {Shankar}, N. Udaya and {Srivani}, K.~S. and {Subrahmanyan}, R. and {Tingay}, S.~J. and {Waterson}, M. and {Webster}, R.~L. and {Whitney}, A.~R. and {Williams}, A. and {Williams}, C.~L.},
        title = "{WSCLEAN: an implementation of a fast, generic wide-field imager for radio astronomy}",
      journal = {\mnras},
         year = 2014,
        month = oct,
       volume = {444},
       number = {1},
        pages = {606-619},
          doi = {10.1093/mnras/stu1368},
archivePrefix = {arXiv},
       eprint = {1407.1943},
 primaryClass = {astro-ph.IM},
       adsurl = {https://ui.adsabs.harvard.edu/abs/2014MNRAS.444..606O}
}

@ARTICLE{offringa2017,
       author = {{Offringa}, A.~R. and {Smirnov}, O.},
        title = "{An optimized algorithm for multiscale wideband deconvolution of radio astronomical images}",
      journal = {\mnras},
         year = 2017,
        month = oct,
       volume = {471},
       number = {1},
        pages = {301-316},
          doi = {10.1093/mnras/stx1547},
archivePrefix = {arXiv},
       eprint = {1706.06786},
 primaryClass = {astro-ph.IM},
       adsurl = {https://ui.adsabs.harvard.edu/abs/2017MNRAS.471..301O}
}

@ARTICLE{parker1966,
       author = {{Parker}, E.~N.},
        title = "{The Dynamical State of the Interstellar Gas and Field}",
      journal = {\apj},
         year = 1966,
        month = sep,
       volume = {145},
        pages = {811},
          doi = {10.1086/148828},
       adsurl = {https://ui.adsabs.harvard.edu/abs/1966ApJ...145..811P}
}

@ARTICLE{prada1998,
       author = {{Prada}, F. and {Guti{\'e}rrez}, C.~M. and {McKeith}, C.~D.},
        title = "{The Stellar and Gaseous Kinematics in NGC 253}",
      journal = {\apj},
         year = 1998,
        month = mar,
       volume = {495},
       number = {2},
        pages = {765-773},
          doi = {10.1086/305307},
archivePrefix = {arXiv},
       eprint = {astro-ph/9711262},
 primaryClass = {astro-ph},
       adsurl = {https://ui.adsabs.harvard.edu/abs/1998ApJ...495..765P}
}

@ARTICLE{pence1981,
       author = {{Pence}, W.~D.},
        title = "{A photometric and kinematic study of the barred spiral galaxy NGC 253. II. The velocity field.}",
      journal = {\apj},
         year = 1981,
        month = jul,
       volume = {247},
        pages = {473-483},
          doi = {10.1086/159056},
       adsurl = {https://ui.adsabs.harvard.edu/abs/1981ApJ...247..473P}
}

@ARTICLE{puche1991,
       author = {{Puche}, Daniel and {Carignan}, Claude and {van Gorkom}, Jacqueline H.},
        title = "{H I Studies of the Sculptor Group Galaxies. V. NGC 253}",
      journal = {\aj},
         year = 1991,
        month = feb,
       volume = {101},
        pages = {456},
          doi = {10.1086/115696},
       adsurl = {https://ui.adsabs.harvard.edu/abs/1991AJ....101..456P}
}

@ARTICLE{radburn2011,
       author = {{Radburn-Smith}, D.~J. and {de Jong}, R.~S. and {Seth}, A.~C. and {Bailin}, J. and {Bell}, E.~F. and {Brown}, T.~M. and {Bullock}, J.~S. and {Courteau}, S. and {Dalcanton}, J.~J. and {Ferguson}, H.~C. and {Goudfrooij}, P. and {Holfeltz}, S. and {Holwerda}, B.~W. and {Purcell}, C. and {Sick}, J. and {Streich}, D. and {Vlajic}, M. and {Zucker}, D.~B.},
        title = "{The GHOSTS Survey. I. Hubble Space Telescope Advanced Camera for Surveys Data}",
      journal = {\apjs},
         year = 2011,
        month = aug,
       volume = {195},
       number = {2},
          eid = {18},
        pages = {18},
          doi = {10.1088/0067-0049/195/2/18},
       adsurl = {https://ui.adsabs.harvard.edu/abs/2011ApJS..195...18R}
}

@ARTICLE{rupke2005,
       author = {{Rupke}, David S. and {Veilleux}, Sylvain and {Sanders}, D.~B.},
        title = "{Outflows in Infrared-Luminous Starbursts at z < 0.5. II. Analysis and Discussion}",
      journal = {\apjs},
         year = 2005,
        month = sep,
       volume = {160},
       number = {1},
        pages = {115-148},
          doi = {10.1086/432889},
archivePrefix = {arXiv},
       eprint = {astro-ph/0506611},
 primaryClass = {astro-ph},
       adsurl = {https://ui.adsabs.harvard.edu/abs/2005ApJS..160..115R}
}

@ARTICLE{recchia2016,
       author = {{Recchia}, S. and {Blasi}, P. and {Morlino}, G.},
        title = "{Cosmic ray driven Galactic winds}",
      journal = {\mnras},
         year = 2016,
        month = nov,
       volume = {462},
       number = {4},
        pages = {4227-4239},
          doi = {10.1093/mnras/stw1966},
archivePrefix = {arXiv},
       eprint = {1603.06746},
 primaryClass = {astro-ph.HE},
       adsurl = {https://ui.adsabs.harvard.edu/abs/2016MNRAS.462.4227R}
}

@ARTICLE{romero2018,
       author = {{Romero}, G.~E. and {M{\"u}ller}, A.~L. and {Roth}, M.},
        title = "{Particle acceleration in the superwinds of starburst galaxies}",
      journal = {\aap},
         year = 2018,
        month = aug,
       volume = {616},
          eid = {A57},
        pages = {A57},
          doi = {10.1051/0004-6361/201832666},
archivePrefix = {arXiv},
       eprint = {1801.06483},
 primaryClass = {astro-ph.HE},
       adsurl = {https://ui.adsabs.harvard.edu/abs/2018A&A...616A..57R}
}

@ARTICLE{rodriguez2024,
       author = {{Rodr{\'\i}guez Montero}, Francisco and {Martin-Alvarez}, Sergio and {Slyz}, Adrianne and {Devriendt}, Julien and {Dubois}, Yohan and {Sijacki}, Debora},
        title = "{The impact of cosmic rays on the interstellar medium and galactic outflows of Milky Way analogues}",
      journal = {\mnras},
         year = 2024,
        month = jun,
       volume = {530},
       number = {4},
        pages = {3617-3640},
          doi = {10.1093/mnras/stae1083},
archivePrefix = {arXiv},
       eprint = {2307.13733},
 primaryClass = {astro-ph.GA},
       adsurl = {https://ui.adsabs.harvard.edu/abs/2024MNRAS.530.3617R}
}

@ARTICLE{strickland2000,
       author = {{Strickland}, David K. and {Heckman}, Timothy M. and {Weaver}, Kimberly A. and {Dahlem}, Michael},
        title = "{Chandra Observations of NGC 253: New Insights into the Nature of Starburst-driven Superwinds}",
      journal = {\aj},
         year = 2000,
        month = dec,
       volume = {120},
       number = {6},
        pages = {2965-2974},
          doi = {10.1086/316846},
archivePrefix = {arXiv},
       eprint = {astro-ph/0008182},
 primaryClass = {astro-ph},
       adsurl = {https://ui.adsabs.harvard.edu/abs/2000AJ....120.2965S}
}

@ARTICLE{strickland2009,
       author = {{Strickland}, David K. and {Heckman}, Timothy M.},
        title = "{Supernova Feedback Efficiency and Mass Loading in the Starburst and Galactic Superwind Exemplar M82}",
      journal = {\apj},
         year = 2009,
        month = jun,
       volume = {697},
       number = {2},
        pages = {2030-2056},
          doi = {10.1088/0004-637X/697/2/2030},
archivePrefix = {arXiv},
       eprint = {0903.4175},
 primaryClass = {astro-ph.CO},
       adsurl = {https://ui.adsabs.harvard.edu/abs/2009ApJ...697.2030S}
}

@ARTICLE{strickland2002,
       author = {{Strickland}, David K. and {Heckman}, Timothy M. and {Weaver}, Kimberly A. and {Hoopes}, Charles G. and {Dahlem}, Michael},
        title = "{Chandra Observations of NGC 253. II. On the Origin of Diffuse X-Ray Emission in the Halos of Starburst Galaxies}",
      journal = {\apj},
         year = 2002,
        month = apr,
       volume = {568},
       number = {2},
        pages = {689-716},
          doi = {10.1086/338889},
archivePrefix = {arXiv},
       eprint = {astro-ph/0111511},
 primaryClass = {astro-ph},
       adsurl = {https://ui.adsabs.harvard.edu/abs/2002ApJ...568..689S}
}

@ARTICLE{strickland2004,
       author = {{Strickland}, David K. and {Heckman}, Timothy M. and {Colbert}, Edward J.~M. and {Hoopes}, Charles G. and {Weaver}, Kimberly A.},
        title = "{A High Spatial Resolution X-Ray and H{\ensuremath{\alpha}} Study of Hot Gas in the Halos of Star-forming Disk Galaxies. II. Quantifying Supernova Feedback}",
      journal = {\apj},
         year = 2004,
        month = may,
       volume = {606},
       number = {2},
        pages = {829-852},
          doi = {10.1086/383136},
archivePrefix = {arXiv},
       eprint = {astro-ph/0306598},
 primaryClass = {astro-ph},
       adsurl = {https://ui.adsabs.harvard.edu/abs/2004ApJ...606..829S}
}

@BOOK{Schlickeiser2002,
       author = {{Schlickeiser}, Reinhard},
        title = "{Cosmic Ray Astrophysics}",
         year = 2002,
       adsurl = {https://ui.adsabs.harvard.edu/abs/2002cra..book.....S}
}

@ARTICLE{sakamoto2011,
       author = {{Sakamoto}, Kazushi and {Mao}, Rui-Qing and {Matsushita}, Satoki and {Peck}, Alison B. and {Sawada}, Tsuyoshi and {Wiedner}, Martina C.},
        title = "{Star-forming Cloud Complexes in the Central Molecular Zone of NGC 253}",
      journal = {\apj},
         year = 2011,
        month = jul,
       volume = {735},
       number = {1},
          eid = {19},
        pages = {19},
          doi = {10.1088/0004-637X/735/1/19},
archivePrefix = {arXiv},
       eprint = {1104.2388},
 primaryClass = {astro-ph.CO},
       adsurl = {https://ui.adsabs.harvard.edu/abs/2011ApJ...735...19S}
}

@ARTICLE{sharma2012,
       author = {{Sharma}, Prateek and {McCourt}, Michael and {Quataert}, Eliot and {Parrish}, Ian J.},
        title = "{Thermal instability and the feedback regulation of hot haloes in clusters, groups and galaxies}",
      journal = {\mnras},
         year = 2012,
        month = mar,
       volume = {420},
       number = {4},
        pages = {3174-3194},
          doi = {10.1111/j.1365-2966.2011.20246.x},
archivePrefix = {arXiv},
       eprint = {1106.4816},
 primaryClass = {astro-ph.CO},
       adsurl = {https://ui.adsabs.harvard.edu/abs/2012MNRAS.420.3174S}
}

@ARTICLE{schmidt2019,
       author = {{Schmidt}, Philip and {Krause}, Marita and {Heesen}, Volker and {Basu}, Aritra and {Beck}, Rainer and {Wiegert}, Theresa and {Irwin}, Judith A. and {Heald}, George and {Rand}, Richard J. and {Li}, Jiang-Tao and {Murphy}, Eric J.},
        title = "{CHANG-ES. XVI. An in-depth view of the cosmic-ray transport in the edge-on spiral galaxies NGC 891 and NGC 4565}",
      journal = {\aap},
         year = 2019,
        month = dec,
       volume = {632},
          eid = {A12},
        pages = {A12},
          doi = {10.1051/0004-6361/201834995},
archivePrefix = {arXiv},
       eprint = {1907.03789},
 primaryClass = {astro-ph.GA},
       adsurl = {https://ui.adsabs.harvard.edu/abs/2019A&A...632A..12S}
}

@ARTICLE{stein_2019a,
       author = {{Stein}, Y. and {Dettmar}, R.-J. and {Irwin}, J. and {Beck}, R. and {We{\.z}gowiec}, M. and {Miskolczi}, A. and {Krause}, M. and {Heesen}, V. and {Wiegert}, T. and {Heald}, G. and {Walterbos}, R.~A.~M. and {Li}, J.-T. and {Soida}, M.},
        title = "{CHANG-ES. XIII. Transport processes and the magnetic fields of NGC 4666: indication of a reversing disk magnetic field}",
      journal = {\aap},
         year = 2019,
        month = mar,
       volume = {623},
          eid = {A33},
        pages = {A33},
          doi = {10.1051/0004-6361/201834515},
archivePrefix = {arXiv},
       eprint = {1901.08090},
 primaryClass = {astro-ph.GA},
       adsurl = {https://ui.adsabs.harvard.edu/abs/2019A&A...623A..33S}
}

@ARTICLE{stein2022,
       author = {{Stein}, M. and {Heesen}, V. and {Dettmar}, R.-J. and {Stein}, Y. and {Br{\"u}ggen}, M. and {Beck}, R. and {Adebahr}, B. and {Wiegert}, T. and {Vargas}, C.~J. and {Bomans}, D.~J. and {Li}, J. and {English}, J. and {Chy{\.z}y}, K.~T. and {Paladino}, R. and {Tabatabaei}, F.~S. and {Strong}, A.},
        title = "{CHANG-ES. XXVI. Insights into cosmic-ray transport from radio halos in edge-on galaxies}",
      journal = {\aap},
         year = 2023,
        month = feb,
       volume = {670},
          eid = {A158},
        pages = {A158},
          doi = {10.1051/0004-6361/202243906},
archivePrefix = {arXiv},
       eprint = {2210.07709},
 primaryClass = {astro-ph.GA},
       adsurl = {https://ui.adsabs.harvard.edu/abs/2023A&A...670A.158S}
}

@ARTICLE{strong2007,
       author = {{Strong}, Andrew W. and {Moskalenko}, Igor V. and {Ptuskin}, Vladimir S.},
        title = "{Cosmic-Ray Propagation and Interactions in the Galaxy}",
      journal = {Annual Review of Nuclear and Particle Science},
         year = 2007,
        month = nov,
       volume = {57},
       number = {1},
        pages = {285-327},
          doi = {10.1146/annurev.nucl.57.090506.123011},
archivePrefix = {arXiv},
       eprint = {astro-ph/0701517},
 primaryClass = {astro-ph},
       adsurl = {https://ui.adsabs.harvard.edu/abs/2007ARNPS..57..285S}
}

@ARTICLE{somerville2015,
       author = {{Somerville}, Rachel S. and {Dav{\'e}}, Romeel},
        title = "{Physical Models of Galaxy Formation in a Cosmological Framework}",
      journal = {\araa},
         year = 2015,
        month = aug,
       volume = {53},
        pages = {51-113},
          doi = {10.1146/annurev-astro-082812-140951},
archivePrefix = {arXiv},
       eprint = {1412.2712},
 primaryClass = {astro-ph.GA},
       adsurl = {https://ui.adsabs.harvard.edu/abs/2015ARA&A..53...51S}
}

@ARTICLE{sokolowski2017,
       author = {{Sokolowski}, M. and {Colegate}, T. and {Sutinjo}, A.~T. and {Ung}, D. and {Wayth}, R. and {Hurley-Walker}, N. and {Lenc}, E. and {Pindor}, B. and {Morgan}, J. and {Kaplan}, D.~L. and {Bell}, M.~E. and {Callingham}, J.~R. and {Dwarakanath}, K.~S. and {For}, Bi-Qing and {Gaensler}, B.~M. and {Hancock}, P.~J. and {Hindson}, L. and {Johnston-Hollitt}, M. and {Kapi{\'n}ska}, A.~D. and {McKinley}, B. and {Offringa}, A.~R. and {Procopio}, P. and {Staveley-Smith}, L. and {Wu}, C. and {Zheng}, Q.},
        title = "{Calibration and Stokes Imaging with Full Embedded Element Primary Beam Model for the Murchison Widefield Array}",
      journal = {\pasa},
         year = 2017,
        month = nov,
       volume = {34},
          eid = {e062},
        pages = {e062},
          doi = {10.1017/pasa.2017.54},
archivePrefix = {arXiv},
       eprint = {1710.07478},
 primaryClass = {astro-ph.IM},
       adsurl = {https://ui.adsabs.harvard.edu/abs/2017PASA...34...62S}
}

@INPROCEEDINGS{tingay2013,
       author = {{Tingay}, S.~J. and {Oberoi}, D. and {Cairns}, I. and {Donea}, A. and {Duffin}, R. and {Arcus}, W. and {Bernardi}, G. and {Bowman}, J.~D. and {Briggs}, F. and {Bunton}, J.~D. and {Cappallo}, R.~J. and {Corey}, B.~E. and {Deshpande}, A. and {deSouza}, L. and {Emrich}, D. and {Gaensler}, B.~M. and {R}, Goeke and {Greenhill}, L.~J. and {Hazelton}, B.~J. and {Herne}, D. and {Hewitt}, J.~N. and {Johnston-Hollitt}, M. and {Kaplan}, D.~L. and {Kasper}, J.~C. and {Kennewell}, J.~A. and {Kincaid}, B.~B. and {Koenig}, R. and {Kratzenberg}, E. and {Lonsdale}, C.~J. and {Lynch}, M.~J. and {McWhirter}, S.~R. and {Mitchell}, D.~A. and {Morales}, M.~F. and {Morgan}, E. and {Ord}, S.~M. and {Pathikulangara}, J. and {Prabu}, T. and {Remillard}, R.~A. and {Rogers}, A.~E.~E. and {Roshi}, A. and {Salah}, J.~E. and {Sault}, R.~J. and {Udaya-Shankar}, N. and {Srivani}, K.~S. and {Stevens}, J. and {Subrahmanyan}, R. and {Waterson}, M. and {Wayth}, R.~B. and {Webster}, R.~L. and {Whitney}, A.~R. and {Williams}, A. and {Williams}, C.~L. and {Wyithe}, J.~S.~B.},
        title = "{The Murchison Widefield Array: solar science with the low frequency SKA Precursor}",
    booktitle = {Journal of Physics Conference Series},
         year = 2013,
       series = {Journal of Physics Conference Series},
       volume = {440},
        month = jun,
    publisher = {IOP},
          eid = {012033},
        pages = {012033},
          doi = {10.1088/1742-6596/440/1/012033},
archivePrefix = {arXiv},
       eprint = {1301.6414},
 primaryClass = {astro-ph.IM},
       adsurl = {https://ui.adsabs.harvard.edu/abs/2013JPhCS.440a2033T}
}

@ARTICLE{tharakkal2023,
       author = {{Tharakkal}, Devika and {Shukurov}, Anvar and {Gent}, Frederick A. and {Sarson}, Graeme R. and {Snodin}, Andrew P. and {Rodrigues}, Luiz Felippe S.},
        title = "{Steady states of the Parker instability}",
      journal = {\mnras},
         year = 2023,
        month = nov,
       volume = {525},
       number = {4},
        pages = {5597-5613},
          doi = {10.1093/mnras/stad2610},
archivePrefix = {arXiv},
       eprint = {2212.03215},
 primaryClass = {astro-ph.GA},
       adsurl = {https://ui.adsabs.harvard.edu/abs/2023MNRAS.525.5597T}
}

@ARTICLE{tumlinson2017,
       author = {{Tumlinson}, Jason and {Peeples}, Molly S. and {Werk}, Jessica K.},
        title = "{The Circumgalactic Medium}",
      journal = {\araa},
         year = 2017,
        month = aug,
       volume = {55},
       number = {1},
        pages = {389-432},
          doi = {10.1146/annurev-astro-091916-055240},
archivePrefix = {arXiv},
       eprint = {1709.09180},
 primaryClass = {astro-ph.GA},
       adsurl = {https://ui.adsabs.harvard.edu/abs/2017ARA&A..55..389T}
}

@ARTICLE{thompson2024,
       author = {{Thompson}, Todd A. and {Heckman}, Timothy M.},
        title = "{Theory and Observation of Winds from Star-Forming Galaxies}",
      journal = {\araa},
         year = 2024,
        month = sep,
       volume = {62},
       number = {1},
        pages = {529-591},
          doi = {10.1146/annurev-astro-041224-011924},
archivePrefix = {arXiv},
       eprint = {2406.08561},
 primaryClass = {astro-ph.GA},
       adsurl = {https://ui.adsabs.harvard.edu/abs/2024ARA&A..62..529T}
}

@ARTICLE{tabatabaei2022,
       author = {{Tabatabaei}, F.~S. and {Cotton}, W. and {Schinnerer}, E. and {Beck}, R. and {Brunthaler}, A. and {Menten}, K.~M. and {Braine}, J. and {Corbelli}, E. and {Kramer}, C. and {Beckman}, J.~E. and {Knapen}, J.~H. and {Paladino}, R. and {Koch}, E. and {Camps Fari{\~n}a}, A.},
        title = "{Cloud-scale radio surveys of star formation and feedback in Triangulum Galaxy M 33: VLA observations}",
      journal = {\mnras},
         year = 2022,
        month = dec,
       volume = {517},
       number = {2},
        pages = {2990-3007},
          doi = {10.1093/mnras/stac2514},
archivePrefix = {arXiv},
       eprint = {2209.01389},
 primaryClass = {astro-ph.GA},
       adsurl = {https://ui.adsabs.harvard.edu/abs/2022MNRAS.517.2990T}
}

@ARTICLE{vogler1999,
       author = {{Vogler}, A. and {Pietsch}, W.},
        title = "{X-ray observations of the starburst galaxy NGC 253. I. Point sources in the bulge, disk and halo}",
      journal = {\aap},
         year = 1999,
        month = feb,
       volume = {342},
        pages = {101-123},
          doi = {10.48550/arXiv.astro-ph/9811071},
archivePrefix = {arXiv},
       eprint = {astro-ph/9811071},
 primaryClass = {astro-ph},
       adsurl = {https://ui.adsabs.harvard.edu/abs/1999A&A...342..101V}
}

@ARTICLE{veilleux2005,
       author = {{Veilleux}, Sylvain and {Cecil}, Gerald and {Bland-Hawthorn}, Joss},
        title = "{Galactic Winds}",
      journal = {\araa},
         year = 2005,
        month = sep,
       volume = {43},
       number = {1},
        pages = {769-826},
          doi = {10.1146/annurev.astro.43.072103.150610},
archivePrefix = {arXiv},
       eprint = {astro-ph/0504435},
 primaryClass = {astro-ph},
       adsurl = {https://ui.adsabs.harvard.edu/abs/2005ARA&A..43..769V}
}

@ARTICLE{vargas2018,
       author = {{Vargas}, Carlos J. and {Mora-Partiarroyo}, Silvia Carolina and {Schmidt}, Philip and {Rand}, Richard J. and {Stein}, Yelena and {Walterbos}, Ren{\'e} A.~M. and {Wang}, Q. Daniel and {Basu}, Aritra and {Patterson}, Maria and {Kepley}, Amanda and {Beck}, Rainer and {Irwin}, Judith and {Heald}, George and {Li}, Jiangtao and {Wiegert}, Theresa},
        title = "{CHANG-ES X: Spatially Resolved Separation of Thermal Contribution from Radio Continuum Emission in Edge-on Galaxies}",
      journal = {\apj},
         year = 2018,
        month = feb,
       volume = {853},
       number = {2},
          eid = {128},
        pages = {128},
          doi = {10.3847/1538-4357/aaa47f},
archivePrefix = {arXiv},
       eprint = {1801.01892},
 primaryClass = {astro-ph.GA},
       adsurl = {https://ui.adsabs.harvard.edu/abs/2018ApJ...853..128V}
}

@ARTICLE{vargas2019,
       author = {{Vargas}, Carlos J. and {Walterbos}, Ren{\'e} A.~M. and {Rand}, Richard J. and {Stil}, Jeroen and {Krause}, Marita and {Li}, Jiang-Tao and {Irwin}, Judith and {Dettmar}, Ralf-J{\"u}rgen},
        title = "{CHANG-ES. XVII. H{\ensuremath{\alpha}} Imaging of Nearby Edge-on Galaxies, New SFRs, and an Extreme Star Formation Region{\textemdash}Data Release 2}",
      journal = {\apj},
         year = 2019,
        month = aug,
       volume = {881},
       number = {1},
          eid = {26},
        pages = {26},
          doi = {10.3847/1538-4357/ab27cb},
archivePrefix = {arXiv},
       eprint = {1906.07763},
 primaryClass = {astro-ph.GA},
       adsurl = {https://ui.adsabs.harvard.edu/abs/2019ApJ...881...26V}
}

@ARTICLE{wayth2015,
       author = {{Wayth}, R.~B. and {Lenc}, E. and {Bell}, M.~E. and {Callingham}, J.~R. and {Dwarakanath}, K.~S. and {Franzen}, T.~M.~O. and {For}, B.-Q. and {Gaensler}, B. and {Hancock}, P. and {Hindson}, L. and {Hurley-Walker}, N. and {Jackson}, C.~A. and {Johnston-Hollitt}, M. and {Kapi{\'n}ska}, A.~D. and {McKinley}, B. and {Morgan}, J. and {Offringa}, A.~R. and {Procopio}, P. and {Staveley-Smith}, L. and {Wu}, C. and {Zheng}, Q. and {Trott}, C.~M. and {Bernardi}, G. and {Bowman}, J.~D. and {Briggs}, F. and {Cappallo}, R.~J. and {Corey}, B.~E. and {Deshpande}, A.~A. and {Emrich}, D. and {Goeke}, R. and {Greenhill}, L.~J. and {Hazelton}, B.~J. and {Kaplan}, D.~L. and {Kasper}, J.~C. and {Kratzenberg}, E. and {Lonsdale}, C.~J. and {Lynch}, M.~J. and {McWhirter}, S.~R. and {Mitchell}, D.~A. and {Morales}, M.~F. and {Morgan}, E. and {Oberoi}, D. and {Ord}, S.~M. and {Prabu}, T. and {Rogers}, A.~E.~E. and {Roshi}, A. and {Shankar}, N. Udaya and {Srivani}, K.~S. and {Subrahmanyan}, R. and {Tingay}, S.~J. and {Waterson}, M. and {Webster}, R.~L. and {Whitney}, A.~R. and {Williams}, A. and {Williams}, C.~L.},
        title = "{GLEAM: The GaLactic and Extragalactic All-Sky MWA Survey}",
      journal = {\pasa},
         year = 2015,
        month = jun,
       volume = {32},
          eid = {e025},
        pages = {e025},
          doi = {10.1017/pasa.2015.26},
archivePrefix = {arXiv},
       eprint = {1505.06041},
 primaryClass = {astro-ph.IM},
       adsurl = {https://ui.adsabs.harvard.edu/abs/2015PASA...32...25W}
}

@ARTICLE{wayth2018,
       author = {{Wayth}, Randall B. and {Tingay}, Steven J. and {Trott}, Cathryn M. and {Emrich}, David and {Johnston-Hollitt}, Melanie and {McKinley}, Ben and {Gaensler}, B.~M. and {Beardsley}, A.~P. and {Booler}, T. and {Crosse}, B. and {Franzen}, T.~M.~O. and {Horsley}, L. and {Kaplan}, D.~L. and {Kenney}, D. and {Morales}, M.~F. and {Pallot}, D. and {Sleap}, G. and {Steele}, K. and {Walker}, M. and {Williams}, A. and {Wu}, C. and {Cairns}, Iver. H. and {Filipovic}, M.~D. and {Johnston}, S. and {Murphy}, T. and {Quinn}, P. and {Staveley-Smith}, L. and {Webster}, R. and {Wyithe}, J.~S.~B.},
        title = "{The Phase II Murchison Widefield Array: Design overview}",
      journal = {\pasa},
         year = 2018,
        month = nov,
       volume = {35},
          eid = {e033},
        pages = {e033},
          doi = {10.1017/pasa.2018.37},
archivePrefix = {arXiv},
       eprint = {1809.06466},
 primaryClass = {astro-ph.IM},
       adsurl = {https://ui.adsabs.harvard.edu/abs/2018PASA...35...33W}
}

@ARTICLE{yamagishi2011,
       author = {{Yamagishi}, Mitsuyoshi and {Kaneda}, Hidehiro and {Ishihara}, Daisuke and {Oyabu}, Shinki and {Onaka}, Takashi and {Shimonishi}, Takashi and {Suzuki}, Toyoaki},
        title = "{AKARI Near-infrared Spectroscopic Observations of Interstellar Ices in the Edge-on Starburst Galaxy NGC 253}",
      journal = {\apjl},
         year = 2011,
        month = apr,
       volume = {731},
       number = {1},
          eid = {L20},
        pages = {L20},
          doi = {10.1088/2041-8205/731/1/L20},
archivePrefix = {arXiv},
       eprint = {1102.4430},
 primaryClass = {astro-ph.GA},
       adsurl = {https://ui.adsabs.harvard.edu/abs/2011ApJ...731L..20Y}
}

@ARTICLE{ye2022,
       author = {{Ye}, Haoyang and {Gull}, Stephen F. and {Tan}, Sze M. and {Nikolic}, Bojan},
        title = "{High accuracy wide-field imaging method in radio interferometry}",
      journal = {\mnras},
         year = 2022,
        month = mar,
       volume = {510},
       number = {3},
        pages = {4110-4125},
          doi = {10.1093/mnras/stab3548},
archivePrefix = {arXiv},
       eprint = {2101.11172},
 primaryClass = {astro-ph.IM},
       adsurl = {https://ui.adsabs.harvard.edu/abs/2022MNRAS.510.4110Y}
}

@article{zweibel2017,
  title={The basis for cosmic ray feedback: Written on the wind},
  author={Zweibel, Ellen G},
  journal={Physics of Plasmas},
  volume={24},
  number={5},
  year={2017},
  publisher={AIP Publishing}
}

\begin{appendix}

\section{Synchrotron emission profile}
\label{sec:intens_mode}

To account for the broadening of vertical intensity profiles by the projection of the disk and the beamwidth, an effective beamwidth $\Theta_c$ is introduced~\citep{muller2017,heesen_2018b}. An intrinsic vertical emission profile is then convolved with a Gaussian function with the effective beamwidth for each strip and then compared with the observed profiles by means of least-squares fitting. Following \citet{dumke1995}, we adopt an intrinsic exponential profile
\begin{equation}
    w_{\rm exp}(z)= w_0\,\mathrm{exp}(-z/h),
	\label{eq:exp_one}
\end{equation}
or an intrinsic Gaussian profile
\begin{equation}
    w_{\rm gauss}(z)= w_0\,\mathrm{exp}(-z^2/h^2).
	\label{eq:gau_one}
\end{equation}
Here, $w_0$ is the peak intensity and $h$ is the scale height. This is convolved with the effective beam $\Theta_c=2\,\sqrt{2\,\mathrm{ln}\,2}\,\sigma_{\rm eff}$,
\begin{equation}
    g(z)= \frac{1}{\sqrt{2\pi \sigma_{\rm eff}^2}}\rm exp(\frac{-z^2}{2\sigma_{eff}^2}),
	\label{eq:gauss_kernel}
\end{equation}
which describes the contribution of the telescope beam and the projected emission of the disk. The convolved emission profile has the form
\begin{equation}
\label{eq:exp_one_beam}
\begin{aligned}
W_{\rm exp}(z) =\;& \frac{w_0}{2}\,
\exp\!\left(-\frac{z^2}{2\sigma_{\rm eff}^2}\right) \\
&\times \Bigg[
\exp\!\left(\frac{(\sigma_{\rm eff}^2 - z h)^2}{\sqrt{2}\sigma_{\rm eff} h}\right)
\,\mathrm{erfc}\!\left(\frac{\sigma_{\rm eff}^2 - z h}{\sqrt{2}\sigma_{\rm eff} h}\right) \\
&\quad + 
\exp\!\left(\frac{(\sigma_{\rm eff}^2 + z h)^2}{\sqrt{2}\sigma_{\rm eff} h}\right)
\,\mathrm{erfc}\!\left(\frac{\sigma_{\rm eff}^2 + z h}{\sqrt{2}\sigma_{\rm eff} h}\right)
\Bigg],
\end{aligned}
\end{equation}
where erfc is the complementary error function, defined as
\begin{equation}
    \rm erfc(x)=1-erf(x)=\frac{2}{\sqrt{\pi}}\int_{x}^{\infty}\rm exp(-r^2) dr.
	\label{eq:erfc}
\end{equation}
For a Gaussian intrinsic distribution, the convolution yields
\begin{equation}
    W_{\rm Gauss}(z)=\frac{w_0h_0}{\rm \sqrt{2\sigma_{\rm eff}^2+h_0^2}} \mathrm{exp}\left(-\frac{z^2}{2\sigma_{\rm eff}^2 + h_0^2}\right).
	\label{eq:gauss_one_beam}
\end{equation}
The vertical radio scale heights $h$ can be determined by fitting one or two convolved exponential or Gaussian functions of the form given in Eqs.~\ref{eq:exp_one_beam} or~\ref{eq:gauss_one_beam} to the averaged radio intensities along each strip of a galaxy.

\onecolumn
\section{Intensity profiles and fits}
\label{sec:intens_prof}

\begin{figure*}[!h]	\includegraphics[width=\linewidth]{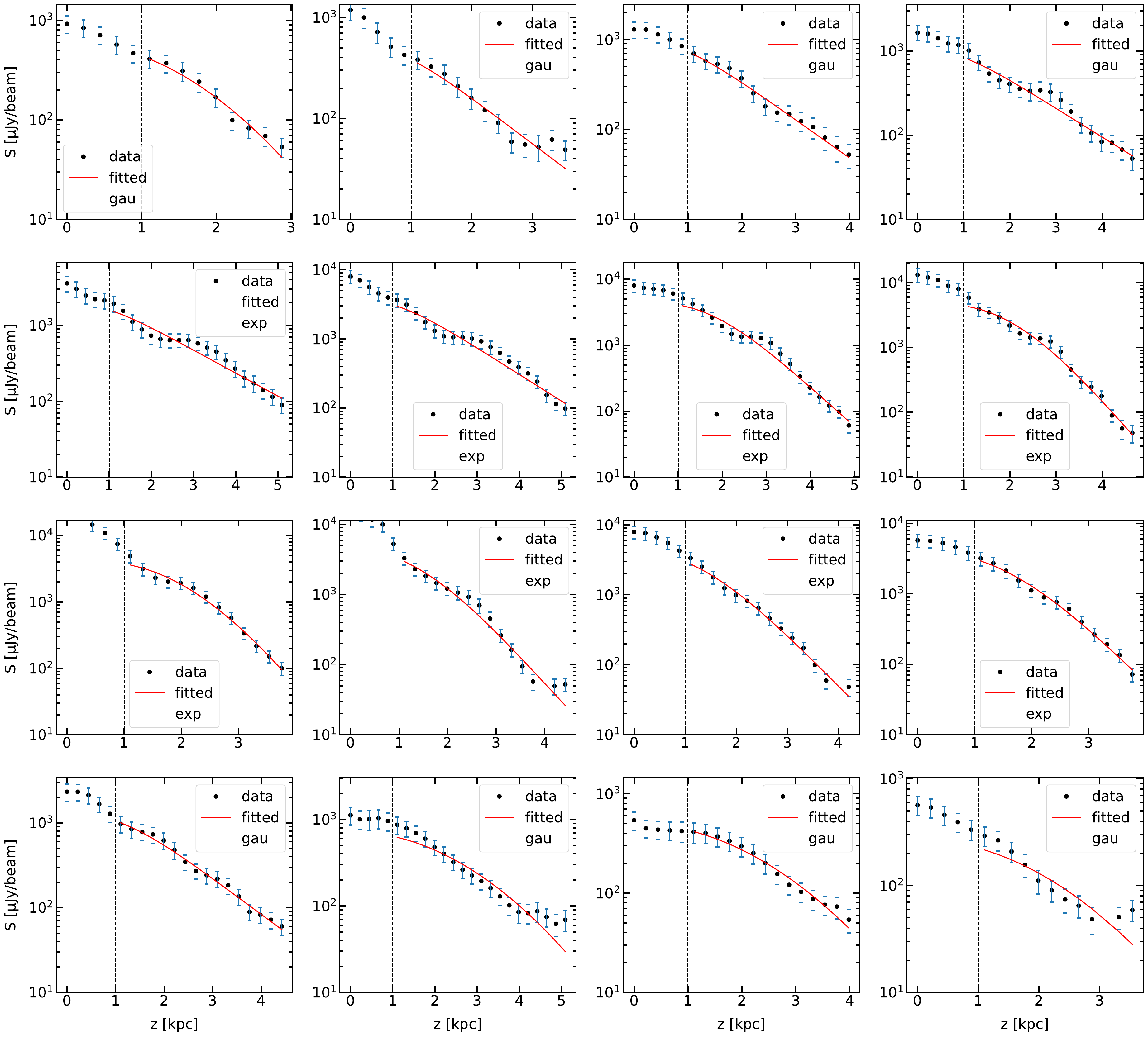}
    \caption{Vertical profiles derived from the source-subtracted ASKAP synchrotron emission intensity image at the original resolution of $13\arcsec$. From left to right and top to bottom are the strips from left to right in the northern part ($z>0$) in Fig.~\ref{fig:cover}. The red solid line represents the fit to data points located in the halo region to the right of the dashed line and above a $3\,\sigma$ threshold, using either an exponential (exp) or a Gaussian (gau) function.}
\label{fig:intensity-fit-north}
\end{figure*}

\begin{figure*}[!h]	\includegraphics[width=\linewidth]{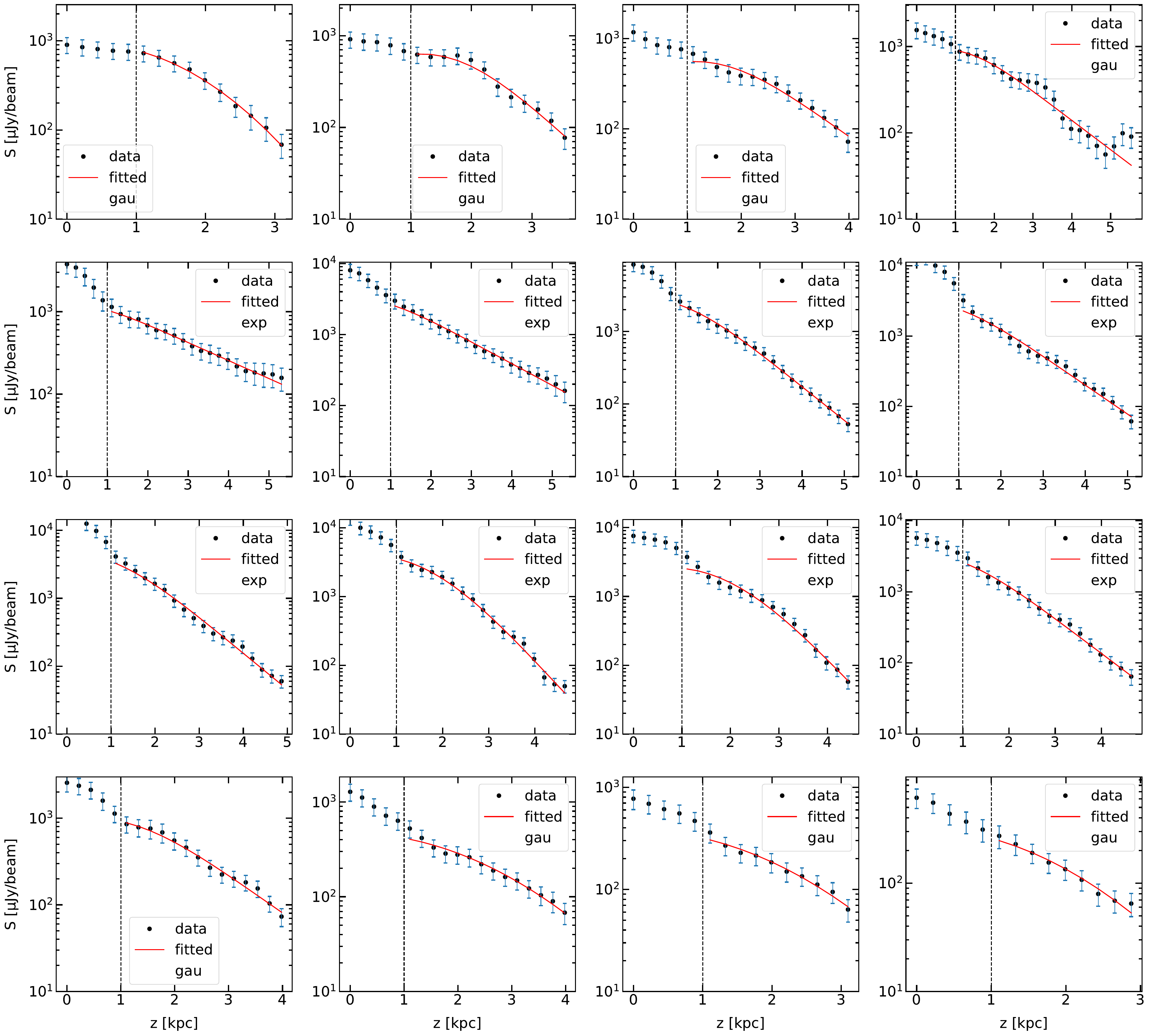}
    \caption{The same as Fig.~\ref{fig:intensity-fit-north} but for the strips in the southern part ($z<0$) in Fig.~\ref{fig:cover}.}
\label{fig:intensity-fit-south}
\end{figure*}

\end{appendix}
\end{document}